\documentclass[longauth]{aa}

\usepackage{multicol}
\usepackage{natbib}
\bibpunct{(}{)}{;}{a}{}{,}

\usepackage{pdfpages}
\usepackage{longtable}
\usepackage{multirow}

\newcommand{\msun}{\mbox{$M_\odot$}}
\newcommand{\lsun}{\mbox{$L_\odot$}}
\newcommand{\hii}{H\mbox{\sc ~ii} }

\def\arcsec{\hbox{$^{\prime\prime}$}}
\def\degree{\ensuremath{^\circ}}

\begin{document}

   \title{The earliest phases of high-mass star formation, as seen in NGC~6334 by 
   \emph{Herschel}\thanks{
        \emph{Herschel} is an ESA space observatory with science instruments provided by 
        European-led Principal Investigator consortia and with important participation from NASA.
        }-HOBYS\thanks{
        Tables~\ref{tab_getsources_coordinates}--\ref{tab_getsources_1200}, Figs.~\ref{MDC_1}--\ref{MDC_43}, 
and Appendices A, B, and C are only available in electronic form at http://www.aanda.org. 
Catalogs built from Tables~\ref{tab_getsources_coordinates}--\ref{tab_getsources_1200} can be queried from the CDS soul67.tex
via anonymous ftp to cdsarc.u-strasbg.fr (130.79.128.5) or via http://cdsweb.u-strasbg.fr/cgi-bin/qcat?J/A+A/.}
        }

   \titlerunning{The earliest phases of high-mass star formation in NGC~6334}

\author{J. Tig\'e\inst{1},
F. Motte\inst{2,3},
D. Russeil\inst{1},
A. Zavagno\inst{1},
M. Hennemann\inst{4},
N. Schneider\inst{5,6},
T. Hill\inst{7},
Q. Nguyen Luong\inst{8,9},
J. Di Francesco\inst{10,11}, 
S. Bontemps\inst{5},
F. Louvet\inst{12},
P. Didelon\inst{3},
V. K\"onyves\inst{3},
Ph. Andr\'e\inst{3},
G. Leuleu\inst{1},
J. Bardagi\inst{1},
L. D. Anderson\inst{13,14}, 
D. Arzoumanian\inst{3}, 
M. Benedettini\inst{15},
J.-P. Bernard\inst{16}, 
D. Elia\inst{15},
M. Figueira\inst{1},
J. Kirk\inst{17},
P. G. Martin\inst{18},
V. Minier\inst{3},
S. Molinari\inst{15},
T. Nony\inst{2},
P. Persi \inst{15},
S. Pezzuto\inst{15},
D. Polychroni\inst{19},
T. Rayner\inst{20},
A. Rivera-Ingraham\inst{21},
H. Roussel\inst{22},
K. Rygl\inst{23},
L. Spinoglio\inst{13},
G. J. White\inst{24, 25}
}

\authorrunning{J. Tig\'e et al.}

\institute{Aix Marseille Univ., CNRS, LAM, Laboratoire d'Astrophysique de Marseille, Marseille, France\\
\email{jeremy.tige@lam.fr} 
\and Universit\'e Grenoble Alpes, CNRS-INSU, Institut de Plan\'etologie et d'Astrophysique de Grenoble, F-38000 Grenoble, France
\and Laboratoire AIM Paris-Saclay, CEA/IRFU - CNRS/INSU - Universit\'e Paris Diderot, Service d'Astrophysique, B\^at. 709, CEA-Saclay, 91191, Gif-sur-Yvette Cedex, France
\and Max-Planck-Institut f\"ur Astronomie, K\"onigsstuhl 17, 69117, Heidelberg, Germany
\and Laboratoire d'Astrophysique de Bordeaux, Univ. Bordeaux, CNRS, B18N, all\'ee G. Saint-Hilaire, 33615 Pessac, France
\and I. Physik. Institut, University of Cologne, 50937 Cologne, Germany
\and Joint ALMA Observatory, 3107 Alonso de Cordova, Vitacura, Santiago, Chile
\and Korea Astronomy and Space Science Institute, 776 Daedeok daero, Yuseoung, Daejeon 34055, Korea
\and NAOJ Chile Observatory, National Astronomical Observatory of Japan, 2-21-1 Osawa, Mitaka, Tokyo 181-8588, Japan
\and National Research Council Canada, Herzberg Institute of Astrophysics, 5071 West Saanich Road, Victoria, BC V9E 2E7, Canada
\and Department of Physics and Astronomy, University of Victoria, P.O. Box 355, STN CSC, Victoria, BC V8W 3P6, Canada
\and Department of Astronomy, Universidad de Chile, Las Condes, Santiago, Chile
\and Department of Physics and Astronomy, West Virginia University, Morgantown, WV 26506, USA
\and National Radio Astronomy Observatory, P.O. Box 2, Green Bank, WV 24944, USA 
\and INAF-Istituto di Astrofisica e Planetologia Spaziali, via Fosso del Cavaliere 100, I-00133 Rome, Italy
\and Universit\'e de Toulouse, UPS-OMP, CNRS, IRAP, 31028, Toulouse Cedex 4, France
\and Jeremiah Horrocks Institute, University of Central Lancashire, Preston, Lancashire, PR1 2HE, UK
\and Canadian Institute for Theoretical Astrophysics, University of Toronto, 60 St. George Street, Toronto, ON M5S 3H8, Canada
\and Departamento de Fisica, Universidad de Atacama, Copiapo, Chile
\and Cardiff School of Physics and Astronomy, Cardiff University, Queen's Buildings, The Parade, Cardiff, Wales, CF24 3AA, UK
\and European Space Astronomy Centre ESA/ESAC, 28691 Villanueva de la Ca\~nada, Madrid, Spain
\and Institut d'Astrophysique de Paris, Sorbonne Universit\'es, UPMC Univ. Paris 06, CNRS UMR 7095, 75014 Paris, France
\and INAF-IRA, Via P. Gobetti 101, 40129 Bologna, Italy
\and Department of Physics and Astronomy, The Open University, Walton Hall, Milton Keynes, MK7 6AA, UK
\and RAL Space, STFC Rutherford Appleton Laboratory, Chilton, Didcot, Oxfordshire, OX11 0QX, UK
}

   \date{ }

 \abstract
  {}
  {To constrain models of high-mass star formation, the \emph{Herschel}/HOBYS key program 
  aims at discovering massive dense cores (MDCs) able to host the high-mass analogs of low-mass prestellar cores, which have been searched for over the past decade. We here focus on 
  NGC~6334, one of the best-studied HOBYS molecular cloud complexes.}
   %
  {We used \emph{Herschel} PACS and SPIRE $70-500~\mu$m images of the NGC~6334 complex 
  complemented with (sub)millimeter and mid-infrared data. 
  We built a complete procedure to extract $\sim$0.1~pc dense cores 
  with the \emph{getsources} software, which simultaneously measures their far-infrared to millimeter fluxes. 
  We carefully estimated the temperatures and masses of these dense cores from their spectral energy distributions (SEDs). We also identified the densest pc-scale cloud structures of NGC~6334, one 2~pc~$\times$~1~pc ridge and two 0.8~pc~$\times$~0.8~pc hubs, with volume-averaged densities of $\sim$$10^5$~cm$^{-3}$.
  } 
  %
  {A cross-correlation with high-mass star formation signposts suggests a mass threshold of $75~\msun$ for MDCs in NGC~6334. MDCs have temperatures of $9.5-40$~K, masses of $75-1000~\msun$, 
  and densities of $1\times 10^5-7\times 10^7$~cm$^{-3}$. Their mid-infrared emission is used 
  to separate 6 IR-bright and 10 IR-quiet protostellar MDCs while their $70~\mu$m emission strength, 
  with respect to fitted SEDs, helps identify 16 starless MDC candidates. 
  The ability of the latter to host high-mass prestellar cores is investigated here and remains questionable.
  An increase in mass and density from the starless to the IR-quiet and IR-bright phases suggests that 
the protostars and MDCs simultaneously grow in mass. 
The statistical lifetimes of the high-mass prestellar and protostellar core phases, 
estimated to be $1-7\times~10^4$~yr and at most $3\times~10^5$~yr respectively, 
suggest a dynamical scenario of high-mass star formation.} 
   {The present study provides good mass estimates for a statistically significant sample, covering the earliest phases of high-mass star formation.
High-mass prestellar cores may not exist in NGC~6334, favoring a scenario presented here, which simultaneously forms clouds and high-mass protostars.}

    \keywords{dust -- ISM: clouds -- stars: formation -- submillimeter: ISM -- stars: protostars -- ISM: individual objects: NGC~6334}
   \maketitle

\section{Introduction}

High-mass (O-B3 type, $M_\star>8$~\msun) stars form in massive dense cores 
(MDCs, $\sim$0.1~pc and $>10^5$~cm$^{-3}$, see \citealt{Motte07}) by the accretion of gas onto stellar embryos \citep[e.g.,][]{Beuther04,Duarte13}. 
The details of this process are under investigation \citep[see recent reviews by, e.g.,][]{Tan14, Krumholz15,Motte17} 
and an increasing number of studies have suggested that they form through dynamical processes initiated by cloud formation \citep[e.g.,][]{Schneider10b,Csengeri11a,Peretto13,Nguyen13}. 

Two main theoretical scenarios have been proposed to explain the formation of high-mass stars: \emph{(1)} powerful accretion driven by a high degree of turbulence (e.g., \citealt{McKee02}; \citealt{Krumholz07}) or \emph{(2)} colliding flows initiated by competitive accretion or cloud formation (e.g., \citealt{Bonnell06}; \citealt{Hartmann12}). These two scenarios lead to distinct characteristics for the initial stages of high-mass star formation. The first model implies the existence of high-mass prestellar cores, which display high degrees of micro-turbulence. In the second model, high-mass prestellar cores never develop. When favorably located at the centers of massive reservoirs of gas, low-mass prestellar cores turn into protostars, attracting gas from further distances and eventually become high-mass protostars. Therefore, one difference between the two scenarios is the presence or absence of high-mass prestellar cores.

A decade ago, \cite{Motte07} made the first unbiased statistical survey of MDCs. They did not find any bona-fide starless MDCs and proposed that this phase would be transitory if existent. Since then, ground-based studies have only been able to identitfy a handful of starless MDCs in molecular complexes \citep[e.g.,][]{Russeil10,Butler12} and only a few hundred starless clumps in the Milky Way \citep{Ginsburg12,Tackenberg12,Csengeri16,Svoboda16}. A few high-resolution studies have been performed with (sub)millimeter interferometers with the aim of identifying individual prestellar cores and protostars within MDCs. These attempts revealed high-mass cores that are protostellar in nature (e.g., \citealt{Duarte13} within protostellar MDCs) or low-mass prestellar fragments (e.g., \citealt{Tan13} within starless clumps). The only prestellar core candidate of \cite{Duarte14} (CygX-N53-MM2, $25~\msun$ mass within 0.025~pc) and the single case found by \cite{Wang14} (G11P6-SMA1, $\sim$$30~\msun$ mass within 0.02~pc) are amongst the very few good examples to date of the massive prestellar cores predicted by \cite{McKee02}. This fact alone argues for a high-mass  star formation scenario with no high-mass prestellar core phase \citep[see][]{Motte17} but the debate continues.

As a first step to resolving the debate, it is crucial to find and characterize MDCs that are potential sites for the formation of high-mass stars, in  molecular cloud complexes. The \emph{Herschel} imaging survey of OB Young Stellar Objects (HOBYS) is dedicated to this purpose  \citep[][see http://hobys-herschel.cea.fr]{Motte10}. It is the first systematic survey of a complete sample of nearby (within 3~kpc) high-mass star progenitors at 0.1~pc scales \citep[see initial catalogs in][]{Nguyen11,Fallscheer13}.   
At five wavelengths from 70~$\mu$m to 500~$\mu$m, the HOBYS program targeted ten molecular cloud complexes forming OB-type stars, amongst which NGC~6334-6357 is one of the most massive \citep[$\sim$7$\times$10$^5~\msun$, see][]{Motte17}.

NGC~6334-6357 is a molecular cloud complex that belongs to the Sagittarius-Carina arm and lies at a 1.75~kpc distance from the Sun (\citealt{Matthews08}; \citealt{Russeil12}). Its central part hosts a 15~pc-long filament \citep{Russeil13,Schneider15}, whose highest density parts consist of a ridge according to HOBYS terminology \citep{Hill11}. NGC~6334 is itself a very active high-mass-star-forming region, as advocated by its numerous \hii regions, maser sources, and molecular outflows \citep[see][]{Loughran86,Persi08,Carral02}. \cite{Willis13} identified approximately 2000 young stellar objects with \emph{Spitzer} and found a heavy concentration of protostars along the NGC~6334 ridge, suggesting a mini-starburst event (in agreement with the ridge definition, see \citealt{Nguyen11}). Using a 1.2~mm continuum map, \cite{Russeil10} extracted in NGC~6334 a sample of 11 clumps, including the well-known sources I and I(N) \citep[e.g.,][]{Gezari82}, however, only one starless candidate was identified. 

In this paper, we search the NGC~6334 complex for MDCs harboring young stellar objects at the earliest phases of high-mass-star formation, that is, high-mass young protostars and high-mass prestellar cores. 
From \emph{Herschel} and complementary images described in Sect.~\ref{section_observations}, 
we extracted ridges, hubs, and dense cores (see Sect.~\ref{section_extraction}). We describe the SED characterization of these cores in Sect.~\ref{section_dense_cores}. 
Section~\ref{section_massive_dense_cores} provides a complete sample of 0.1~pc MDCs with robust mass estimates. 
Finally, in Sect.~\ref{section_discussions}, we discuss the existence of starless MDCs and prestellar cores, 
the lifetimes of high-mass star formation phases, and the favored scenario for high-mass-star formation.

\begin{figure*}[]
\begin{center}
\includegraphics[width=18cm,angle=0,]{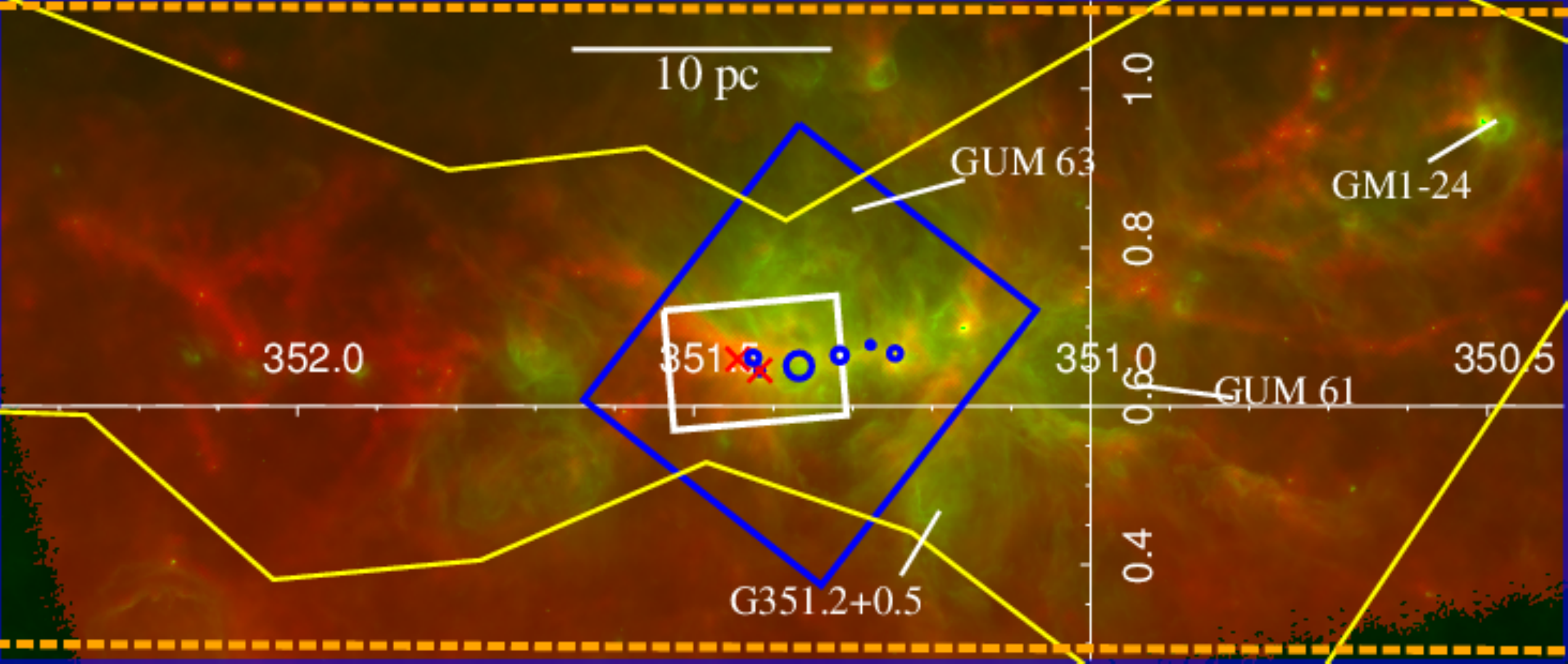} 
\end{center}
\caption{\emph{Herschel}-color image of NGC~6334: 70~$\mu$m (green, $5.9\arcsec$ resolution) and column density (red, $36.6\arcsec$ resolution).
Blue circles are young \hii regions (seen as compact radio sources; \citealt{Persi10}) 
while crosses are I and I(N), the two highest column density peaks of NGC~6334. 
The yellow polygon represents the mapped area of SIMBA-1200~$\mu$m observations, 
the white box represents the extent of SCUBA2 450/850~$\mu$m observations, 
the blue box highlights the highest-density parts of NGC~6334 
presented in Sect.~\ref{section_ridge} and Fig.~\ref{ridge_hubs}, and dashed orange lines outline 
the Galactic latitude range we associated with NGC~6334 
(0.3\degree $<$ latitude $<$ 1.1\degree) 
to build the HOBYS catalog.}
\label{F70_coldens}
\end{figure*}

\section{Observations}
\label{section_observations}

\subsection {Herschel observations, data reduction, and column density images}
\label{section_HerschelObs}

NGC~6334 was observed by the \emph{Herschel} space observatory with the PACS (\citealt{Poglitsch10}) and SPIRE (\citealt{Griffin10}) instruments\footnote{
        Instrument parameters and calibration are given in the PACS and SPIRE observers manuals. See
        \url{http://Herschel.esac.esa.int/Docs/PACS/html/pacs_om.html} for PACS and \url{http://Herschel.esac.esa.int/Docs/SPIRE/html/spire_handbook.html} for SPIRE.}
as part of the HOBYS (\citealt{Motte10}) key program (OBSIDs: 1342204421 and 1342204422). Data were taken in five bands: 70~$\mu$m and 160~$\mu$m for PACS at FWHM resolutions or half power beam widths (HPBWs) of 5.9\arcsec~and 11.7\arcsec, respectively, and 250~$\mu$m, 350~$\mu$m, and 500~$\mu$m for SPIRE at FWHM resolutions of 18.2\arcsec, 24.9\arcsec, and 36.3\arcsec, respectively (see Table~\ref{tab_images_properties}). Observations were performed in parallel mode, using both instruments simultaneously, with a scanning speed of 20\arcsec/s. Two perpendicular scans were taken, quoted as nominal and orthogonal. The size of the observed field is 2.0\degree~$\times$~0.5\degree, which corresponds to 60~pc$~\times~$15~pc at a distance of 1.75~kpc. 
\emph{Herschel} images of NGC~6334 are thus sensitive to cloud structures with $\sim$0.1~pc to $\sim$30~pc typical sizes, which correspond to MDCs, clumps, and clouds following definitions by, for example, \cite{Motte17}.

\begin{table}[h]
\begin{center}
\caption{Main observational parameters of the present paper}
\label{tab_images_properties}
\begin{tabular}{| c | c | c| c |}
\hline
Data                        & $\lambda$        & HPBW      &   rms ($1\sigma$)      \\
                             & ($\mu$m)        &  (arcsec) &   (mJy)     \\
\hline
\multirow{2}{*}{\emph{Herschel}/PACS}   & 70   & 5.9       &   90      \\                            & 160              & 11.7      &   650       \\
\multirow{3}{*}{and SPIRE}  & 250              & 18.2      &   1100        \\
                            & 350              & 24.9      &   860       \\
                            & 500              & 36.3      &   680        \\  
\hline
\hline
\multirow{2}{*}{JCMT/SCUBA-2} & 450            &  8.5      &   720        \\
                            & 850              &  15.0     &   470       \\
\hline
APEX/LABOCA                 & 870              & 19.2      &   90         \\
\hline
SEST/SIMBA                  & 1200             & 24.0      &   10          \\
\hline
\hline
\multirow{3}{*}{\emph{Spitzer}/IRAC}  & 3.6    & 1.5       &   0.6          \\
                            & 4.5              & 1.7       &   0.6          \\
                            & 5.8              & 1.7       &   0.7         \\
\multirow{2}{*}{and MIPS}   & 8                & 2.0       &   5        \\
                            & 24               & 6.0       &   5         \\ 
\hline                             
\emph{WISE}                 & 22               & 12.0      &   10         \\    
\hline                         
\emph{MSX}                  & 21               & 18.3      &   60         \\
\hline  
\hline
\end{tabular}
\end{center}
\end{table}

We reduced \emph{Herschel} data using the \emph{Herschel} 
Interactive Processing Environment (HIPE, \citealt{Ott10})\footnote{
        HIPE has been jointly developed by the Herschel Science Ground Segment Consortium, consisting of ESA, the NASA Herschel Science Center, and the HIFI, PACS, and SPIRE consortia.} 
software, version 10.0.2751. Versions 7.0 onwards contain a module which significantly removes the striping effects that have been observed in SPIRE maps produced with previous HIPE versions. SPIRE nominal and orthogonal maps were separately processed and subsequently combined and reduced for de-striping, relative gains, and color correction with HIPE. PACS maps were reduced with HIPE up to Level~1 and, from there up to  their final version (Level~3) using Scanamorphos v21.0 (\citealt{Roussel13}). Final PACS images are shown in Figs.~\ref{appendix_herschel70}~and~\ref{appendix_herschel160}.

To correct for a few saturated pixels observed in SPIRE images, individual SPIRE mappings of 
small fields were performed with a different gain setting. The original map with saturated pixels and the small field were then combined following the method described in Appendix B of \cite{Nguyen13} to produce full images without saturation (see Figs.~\ref{appendix_herschel250}--\ref{appendix_herschel500}). Calibration for all \emph{Herschel} maps was done using the offsets obtained following the procedure described in \cite{Bernard10}. This method makes use of IRIS (IRAS survey update) and \emph{Planck} HFI DR2 data (\citealt{Planck11}) to predict an expected flux, which is compared to the median values from PACS and SPIRE maps, smoothed to the adequate resolution. 

We built column density maps both at the $36.3\arcsec$ and $18.2\arcsec$ resolutions of SPIRE 500~$\mu$m and 250~$\mu$m data (see Figs.~\ref{F70_coldens} and \ref{appendix_nh2}). The procedure used to construct the $36.3\arcsec$ resolution image uses the SED fit method fully described in \cite{Hill11,Hill12}. The one used to construct the high-resolution column-density map is based on a multi-scale decomposition of the imaging data described in detail in \cite{Palmeirim13}. We used a dust opacity law similar to that of \cite{Hildebrand83} but with $\beta=2$ instead of $\beta = 1$ and assumed a gas-to-dust ratio of 100: $\kappa_{0}= 0.1 \times (\rm \nu / 1000\,GHz)^{2}$~cm$^{2}$~g$^{-1}$.

Figure~\ref{F70_coldens} presents a combined image of \emph{Herschel} 70~$\mu$m and column density images of the NGC~6334 molecular complex. While the column density is concentrated along a filamentary structure, described in \cite{Russeil12} for example, the 70~$\mu$m emission traces warm dust toward star-forming sites such as \hii regions and young stars.

\subsection {Ancillary data}
\label{section_ancillary}

We complemented our \emph{Herschel} observations with submillimeter and mid-infrared data (see Table~\ref{tab_images_properties}). Multiple SCUBA-2/JCMT 450~$\mu$m and 850~$\mu$m images \citep{Holland13} toward the I and I(N) sources of NGC~6334 were taken from the JCMT Science Archive\footnote{
        SCUBA-2 data from the JCMT Science Archive are available at \url{http://www.cadc-ccda.hia-iha.nrc-cnrc.gc.ca/en/jcmt/}.}
(Proposal ID: JCMTCAL).
We reduced the data following the standard pipeline using the makemap command of the SMURF software (\citealt{Chapin13}), bundled with the gaia package. We made two mosaics from all calibrated 450~$\mu$m and 850~$\mu$m data 
and obtained a $0.2\degree\times0.2\degree$ coverage (see Figs.~\ref{F70_coldens} and \ref{appendix_450}--\ref{appendix_850}). The angular resolutions were $8.5\arcsec$ at 450~$\mu$m and $15.0\arcsec$ at 850~$\mu$m. 
The ATLASGAL survey\footnote{
        The APEX Telescope Large Area Survey of the GALaxy (ATLASGAL) provides 870~$\mu$m images of the inner Galactic plane. Detailed information and reduced images are available at \url{http://www3.mpifr-bonn.mpg.de/div/atlasgal/}.}
\citep{Schuller09}, using the LABOCA/APEX camera at 870~$\mu$m itself covered the NGC~6334 molecular cloud with $19.2\arcsec$ resolution. 
Dedicated SIMBA/SEST 1.2~mm observations presented by \cite{Munoz07} and \cite{Russeil10} also covered the main filamentary structures of NGC~6334, with $24\arcsec$ angular resolution (see coverage in Fig.~\ref{F70_coldens}). 

At mid-infrared wavelengths, NGC~6334 was imaged by \emph{Spitzer}/IRAC and MIPS at $3.6-24~\mu$m with $1.5\arcsec-6\arcsec$ resolutions, as part of the GLIMPSE and MIPSGAL surveys\footnote{
         Detailed information on the GLIMPSE and MIPSGAL surveys of the inner Galactic plane with \emph{Spitzer} and reduced images are available at \url{http://irsa.ipac.caltech.edu/data/SPITZER/}.}
\citep{Benjamin03,Carey09}. 
It was also imaged by the two mid-infrared space observatories\footnote{
        Detailed information on the \emph{WISE} full-sky and \emph{MSX} Galactic Plane surveys and their reduced images are available at \url{http://irsa.ipac.caltech.edu/Missions/wise.html} and \url{http://irsa.ipac.caltech.edu/Missions/msx.html}, respectively.} 
that are called \emph{WISE}, at four bands and notably at 22~$\mu$m with an angular resolution of $12\arcsec$, and \emph{MSX}, in four bands, including 21.3~$\mu$m with spatial resolution of $18.3\arcsec$ \citep{Wright10,Price01}.

\section{Identification of dense cloud structures} 
\label{section_extraction}

We investigated the cloud structure of NGC~6334 from parsec-scale clumps (see Sect.~\ref{section_ridge}) to 0.1~pc-scale dense cores (see Sect.~\ref{section_extraction_getsources}).

\subsection{The densest pc-scale clumps: ridges and hubs}
\label{section_ridge}

\begin{figure}[]
\begin{center}
\includegraphics[width=9cm]{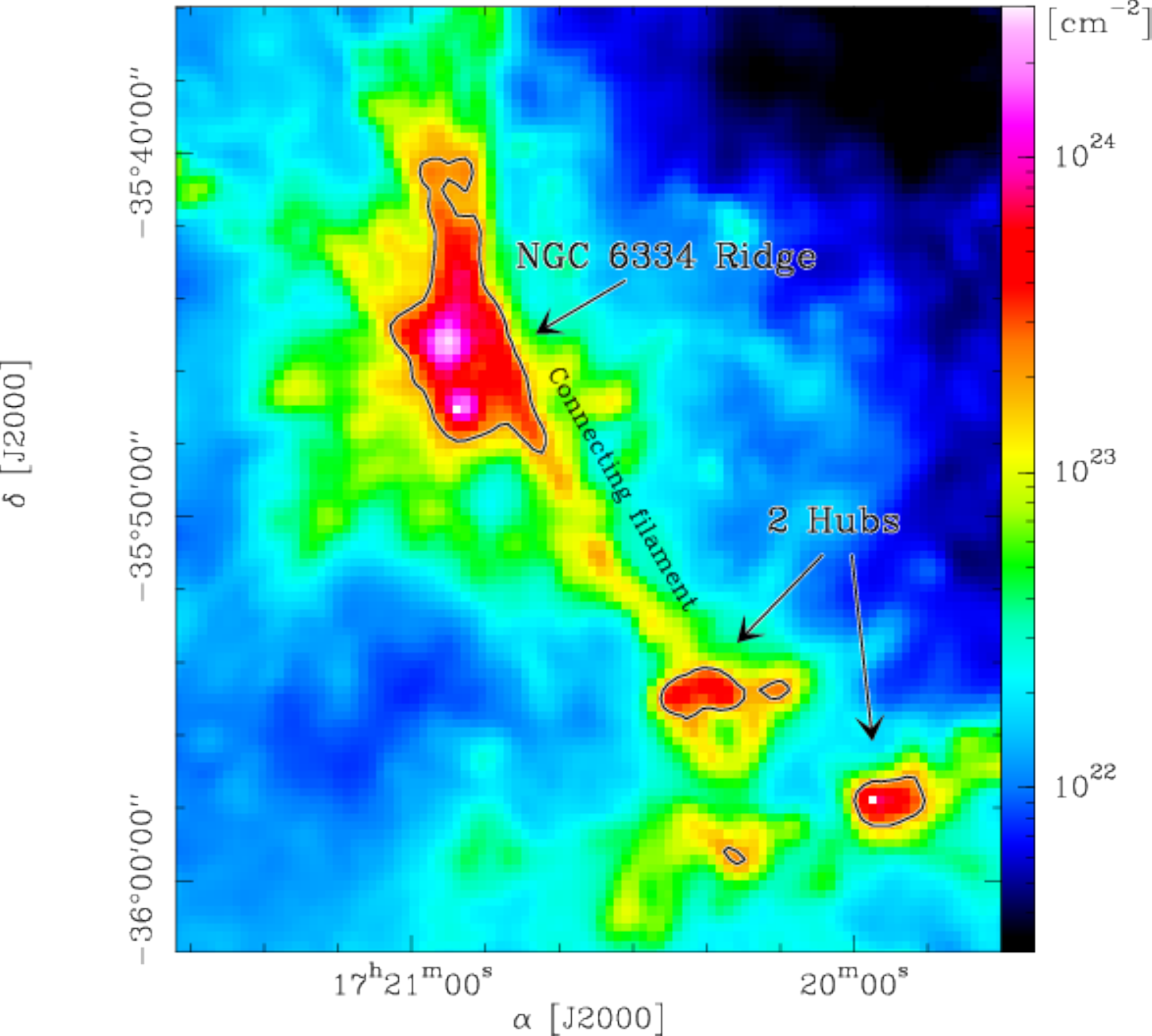} 
\end{center}
\caption{NGC~6334 highest density clumps delineated in the column density image (color and contours) derived from \emph{Herschel} images. The contour is the $2\times 10^{23}$~cm$^{-2}$ level, outlining the NGC~6334 ridge and hubs.}
\label{ridge_hubs}
\end{figure}

Among parsec-scale clumps, the densest ones qualify as ridges and hubs. 
\cite{Hill09} first proposed to outline ridges, that is, filamentary clumps forming high-mass stars, with a $10^{23}$~cm$^{-2}$ column density lower limit. In contrast to threshold claims by, for example, \cite{Krumholz08}, \cite{Hill09} did not find any objective reason to set a definite threshold for a clump to be able to form high-mass stars. 
The probability distribution function (PDF) of column density in NGC~6334, studied by \cite{Schneider15}, then revealed a deviation\footnote{
The $1\pm 0.5\times 10^{23}$~cm$^{-2}$ threshold measured in the column density map of \cite{Schneider15} becomes $2\pm 1\times 10^{23}$~cm$^{-2}$ in the image of Fig.~\ref{ridge_hubs}. This factor 2 increase arises from the change in the SED fit procedure from a linear to a logarithmic fit, which allows a better measurement of the temperature and column density of clumps and filaments strongly heated from outside. The maximum increase, 2.5, is observed along the filament connecting the NGC~6334 ridge and main hub studied in \cite{Andre16}.
} 
from the first power-law tail at $\sim$2$\times 10^{23}$~cm$^{-2}$, interpreted as linked to (high-mass) star formation. Figure~\ref{ridge_hubs} highlights the highest-density parts of NGC~6334 on a zoomed column density image. 
The $2\times 10^{23}$~cm$^{-2}$ contours outline the NGC~6334 ridge spanning a 2~pc~$\times$~1~pc area and containing the I and I(N) sources as well as two smaller hubs with $\sim$0.8~pc diameters. A narrow filament discussed at length in \cite{Matthews08} and \cite{Andre16} connects the ridge and largest hub (see Fig.~\ref{ridge_hubs}).

We used a $2\times 10^{23}$~cm$^{-2}$ column density background and Eqs.~1--2 of \cite{Nguyen13} to derive a mass of $\sim$15\,000$~\msun$ and a density of $2.5\times10^5$~cm$^{-3}$ for the NGC~6334 ridge. With the same assumptions, the NGC~6334 hubs have masses of $\sim$$2\,200$ and $\sim$$2\,800~\msun$ and densities around $10^5$~cm$^{-3}$.

\subsection{Compact sources}
\label{section_extraction_getsources}

Compact sources were extracted using \emph{getsources} (v1.131121), 
a multi-scale, multi-wavelength source-extraction algorithm \citep{Menshchikov12,Menshchikov13}. 
Extraction is based on the decomposition of images across finely spaced spatial scales. 
Each of these single-scale images is cleaned of noise and background emission by iterating on the appropriate cut-off levels. 
Images are then re-normalized and summed over all wavelengths into a combined
single-scale detection image. The main advantage of this algorithm is its multi-wavelength design. 
Namely, using the same combined detection image across all wavelengths eliminates the need for matching multiple catalogs obtained at different wavelengths and corresponding angular resolutions.
Also, the generally non-Gaussian noise spectrum of the initial maps becomes more Gaussian 
on single-scale images, improving detection of the faintest objects and the deblending of sources in crowded regions. The cleaning of background emission, fully explained in \cite{Menshchikov12}, starts from global image flattening and is iteratively refined to determine precisely the local background of each individual core.

Following the procedure established by the HOBYS consortium 
and fully described in this paper, the detection of compact sources 
is based on \emph{Herschel} data plus a high-resolution column-density map. 
The PACS-160~$\mu$m and SPIRE-250~$\mu$m images are temperature-corrected to minimize the effects 
of temperature gradients. The departure to an intensity map arising from a constant 
temperature (17~K) is used to scale the original PACS-160~$\mu$m and SPIRE-250~$\mu$m images, with the goal of
emphasizing the cold structures. The procedure is fully explained in Appendix~C and images are shown in 
Figs.~\ref{appendix_161}-\ref{appendix_temperature_corrected_multi}.
The purpose of temperature-corrected maps at 160~$\mu$m and 250~$\mu$m is to weaken 
the impact of external heating, especially at the periphery of photo-dissociation regions (PDRs) 
to help \emph{getsources} focus on cold peaks rather than temperature peaks created by 
local heating (see Fig.~\ref{appendix_temperature_corrected_multi}).

During the detection step, we simultaneously applied \emph{getsources} toall images to define a catalog of sources with a unique position. 
At the measurement stage, we use the original (not temperature-corrected) \emph{Herschel} maps 
from 70~$\mu$m to 500~$\mu$m plus all available submillimeter maps 
listed in Sect.~\ref{section_ancillary}, that is, the 450~$\mu$m and 850~$\mu$m SCUBA-2 mosaics and
the 870~$\mu$m LABOCA and 1200~$\mu$m SIMBA images. 
These measurements go beyond simple aperture photometry
since they are done together with background subtraction and
deblending of overlapping sources. The resulting source catalog contains a monochromatic significance index (Sig$_{\rm mono}$), peak and 
integrated fluxes (with errors), full width at half maximum (FWHM) major and minor sizes, 
and the position angle of the elliptical footprint
for each extracted source in each far-infrared to submillimeter band. 
A subset of these catalogs is given in Tables~\ref{tab_getsources_70}--\ref{tab_getsources_1200}.

The \emph{Herschel}-HOBYS imaging encompassing NGC~6334 extends 
from $0\degree$ to $+1.3\degree$ in Galactic latitude. 
On the basis of H$\alpha$, radio continuum, and H{\small I} surveys, \cite{Russeil16} defined 
the velocity structure and thus the spatial area associated with the complex. 
Since we cannot confirm the association of NGC~6334 
with gas located at the edges of the complex, 
we masked areas between +0.0\degree$\:$\,and +0.3\degree$\:$\,and 
between +1.1\degree$\:$\,and +1.3\degree$\:$\,in Galactic latitude (see Fig.~\ref{F70_coldens}). 
In the following, we build the first-generation catalog of NGC~6334 MDCs from the corresponding \emph{getsources} catalog of 4733 sources found within the unmasked area. 

Sources extracted by \emph{getsources} are far-infrared fluctuations among which compact dense cores persist across wavelengths. Intermediate- to high-mass dense cores\footnote{
0.1~pc dense cores are larger-scale cloud structures than 0.02~pc low-mass (prestellar or protostellar) cores assumed to form individual solar-type stars \citep{Motte98,Konyves15}. 
} 
are expected to be cold ($10-30$~K), dense cloud fragments of approximately 0.1~pc in size and several times 1-100~\msun~in mass. Given their expected densities, their thermal emission should mainly be optically thin for wavelengths larger than 100~$\mu$m. 

Dense cores have spectral energy distributions (SEDs) peaking between 100~$\mu$m and 300~$\mu$m 
and should therefore have reliable flux measurements 
at either \emph{Herschel}-160~$\mu$m or \emph{Herschel}-250~$\mu$m 
or, more generally, at both. We define reliable flux measurements as those which have 
1) signal-to-noise ratios greater than 2 for both the peak and integrated fluxes and monochromatic significance index greater than 5,
and
2) deconvolved sizes less than 0.3~pc and aspect ratios less than 2.
Sources in the \emph{getsources} catalog which do not follow these criteria are either 
PDR features whose 70~$\mu$m emission is not associated with density peaks, filaments with ellipticity $>$2, or even $\sim$1~pc clumps. This step selects 2063 sources out of the 4733 initial \emph{getsources} sources. Roughly 90\%, 6\%, and 4\% of the sources excluded by these criteria are 70~$\mu$m-only sources, filaments, and clumps, respectively. 

160~$\mu$m is the highest-resolution \emph{Herschel} wavelength ($\sim$12\arcsec)
that should correctly represent the optically thin dust emission of dense cores. 
In some cases, however, thermal emission at 160~$\mu$m can be contaminated by emission of small grains heated by PDRs at the peripheries of dense cores. 
One direct consequence of this effect is that deconvolved sizes 
can be larger at 160~$\mu$m than at 250~$\mu$m. 
The size at 160~$\mu$m or, if its flux is unreliable or contaminated, the size at 250~$\mu$m 
was chosen for reference to estimate the physical size of the source. 
Of the dense cores, 35\%  have their reference wavelength set to 160~$\mu$m. 
Another 40\% of the dense cores have reliable measurements at 160~$\mu$m, but with large deconvolved sizes their reference wavelength is set to 250~$\mu$m. The remaining 25\% of the dense cores only 
have reliable measurements at 250~$\mu$m, which thus is their reference wavelength.  

A minimum number of three reliable fluxes at wavelengths larger than 100~$\mu$m are necessary to correctly constrain the SEDs of dense gas fragments. We therefore enforced a minimum of three reliable flux measurements: 1) one at either \emph{Herschel}-160~$\mu$m 
or \emph{Herschel}-250~$\mu$m, which we call the reference wavelength (see above), 
2) accompanied by  at least a second \emph{Herschel} flux measurement 
at a wavelength larger than 100~$\mu$m, and 3) a third flux measurement 
taken at $\lambda>100$~$\mu$m with either \emph{Herschel}/SPIRE, 
APEX/LABOCA, or SEST/SIMBA. 
In NGC~6334, this step removes more than half of the remaining \emph{getsources} sources, 
leading to a sample of 654 dense cores. 
Sources excluded by this process are at the limit 
of our detection level in \emph{Herschel}-SPIRE wavelengths and 
thus correspond to dense cores with very low mass.

\section{Physical characterization of dense cores}
\label{section_dense_cores}
\subsection{Coherent SEDs: fluxes scaled to the same aperture}
\label{section_scaling}

We aim to construct SEDs with fluxes measured within one aperture of the same size. 
Since the \emph{Herschel} beam increases with wavelength, 
source sizes and thus the flux in an aperture often increase as well \citep[see, e.g., Table~2 of][]{Nguyen11}.
We correct for this bias by: 
1) adding submillimeter fluxes with better angular resolution than our 500~$\mu$m \emph{Herschel} data 
(e.g., 19.2\arcsec~for \emph{APEX}-870~$\mu$m and 24\arcsec~for \emph{SIMBA}-1200~$\mu$m 
versus 36.6\arcsec~for \emph{Herschel}-500~$\mu$m) 
and 2) scaling fluxes appropriately. 
The flux-scaling procedure primarily used by \cite{Motte10} is described in more detail in 
\cite{Nguyen11}, \cite{Giannini12}, and \cite{Fallscheer13}. 
In short, these HOBYS papers incorporated a simple linear scaling method that relies on the assumption that MDCs have intensity versus aperture radius profiles close to those found for high-mass protostellar dense cores, $I\propto \theta^{-1}$ (e.g., \citealt{Beuther02}). Scaling thus provides flux estimates as they should be when integrated within the smaller aperture of the reference wavelength, that is, 160~$\mu$m or 250~$\mu$m. 
It aims at correcting integrated fluxes which mainly arise from optically thin thermal emission, 
in our case at $\lambda>100$~$\mu$m. 
The procedure uses deconvolved sizes, ${\it FWHM}^{\rm dec}$, 
integrated fluxes, $F_{\rm \lambda > 100\mu m}$, and the formula:
\begin{equation}
F^{\rm corrected}_{\rm \lambda > 100\mu m} = F_{\rm \lambda > 100\mu m} \times 
\frac{{\it FWHM}^{\rm dec}_{\rm Ref~\lambda}}{{\it FWHM}^{\rm dec}_{\rm \lambda > 100\mu m}}.
\end{equation}

The FWHM sizes estimated using \emph{getsources} are equivalent to a Gaussian FWHM, and have been deconvolved assuming that the \emph{Herschel} beams themselves are Gaussian. 
Simple Gaussian deconvolution was stopped when sources were smaller than 0.5\,$\times$\,{\it HPBW}$_\lambda$, 
which translates into an observed {\it FWHM}$^{\rm obs}_\lambda \le 1.12\,\times$\,{\it HPBW}$_\lambda$. 
Unresolved sources were assigned at their reference wavelength 
a minimum physical size of 
{\it FWHM}$^{\rm dec}_{\rm 160\mu m} = 0.5 \times${\it HPBW}$_{\rm 160\mu m} \simeq 5.85\arcsec$ 
or {\it FWHM}$^{\rm dec}_{\rm 250\mu m} \simeq 9.1\arcsec$, 
corresponding to 0.05~pc or 0.08~pc at 1.75~kpc. Dense cores are generally resolved at their reference wavelength but 36\% are not. 

When fluxes remain uncorrected, SED fits of \emph{Herschel} data alone lead to, on average, twice larger masses \citep[see, e.g., ][]{Nguyen11}. The reason behind this increase is that fluxes are measured within sizes varying from the reference wavelength size to approximately twice its value when measured at $500~\mu$m. Since NGC~6334 MDCs are imaged in the submillimeter regime at higher-angular resolutions than SPIRE, adding these fluxes limits this mass increase to $\sim$$50\%$.
As mentioned by \cite{Nguyen11}, the fitted reduced $\chi^{2}$ value is largely decreased when \emph{Herschel} fluxes are linearly corrected, showing that fluxes measured in single apertures are better fitted by modified blackbody models. The linear correction is a first-order correction method of dense core fluxes measured within varying apertures.
 Indeed, in the case of starless dense cores, which are less centrally concentrated than protostellar dense cores, a stronger flux-scaling correction should be applied. Since there are no theoretical laws to describe the low gas concentration of a starless structure whose fragmentation is largely governed by turbulence, for consistency we decided to keep the same flux correction for all types of dense cores. In the following, the number of starless dense cores with a given mass is probably overestimated because a mass concentration steeper than $Mass(<r) \propto r$ (see Fig.~\ref{M_r}) would lead to much smaller corrected 350~$\mu$m and 500~$\mu$m fluxes, higher fitted SED dust temperatures and, finally, lower derived masses.

\subsection{SED fitting procedure}
\label{section_fitting}

We used the \emph{MPFIT} least-squares minimisation routine in IDL (\citealt{Markwardt09}) 
to fit the SEDs with modified blackbody models such as 
\begin{equation}
S_{\nu} = A\, \nu^{\beta}\, B_{\nu}(T_{\rm dust}),
\end{equation}
where $S_{\nu}$ is the observed flux distribution, $A$ is a scaling factor, 
$B_{\nu}(T_{\rm dust})$ is the Planck function for the dust temperature $T_{\rm dust}$~, and 
$\beta$ is the dust emissivity spectral index here set to 2. 

We assumed, for the representative error of the reliable fluxes, the quadratic sum of \emph{getsources} flux errors, image calibration errors\footnote{
In the observers manuals for PACS and for SPIRE (see Footnote~1), 
calibration errors are 3$\%$, 5$\%$, 10$\%$, 10$\%$, and 10$\%$ 
at 70~$\mu$m, 160~$\mu$m, 250~$\mu$m, 350~$\mu$m, and 500~$\mu$m, respectively.
}, 
and background extraction errors. The two latter factors combined are estimated to be 
of the order of 10\% for $70~\mu$m and 20\% for fluxes at $\lambda>100~\mu$m.
 
The $70~\mu$m flux was only used for SED fitting when the temperature fitted to $\lambda>100~\mu$m fluxes rises above 32~K, corresponding to a modified blackbody peaking at $100~\mu$m. 
When cloud fragments are hotter than 32~K, their SEDs are incorrectly constrained by $\lambda>160~\mu$m data alone. The $70~\mu$m emission from the inner part of the protostellar envelope becomes larger than the extinction depth at $70~\mu$m of the modified blackbody fitted at $\lambda>100~\mu$m. 
In such cases, the $70~\mu$m flux was therefore used as an upper limit. 
Throughout our sample, 20\% of the sources needed the $70~\mu$m flux in their SEDs, corresponding to sources that are either IR-bright protostellar dense cores or 
weak sources with poorly constrained $\lambda>100~\mu$m SED portions.

Since dense cores do not have flat spectra across far-infrared to submillimeter bands, 
we applied color-correction factors to all \emph{Herschel} fluxes.
SPIRE color-correction factors depend only on the dust emissivity spectral index $\beta$, here fixed to 2. They are $4\%$, $4\%$, and $5\%$ at 250~$\mu$m, 350~$\mu$m, and 500~$\mu$m respectively (cf. SPIRE observers manual). PACS color-correction factors themselves vary with $T_{\rm dust}$, especially at 70~$\mu$m \citep{Poglitsch10}.
In our sample of 654 dense cores, the mean values of these corrections 
are $13\%$ and $3\%$ but they can go up to 
60$\%$ and $10\%$, at 70~$\mu$m and 160~$\mu$m, respectively. 

Since we attempted to fit all the SEDs of the 654 dense cores, even when they were very low-mass and have uncertain fluxes, 
we enforced fits with temperatures between 8~K and 70~K. 
We used the fitted reduced $\chi^2$ value to discriminate 
between coherent sources with robust fits and inhomogeneous cloud structures, 
and qualified those with reduced $\chi^2 < 10 $  as robust SEDs, and
those with reduced $\chi^2 > 10$  as dubious. We found that $25\%$ of the dense core SEDs are not robust. 
We excluded them in the following analysis because their parameters are too uncertain.
Our fitting procedure leads to a sample of 490 dense cores 
for which envelope mass can be properly determined. For the remainder of the study, we consider only this sample of dense cores.  

We determined the total gas $+$ dust mass of dense cores, $Mass$, 
from their dust continuum emission, $S_{\nu}$, and the following equation: 
\begin{equation}
{\rm\it Mass} = \frac{S_{\nu} \,d^{2}}{\kappa_{\nu} \, B_{\nu}(T_{dust})},
\end{equation}
where $d$ is the distance to the Sun, 1.75~kpc, and $\kappa_{\nu}$ is the dust mass opacity. 
$\kappa_\nu$ depends on the size, shape, chemical composition, and 
temperature of the grains as well as the gas-to-dust mass ratio
\citep[see][]{Ossenkopf94,Martin12,Deharveng12}.
$\kappa_\nu$ thus depends on the density and temperature of the medium considered. 
The main assumptions used are that the source fluxes are optically thin, sources are
homogeneous in density, temperature, and optical properties along the line of sight. 
As in Sect.~\ref{section_HerschelObs}, the HOBYS consortium uses the relation $\kappa_{\nu} = \kappa_{0} (\nu / \nu_{0})^{\beta}$ with 
$\kappa_{0}$= 0.1~cm$^{2}$~g$^{-1}$ assuming a gas-to-dust ratio of 100, $\nu_{0}$= 1000~GHz 
and $\beta$ = 2 for the dust emissivity spectral index 
(\citealt{Hill09,Motte10}). 
If we were to use $\beta = 1.5$, which is the other extreme value appropriate for dense cores, temperatures and masses would differ by $5-10\%$. We estimate that the overall uncertainty on masses measured through this procedure for dense cores with known distance and fixed size is approximately $50\%$ \citep[]{Roy14}.

\subsection{Luminosity and volume-averaged density}
\label{section_luminosities}

We define the bolometric luminosity 
and volume-averaged density for our sample of dense cores. 
The bolometric luminosity, $L_{\rm bol}$, should be calculated by integrating the flux densities 
over the source SED from 2~$\mu$m to 1~mm. 
To get an accurate bolometric luminosity for protostellar sources 
which can have significant mid-infrared emission, 
we used data from the GLIMPSE $3.6-8~\mu$m, MIPSGAL 24~$\mu$m, 
\emph{WISE} 22~$\mu$m, and \emph{MSX} 21~$\mu$m catalogs\footnote{
The catalogs (GLIMPSE (I + II + 3D), MIPSGAL, \emph{WISE} All-Sky Source, 
and MSXPSC v2.3) can be found via \url{http://irsa.ipac.caltech.edu/applications/Gator/}.
}. 
These catalogs were cross-correlated with our MDCs within a radius of 6\arcsec, 
roughly corresponding to the 160~$\mu$m FWHM beam. 
When associations are found with both the MIPSGAL and WISE catalogs, or the MSX catalog, 
we favored that with the higher-resolution measurement. 
We estimated the mid-IR noise level in NGC~6334 as the mean dispersions in several background areas, 
$\sigma_{\rm 24\,\mu m}=5$~mJy/$6\arcsec$-beam. 

Starless dense cores that do not show mid-infrared 
or even 70~$\mu$m emission can theoretically have their luminosity calculated 
by integrating below their respective fitted modified blackbodies. 
Their bolometric luminosities are thus set to the maximum value of either this estimate or the one obtained by integrating below observed fluxes.

\begin{figure*}[htpb!]  
\centering
\includegraphics[width=15cm]{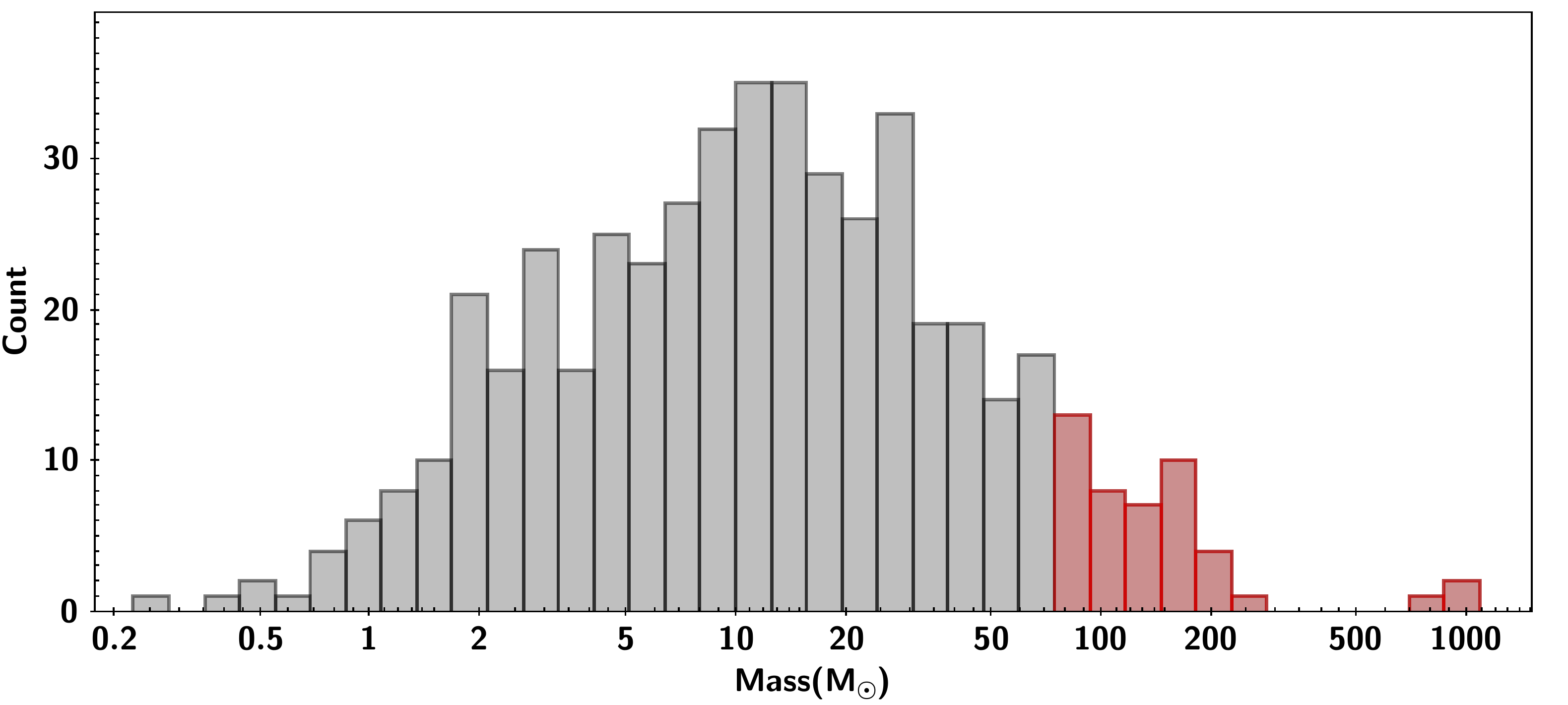}
\vskip 0.5cm
\includegraphics[width=7.9cm]{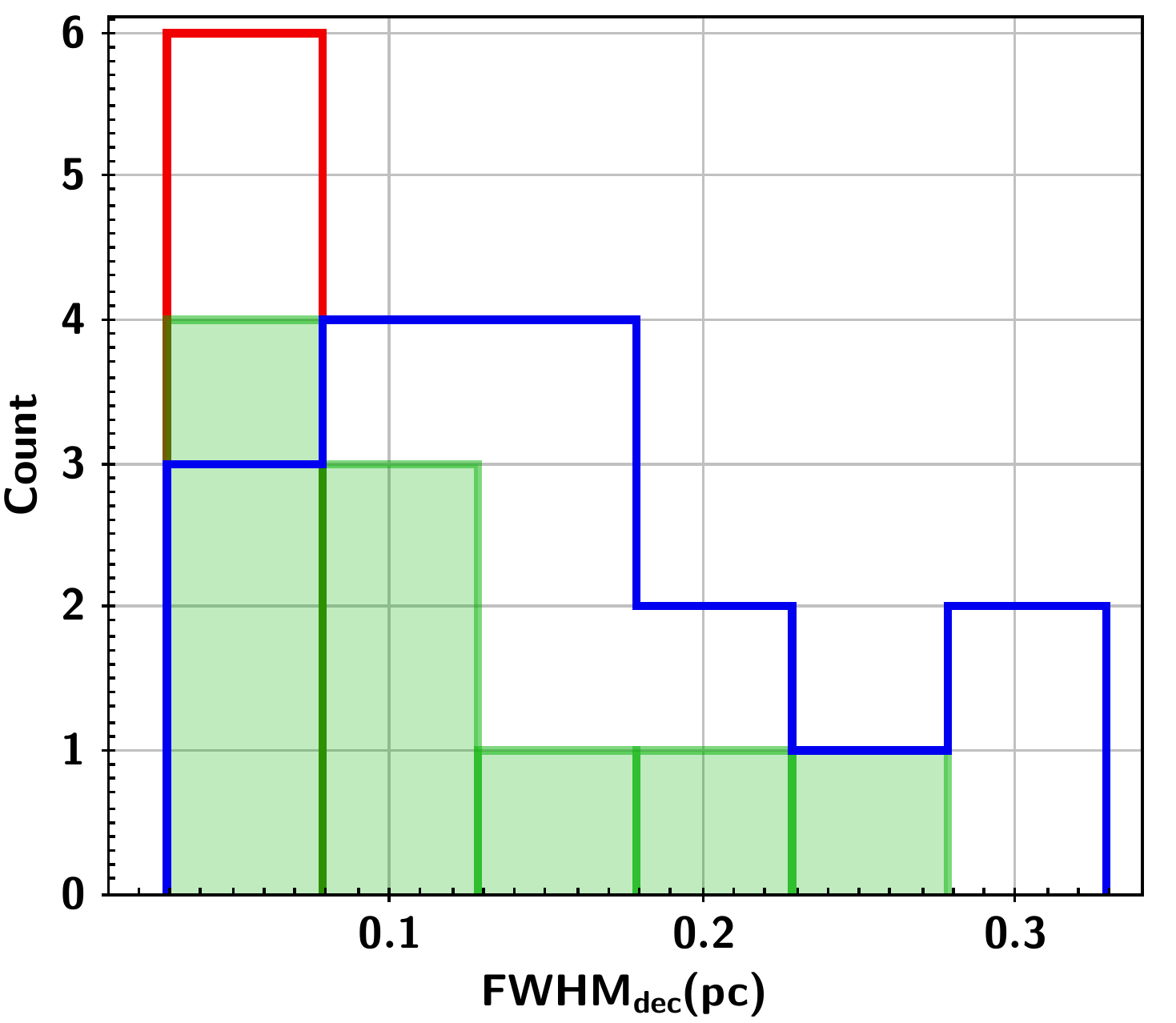}\hskip 0.5cm\includegraphics[width=8cm]{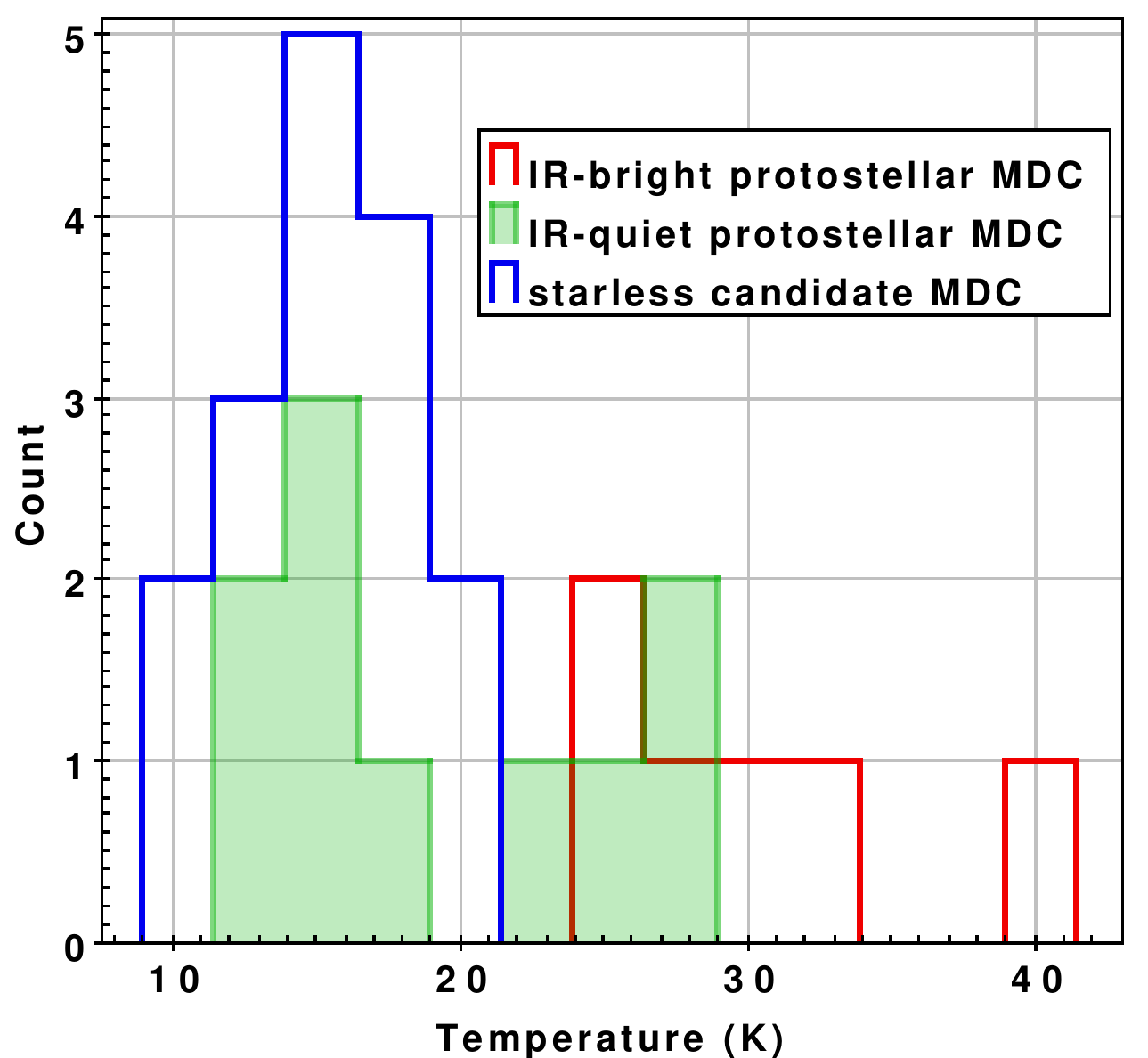}
\vskip 0.4cm
\includegraphics[width=8cm]{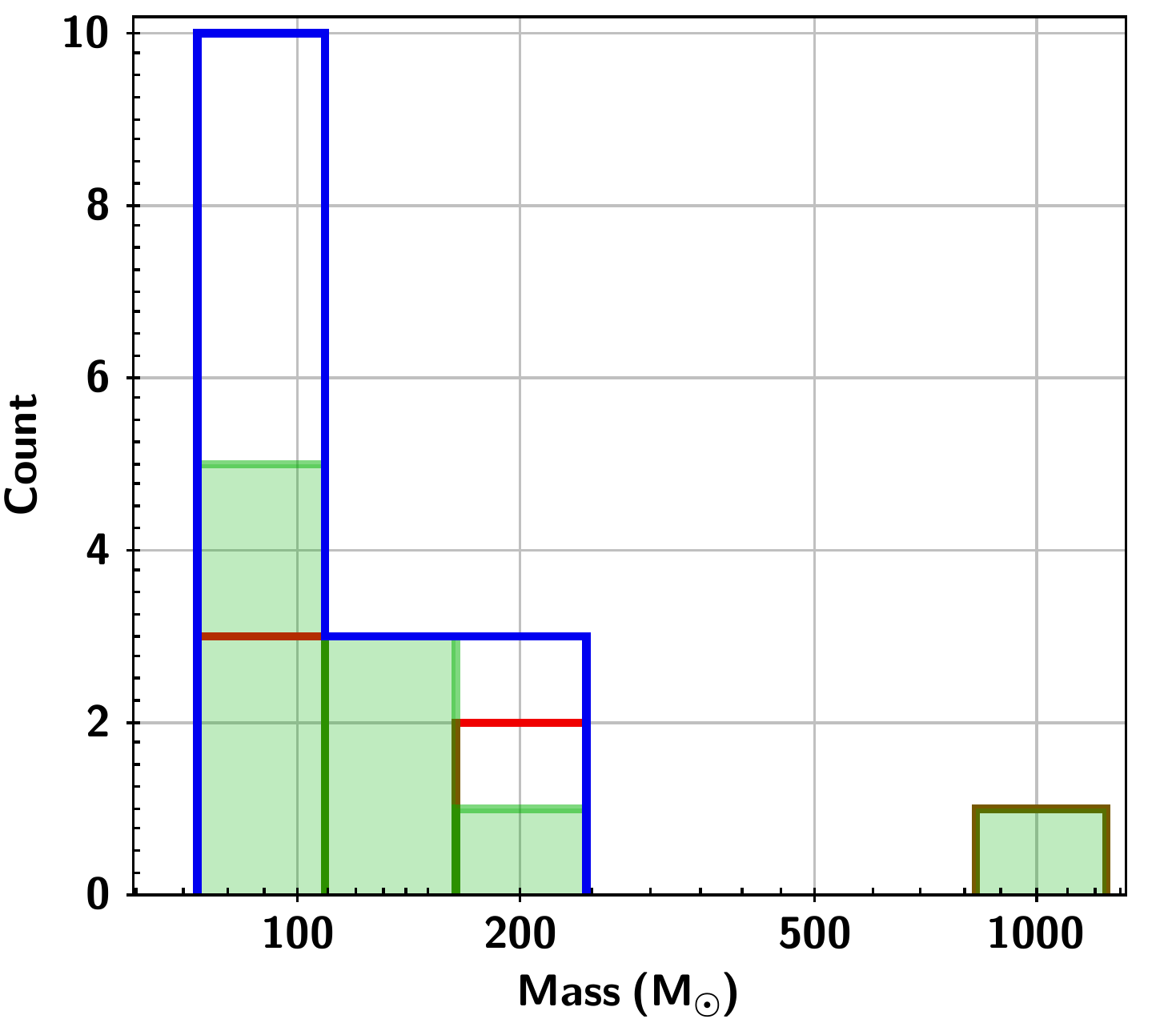}\hskip 0.5cm\includegraphics[width=7.4cm]{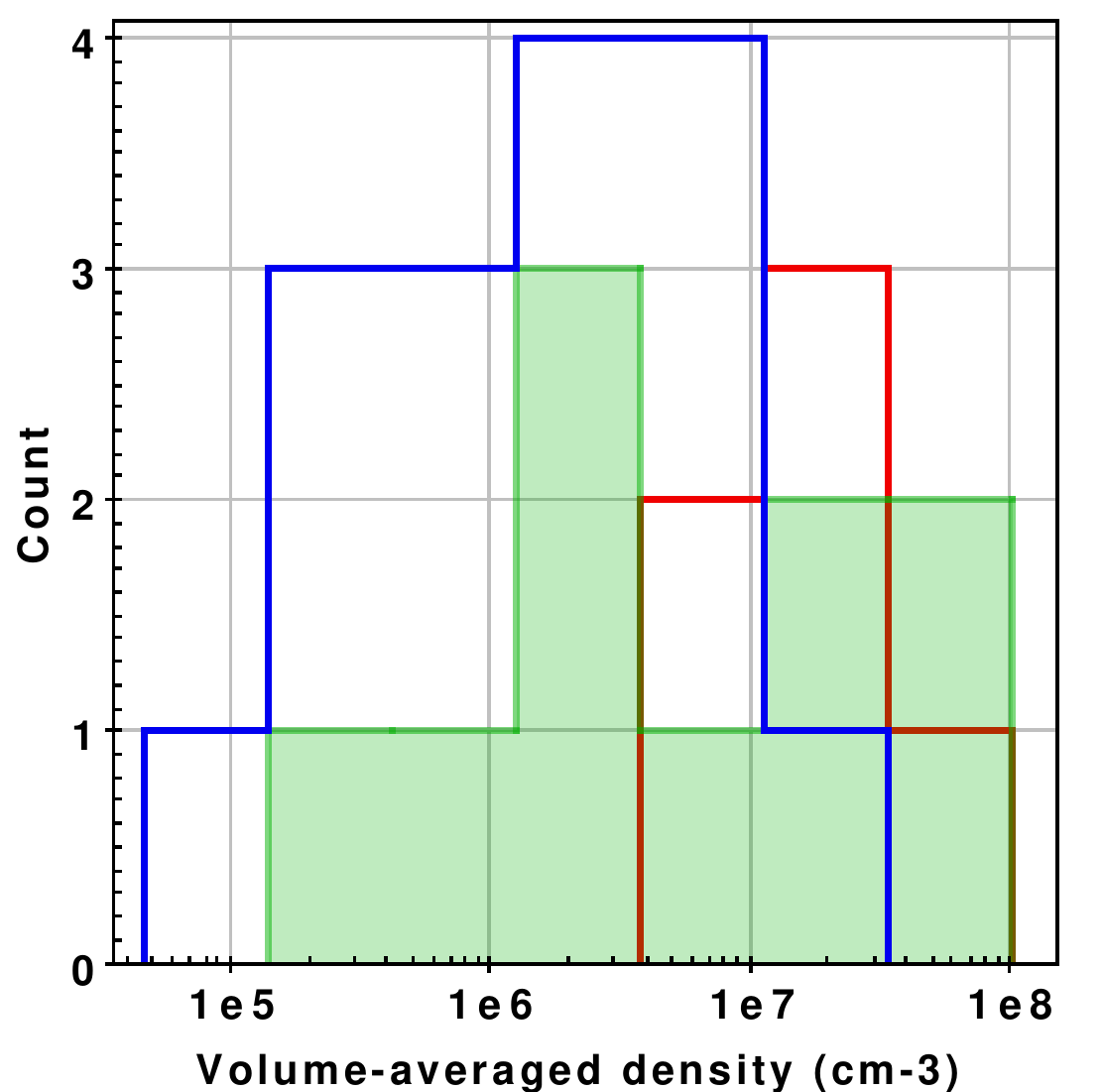}
\caption{
Mass distribution of the 490 dense cores (gray histogram) and 46 MDCs (red histogram) of NGC~6334 {\bf (top)}. Sample is estimated to be complete down to $\sim$15~\msun. Distributions of the deconvolved FWHM {\bf (middle-left)}, temperature {\bf (middle-right)}, mass {\bf (bottom-left)}, and volume-averaged density {\bf (bottom-right)} for the 32 reliable MDCs at various evolutionary phases (blue = starless candidate, green = IR-quiet, red = IR-bright). 
MDCs become more massive, hotter, smaller, and thus denser along the high-mass star formation process.}
\label{distrib_subset}
\end{figure*}

The volume-averaged density is taken as:
\begin{equation}
<n_{H_{2}}> ~=  \frac{Mass}{\frac{4}{3} \pi \, \mu \, m_{\rm H} \, ({\rm\it FWHM}^{\rm dec}_{\rm Ref\lambda}/2)^{3}},
\end{equation}
where $FWHM^{\rm dec}_{\rm Ref\lambda}$ is measured at the reference wavelength 
and deconvolved as explained in Sect.~\ref{section_scaling}, 
$\mu = 2.8$ is the mean molecular weight (\citealt{Kauffmann08}), and $m_{\rm H}$ is the hydrogen mass. 
For unresolved sources at 160~$\mu$m or 250~$\mu$m, their upper limit size of 5.85\arcsec~(resp. 9.1\arcsec) leads to lower limits for their volume-averaged density defined by:
\begin{equation}
<n_{\rm H_{2}\_min}> ~=  \frac{Mass}{\frac{4}{3} \pi \, \mu \, m_{\rm H} \, (0.5\, \times {{\rm \it HPBW}}_{\rm Ref\lambda}/2)^{3}}
.\end{equation}
Table~\ref{cores_properties} summarizes the main properties of our sample of 490 dense cores and Figure~\ref{distrib_subset}~(top) displays their mass distribution.

\begin{table}[h]
\begin{center}
\caption{Main physical properties of dense cores in NGC~6334}
\label{cores_properties}
\begin{tabular}{| c | c | c | c |}
\hline
\multicolumn{4}{|c|}{\multirow{2}{*}{490 dense cores}}    \\
\multicolumn{4}{|c|}{\multirow{2}{*}{}}          \\
\multicolumn{1}{|c|}{}     &   Min & Median  &  Max   \\               
\hline
FWHM$_{\rm dec}$ (pc)      & 0.05  & 0.09 $\pm$ 0.05  & 0.3\\
<$T_{\rm dust}$> (K)       & 9.5   & 20 $\pm$ 6    & 51     \\
$L_{\rm bol}$ (\lsun)      & 4     & 860 $\pm$ 6100   & 87000  \\
Mass (\msun)               & 0.25  & 32 $\pm$ 82    & 1020     \\
<$n_{\rm H_{2}}$> (cm$^{-3}$)  &  1 $\times$ 10$^{4}$   &  $(2\pm6)\times10^{6}$   & $\ge$ 7 $\times$ 10$^{7}$ \\ 
\hline
\end{tabular}
\end{center}
\end{table}

\begin{figure*}[htpb]  
\centering
\vskip -0.5cm
\includegraphics[width=12cm, angle = -90]{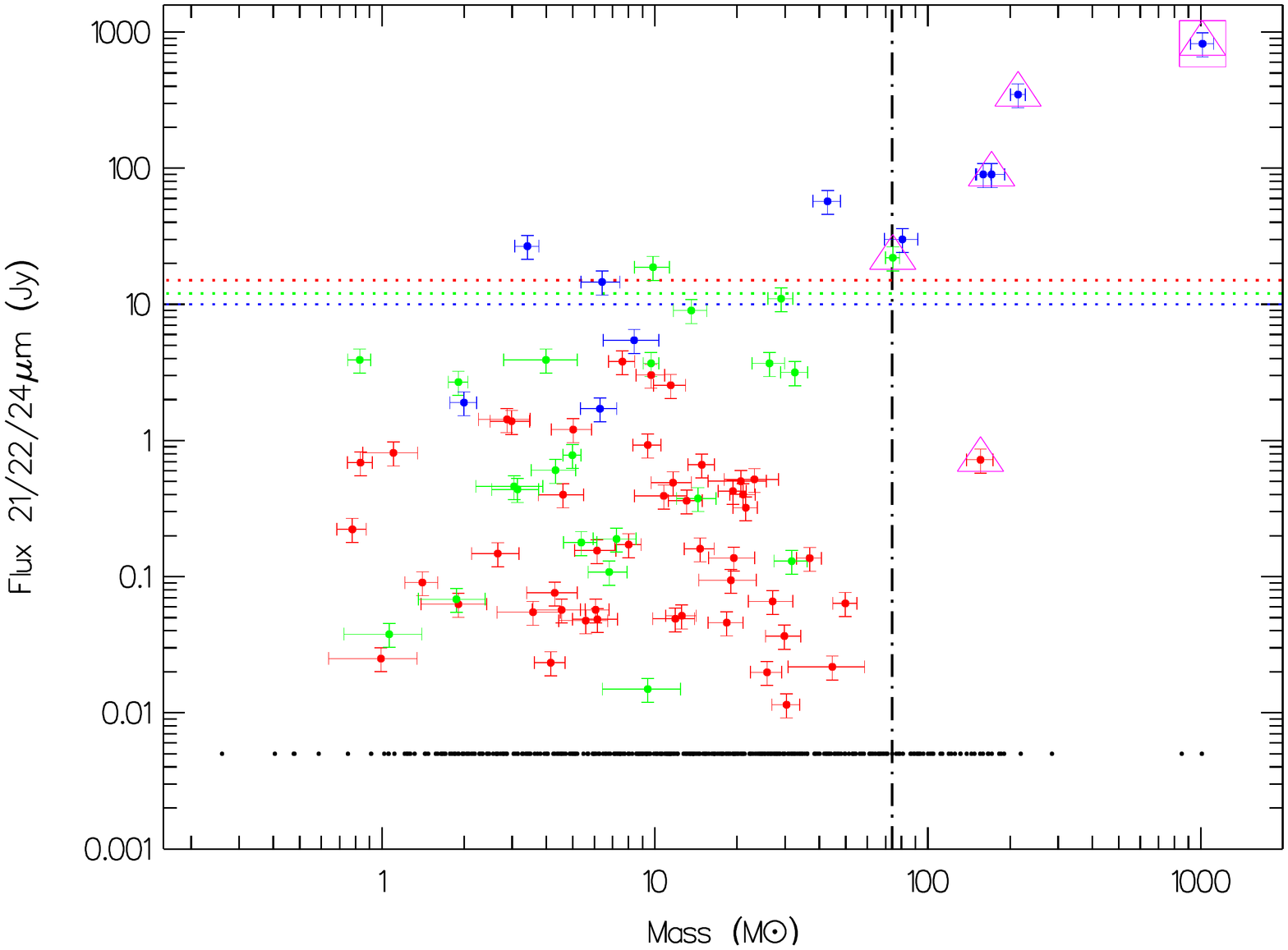}
\vskip -0.5cm
\caption{Determination of the minimum mass for MDCs in NGC~6334 in a diagram displaying their 24~$\mu$m (red), 22~$\mu$m (green), or 21~$\mu$m (blue) detected flux, or their 24~$\mu$m (black) non-detection (set to the $1\times \sigma_{\rm 24\,\mu m}\sim5$~mJy point-source detection level; see Sect.~\ref{section_luminosities}) as a function of their mass. Pink squares and triangles denote dense cores associated with 20~cm compact radio sources and CH$_{3}$OH masers, respectively. Red, green, and blue dotted lines are the 15~Jy, 12~Jy, and 10~Jy limits at 24~$\mu$m, 22~$\mu$m, and 21~$\mu$m for IR-bright sources, respectively. The black dotted-dashed line represents our mass threshold of $75~\msun$ for MDCs. Errors on the mass are taken from \emph{MPFIT} while flux errors are estimated to be 20\%.}
\label{F24_Mass_6334}
\end{figure*}

\section{Complete sample of NGC~6334 MDCs}
\label{section_massive_dense_cores}

The masses of dense cores extracted from \emph{Herschel} data range from 0$.3~\msun$ to approximately $1000~\msun$. If we expect high-mass stars to form in the most massive dense cores, we must establish the lowest mass required by a dense core to form at least one high-mass star.

\subsection{MDCs with known high-mass star formation signposts}
\label{section_signposts}

For this purpose, we used two signposts of on-going high-mass star formation: Class~II CH$_{3}$OH maser and compact radio centimeter sources. Masers result from the interaction of protostars with their envelope. In particular, 6.7~GHz Class~II CH$_{3}$OH masers are found to be exclusively associated with high-mass star formation (\citealt{Breen13}). As for compact radio centimeter sources, they generally arise from ionizing UC\hii regions. Dense cores displaying one of these signposts have already started to form high-mass stars. To pinpoint these high-mass star-forming dense cores, we cross-correlated the positions of our sample of 490 dense cores with those of 6.7~GHz Class~II CH$_{3}$OH masers having appropriate velocities and compact radio centimeter sources listed in published catalogs (e.g., \citealt{White05}, \citealt{Combi11}, \citealt{Pestalozzi05}, \citealt{Caswell10}).

Allowing up to 10\arcsec~offsets, we found six associations. 
MDC \#6 is associated with both 6.7~GHz CH$_{3}$OH maser emission (\citealt{Caswell10}) and an UC\hii region observed at 20~cm (\citealt{White05}), while MDCs \#1 (source I), \#10, \#14 (source I(N)), and \#46 are only associated with 6.7~GHz CH$_{3}$OH maser sources (\citealt{Pestalozzi05}; \citealt{Caswell10}).  
In addition, we used N$_{2}$H$^{+}$ observations published in \cite{Foster11} and \cite{Russeil10}, and checked that dense cores indeed contain dense gas and that their velocities are close to the mean velocity measured for the NGC~6334 cloud complex of $-4~$km\,s$^{-1}$, with a velocity gradient (\citealt{Kraemer99}) from the center ($V_{\rm LSR} \sim 0$~km\,s$^{-1}$) to the edge of the central filament ($V_{\rm LSR} \sim -11$~km\,s$^{-1}$).

\begin{table*}[htbp!] 
\centering
\caption{Physical parameters of the 46 MDCs found in NGC~6334}
\label{tab_parameters_6334}
\begin{tabular}{| c | c | c | c | c |c | c | c |}
\hline
  Number     & FWHM$_{\rm dec}$          & <T$_{\rm dust}$>    &       Mass            & L$_{\rm bol}$  &  $<n_{\rm H_{2}}>$    &  $\chi^2$   \\
    Id.       &         (pc)              &      (K )      &       (\msun)         &    (\lsun)     &      (cm$^{-3}$)  &               \\                       
\hline
\multicolumn{7}{|c|}{\multirow{2}{*}{IR-bright protostellar MDC}}     \\
\multicolumn{7}{|c|}{\multirow{2}{*}{}}          \\
  1    &$\le$ 0.08 & 24.6 $\pm$ 1.1 & 1020 $\pm$ 100 & 42600  &  $\ge$ 7.4  $\times$ 10$^{7}$ & 3\\
  6    &$\le$ 0.08 & 40.1 $\pm$ 0.7 & 200 $\pm$ 10 & 87200  &  $\ge$ 1.5  $\times$ 10$^{7}$ & 2\\
  10    &$\le$ 0.08 & 27.3 $\pm$ 1.5 & 180 $\pm$ 20 & 9600  &  $\ge$ 1.3  $\times$ 10$^{7}$ & 1\\
  32   &$\le$ 0.08 & 30.8 $\pm$ 2.0 & 95 $\pm$ 10 & 8300   &  $\ge$ 6.8  $\times$ 10$^{6}$ & 2\\
  38   &$\le$ 0.08 & 33.1 $\pm$ 2.3 & 90 $\pm$ 10 & 9300   &  $\ge$ 6.4  $\times$ 10$^{6}$ & 3\\
  46    &$\le$ 0.05 & 25.9 $\pm$ 0.3 & 75 $\pm$ 4 & 3200     &  $\ge$ 2.0  $\times$ 10$^{7}$ & 2\\
\hline
\multicolumn{7}{|c|}{\multirow{2}{*}{IR-quiet protostellar MDC}}     \\
\multicolumn{7}{|c|}{\multirow{2}{*}{}}          \\
  2    &$\le$ 0.08 & 15.2 $\pm$ 0.4 & 950 $\pm$ 80 & 940    &  $\ge$ 7.0  $\times$ 10$^{7}$ & 6\\
  7    &$\le$ 0.08 & 27.2 $\pm$ 1.5 & 200 $\pm$ 20 & 6400   &  $\ge$ 1.4  $\times$ 10$^{7}$ & 0.4\\
  14   &$\le$ 0.05 & 26.6 $\pm$ 1.4 & 160 $\pm$ 15 & 7000   &  $\ge$ 4.3  $\times$ 10$^{7}$ & 5\\
  15   &0.11       & 23.5 $\pm$ 1.2 & 160 $\pm$ 20 & 2200   &        4.3  $\times$ 10$^{6}$ & 8\\
  24   & 0.18      &  16.9 $\pm$ 0.5 & 110 $\pm$ 10 & 580   &        6.0  $\times$ 10$^{5}$ & 3\\
  30   &0.13       & 14.4 $\pm$ 0.4 & 100 $\pm$ 10 & 70     &        1.6  $\times$ 10$^{6}$ & 3\\
  33   &0.11       & 12.8 $\pm$ 0.5 & 90 $\pm$ 10 & 30      &        2.0  $\times$ 10$^{6}$ & 5\\
  37   &0.13       & 11.7 $\pm$ 0.6 & 90 $\pm$ 15 & 28      &        1.3  $\times$ 10$^{6}$ & 5\\
  41   &$\le$ 0.05 & 26.2 $\pm$ 1.4 & 85 $\pm$ 10 & 2200    &  $\ge$ 2.3  $\times$ 10$^{7}$ & 1\\
  42   &0.26       & 14.0 $\pm$ 0.4 & 80 $\pm$ 10 & 50      &        1.6  $\times$ 10$^{5}$ & 3\\
\hline
\multicolumn{7}{|c|}{\multirow{2}{*}{Starless MDC candidate}}     \\
\multicolumn{7}{|c|}{\multirow{2}{*}{}}          \\
  5    &$\le$ 0.08 & 14.6 $\pm$ 0.4 & 210 $\pm$ 15 & 160    &  $\ge$ 1.5  $\times$ 10$^{7}$ & 3\\
  11   &0.28       & 14.1 $\pm$ 0.6 & 170 $\pm$ 20 & 110    &        2.4  $\times$ 10$^{5}$ & 3\\
  13   &0.18       & 9.5 $\pm$ 0.4 & 170 $\pm$ 30 & 10      &        1.0  $\times$ 10$^{6}$ & 8\\
  16   &0.15       &  9.9 $\pm$ 0.4 & 150 $\pm$ 20 & 11     &        1.6  $\times$ 10$^{6}$ & 10\\
  17   &0.1        & 16.5 $\pm$ 0.7 & 150 $\pm$ 15 & 230    &        4.8  $\times$ 10$^{6}$ & 4\\
  21   &0.27       & 18.3 $\pm$ 0.6 & 130 $\pm$ 10 & 390    &        2.2  $\times$ 10$^{5}$ & 6\\
  26   &$\le$ 0.08 & 21.1 $\pm$ 2.4 & 100 $\pm$ 30 & 740    & $\ge$  7.5  $\times$ 10$^{6}$ & 6\\
  27   &0.14       & 15.6 $\pm$ 0.6 & 105 $\pm$ 11 & 120    &        1.3  $\times$ 10$^{6}$ & 6\\
  28   &$\le$ 0.08 & 19.2 $\pm$ 2.2 & 100 $\pm$ 24 & 410    &  $\ge$ 7.3  $\times$ 10$^{6}$ & 7\\
  31   &0.17       & 11.9 $\pm$ 0.4 & 100 $\pm$ 14 & 22     &        6.2  $\times$ 10$^{5}$ & 4\\
  35   &0.21       & 16.4 $\pm$ 0.9 & 90 $\pm$ 10 & 140     &        3.4  $\times$ 10$^{5}$ & 4\\
  36   &0.08       & 17.7 $\pm$ 0.7 & 90 $\pm$ 9 & 220      &        5.7  $\times$ 10$^{6}$ & 6\\
  39   &0.19       & 12.1 $\pm$ 0.4 & 85 $\pm$ 15 & 20      &        4.4  $\times$ 10$^{5}$ & 9\\
  40   &0.12       & 18.1 $\pm$ 0.8 & 85 $\pm$ 10 & 240     &        1.7  $\times$ 10$^{6}$ & 10\\
  44   & 0.3       & 12.8 $\pm$ 0.6 & 77 $\pm$ 10 & 27      &        1.0  $\times$ 10$^{5}$ & 3\\
  45   & 0.11      & 14.7 $\pm$ 0.8 & 76 $\pm$ 10 & 60      &        2.0  $\times$ 10$^{6}$ & 4\\
\hline
\multicolumn{7}{|c|}{\multirow{2}{*}{Undefined cloud structure}}     \\
\multicolumn{7}{|c|}{\multirow{2}{*}{}}          \\
  3    &$\le$ 0.08 & 14.7 $\pm$ 0.4 & 820 $\pm$ 70 & 670    &  $\ge$ 5.9  $\times$ 10$^{7}$ & 8\\
  4    &$\le$ 0.08 & 15.4 $\pm$ 1.0 & 280 $\pm$ 40 & 300    &  $\ge$ 2.0  $\times$ 10$^{7}$ & 4\\
  8    &$\le$ 0.08 & 15.3 $\pm$ 1.1 & 180 $\pm$ 35 & 190    &  $\ge$ 1.3  $\times$ 10$^{7}$ & 9\\
  9    &$\le$ 0.08 & 19.2 $\pm$ 0.7 & 180 $\pm$ 20 & 720    &  $\ge$ 1.3  $\times$ 10$^{7}$ & 1\\
  12   &$\le$ 0.08 & 24.1 $\pm$ 1.1 & 170 $\pm$ 20 & 2600   &  $\ge$ 1.2  $\times$ 10$^{7}$ & 3\\
  18   &0.26       & 11.5 $\pm$ 0.5 & 140 $\pm$ 20 & 50     &        2.6  $\times$ 10$^{5}$ & 3\\
  19   &$\le$ 0.08 & 17.7 $\pm$ 1.1 & 140 $\pm$ 20 & 340    &  $\ge$ 9.9  $\times$ 10$^{6}$ & 5\\
  20   &$\le$ 0.08 & 19.4 $\pm$ 0.8 & 130 $\pm$ 15 & 540    &  $\ge$ 9.2  $\times$ 10$^{6}$ & 4\\
  22   &$\le$ 0.08 & 15.8 $\pm$ 0.9 & 125 $\pm$ 20 & 160    &  $\ge$ 9.0  $\times$ 10$^{6}$ & 5\\
  23   &$\le$ 0.08 & 21.9 $\pm$ 1.1 & 120 $\pm$ 15 & 1100   &  $\ge$ 8.7  $\times$ 10$^{6}$ & 3\\
  25   &$\le$ 0.08 & 21.5 $\pm$ 0.9 & 110 $\pm$ 10 & 870    &  $\ge$ 8.0  $\times$ 10$^{6}$ & 4\\
  29   &0.09       & 22.8 $\pm$ 0.4 & 100 $\pm$ 10 & 3300   &        4.5  $\times$ 10$^{6}$ & 7\\
  34   &0.11       & 11.9 $\pm$ 0.8 & 90 $\pm$ 25  & 20     &        2.3  $\times$ 10$^{6}$ & 10\\
  43   &$\le$ 0.08 & 15.2 $\pm$ 0.6 & 80 $\pm$ 10  & 80     &  $\ge$ 5.9  $\times$ 10$^{6}$ & 7\\
\hline
\end{tabular}
\end{table*}

\subsection{Minimum mass for MDCs in NGC~6334}
To determine the minimum mass required for dense cores to form high-mass stars, we followed the method of \cite{Motte07} and compared the dense cores of Table~\ref{cores_properties} with and without signposts of high-mass star formation.

Figure~\ref{F24_Mass_6334} shows the 24~$\mu$m flux as a function of the mass of each source, unless 24~$\mu$m flux is saturated, in which case the 22~$\mu$m or 21~$\mu$m flux is shown instead. The dense cores associated with CH$_{3}$OH masers or compact radio centimeter sources are well separated from the others. Since those dense cores with on-going high-mass star formation exhibit $75-1000~\msun$ masses, we fixed $75~\msun$ as the mass threshold for dense cores able to form high-mass stars in NGC~6334.
 
If one wants to compare the NGC~6334 mass threshold to that obtained in other studies, one must realize that the threshold depends on the typical size of the sources considered. 
For Cygnus~X MDCs that have 30\% smaller mean sizes than NGC~6334 MDCs, \cite{Motte07} found a mass limit of $40~\msun$ (see their Fig.~7), 45\% lower than the one here. In contrast, \cite{Csengeri14} used a much larger threshold, $>$650~\msun, to select, from the ATLASGAL survey, massive clumps with sizes of $0.4-1$~pc, almost ten times larger than MDCs. The threshold found for NGC~6334 MDCs is also consistent with values used to distinguish intermediate-mass dense cores in three other HOBYS regions \citep{Motte10,Nguyen11,Fallscheer13}.

\subsection{MDC characteristics}

Using our lower mass limit of $75~\msun$ and keeping robust dense cores with reduced $\chi^{2} < 10$, we ended up with a selection of 46 MDCs in NGC~6334. Table~\ref{tab_parameters_6334} lists each of these 46 MDCs along with their main physical properties as derived from the SED analysis of Sect.~\ref{section_dense_cores} and presented in Figs.~\ref{MDC_1}--\ref{MDC_43}. 
As given in Table~\ref{mdcs_properties},
MDCs have median FWHM sizes of $\sim$0.08~pc, dust temperatures of $\sim$17~K, total masses around $120~\msun$, and median densities of $6\times 10^6$~cm$^{-3}$. 
Figure~\ref{distrib_subset} displays their size, temperature, mass, and volume-averaged density distributions, with their mass distribution also shown on top of that of the complete sample of 490 dense cores. 
The role of MDCs in the high-mass star formation scenario is to be the spatial and kinematical link between ridges or hubs and individual protostars.

For the most extended MDCs, we checked for infrared associations out to a radius of $11\arcsec$ from the peak. When a mid-infrared counterpart was visible but had not been measured by earlier infrared surveys, we performed aperture photometry directly on the images  using the GLIMPSE and MIPSGAL guidelines for point sources\footnote{
        In GLIMPSE images, aperture, inner, and outer sky annulus radii are recommended to be 3.6\arcsec, 3.6\arcsec, and $9.6\arcsec$ respectively (see \url{https://irsa.ipac.caltech.edu/data/SPITZER/GLIMPSE/doc/glimpse_photometry_v1.0.pdf}). In MIPSGAL images, aperture, inner, and outer sky annulus radii are recommended to be 6.4\arcsec, 7.6\arcsec, and $17.8\arcsec$, respectively (\citealt{Gutermuth15}).
        }.
The resulting bolometric luminosities of NGC~6334 MDCs range from $10~\lsun$ to $9\times 10^4\,\lsun$ (see Table~\ref{mdcs_properties}).

\begin{table}[h]
\begin{center}
\caption{Main physical properties of MDCs in NGC~6334}
\label{mdcs_properties}
\begin{tabular}{| c | c | c | c |}
\hline
\multicolumn{4}{|c|}{\multirow{2}{*}{46 MDCs}}    \\
\multicolumn{4}{|c|}{\multirow{2}{*}{}}          \\
\multicolumn{1}{|c|}{}  &   Min         & Median  &  Max           \\              
\hline
FWHM$_{\rm dec}$ (pc)   &    0.05     & 0.08  & 0.3            \\
<$T_{\rm dust}$> (K)    &    9.5      & 16.7  & 40.1               \\
$L_{\rm bol}$ (\lsun)   &    10        & 320  & 87000            \\
Mass (\msun)            &    75       & 120   & 1020               \\
$<n_{\rm H_{2}}>$   (cm$^{-3}$)  & 1 $\times$ 10$^{5}$   & 6 $\times$ 10$^{6}$   &  $\ge$ 7 $\times$ 10$^{7}$    \\ 
\hline
\end{tabular}
\end{center}
\end{table}

We checked that there is a good agreement between the location of \emph{Herschel} MDCs of Table~\ref{tab_parameters_6334} and the massive clumps extracted at 1.2~mm by \cite{Russeil10}. 
The MDCs found here are four times smaller in size and four times less massive than those clumps, that is, $\sim$0.1~pc and $150~\msun$ versus $\sim$0.4~pc and $600~\msun$, respectively. Most (8 out of 10) of the 1.2~mm clumps are subdivided into several MDCs in our data, showing that the present study allows us to follow the cloud-fragmentation process to smaller scales. The two 1.2~mm clumps which do not contain any \emph{Herschel} MDCs are actually associated with compact \hii regions. Hence, their dense cloud structures could already be ionized. The masses of these two clumps were also overestimated by \cite{Russeil10} since they underestimated their dust temperature and free-free contribution to the 1.2~mm emission. 
MDCs of Table~\ref{tab_parameters_6334} are generally located within the $\ge200~\msun$ clumps of \cite{Russeil10}, with the exception of a few that are found either close ($<$0.3~pc) to them or within intermediate-mass ($100-200~\msun$) clumps of \cite{Russeil10}.
Therefore, the present sample of \emph{Herschel} MDCs in NGC~6334 is consistent with the previous study done at a lower angular resolution and a single wavelength by \cite{Russeil10}.

\begin{table*}[htbp]
\centering
\caption{46 MDCs: Nature and additional comments of $>$$75~\msun$ compact sources: 32 MDCs and 14 undefined cloud structures.}
\label{tab_nature}
\begin{tabular}{| c | c | c | c | c | c |}
\hline
  Number        & Nature &   Comments   &  8~$\mu$m     &  21-24~$\mu$m    &  L$_{\rm submm}$/L$_{\rm bol}$  \\                     
     Id         &             &              &  source        &       flux        &   (\%)                           \\
\hline
  1 &      &  UC\hii region NGC6334-F$^{1}$. NGC6334-I$^{2}$. CH$_{3}$OH Maser$^{3}$.           & Y & Strong & 0.4    \\
  6 &     IR-bright    &  NGC6334-V$^{2}$. EGO. CH$_{3}$OH Maser$^{2}$.                     & Y & Strong & 0.1    \\
  10 & \multirow{2}{*}{protostellar}   &  Close to C\hii region NGC6334-C$^{1}$. CH$_{3}$OH Maser$^{3}$.  & Y & Strong & 0.4    \\ 
  32 &    \multirow{2}{*}{}   &  Partially contains \#38.                                   & Y & Strong & 0.3    \\
  38 &    MDC   &  NGC6334-IV (A)$^{4}$.                                                    & Y & Strong & 0.2    \\
  46 &      &  NGC634-IV (MM3)$^{5}$.  CH$_{3}$OH Maser$^{3}$.                              & Y & Strong & 0.5    \\
\hline
  2 &   &  Poor fit. Not centrally peaked.  partially contains \#14.                & N & None & 6.5      \\
  7 &      &  Close to \#6.                                                            & N & None & 0.6      \\
  14 &       &  Poor fit. NGC6334-I(N)$^{6}$. CH$_{3}$OH Maser$^{3}$.          & N & Weak & 0.5      \\
  15 &  IR-quiet    &  Poor fit. Not centrally peaked. Close to \#6.                            & N & None & 1.2      \\
  24 &  \multirow{2}{*}{protostellar}    &                                                                           & Y & Weak & 1.6      \\
  30 &  \multirow{2}{*}{}    &                                                                           & N & None & 7.9      \\
  33 &  MDC   &  Poor fit.                                                                & N & None & 11.7     \\
  37 &      &  Poor fit.                                                                & Y & Weak & 10.1     \\
  41 &      &                                                                           & N & None & 0.8      \\
  42 &    &                                                                           & N & None & 8.5      \\
\hline
  5$^{**}$ &     &                                                                    & N & None & 7.4      \\
  11$^{*}$ &      &                                                                     & N & None & 8.5      \\
  13 &      &  Poor fit.                                                                & N & None & 27.7     \\
  16 &      &  Poor fit.                                                                & N & None & 24.8     \\
  17 &     &  Close to \#6.                                                            & N & None & 4.8      \\
  21$^{*}$ &     &  Poor fit. Extended 160~$\mu$m.                                     & N & None & 3.2      \\
  26 & starless   &  Poor fit. May contain background emission.                       & N & None & 1.8      \\
  27 &  \multirow{2}{*}{MDC}    &  Poor fit. May contain background emission.            & N & None & 5.8      \\
  28 &  \multirow{2}{*}{}    &  Poor fit. May contain background emission.                & N & None & 2.7      \\
  31$^{*}$ &  candidate    &                                                            & N & None & 14.5     \\
  35$^{*}$ &      &                                                                     & N & None & 4.9      \\
  36 &      &  Poor fit.                                                                & N & None & 3.7      \\
  39 &     &   Poor fit. Missing submm measures                                         & N & None & 13.9     \\
  40 &     &  Poor fit. May contain background emission.                               & N & None & 3.4      \\
  44$^{*}$ &      &                                                                     & N & None & 11.6     \\
  45$^{*}$ &      &  May contain background emission.                                  & N & None & 7.3      \\
\hline
  3 &     &  Poor fit. Not centrally peaked. Close to \#14.                           & N & None & 7.2      \\
  4 &      &  Not centrally peaked. Close to \#14.                                     & N & None & 6.2      \\
  8 &     &  Poor fit. Not centrally peaked.                                            & N & None & 6.2      \\
  9 &    &  Not centrally peaked. Close to \#46.                                      & N & None & 2.7      \\
  12 &        &  Not centrally peaked.                                                    & N & None & 1.1      \\ 
  18 &   undefined  &  Overlap with IR source. Not centrally peaked.                     & Y & Weak & 9.2      \\
  19 &    \multirow{2}{*}{cloud}  &  Poor fit. Not centrally peaked. Close to \#46.          & N & None & 3.7      \\
  20 &    \multirow{2}{*}{}  &  Not centrally peaked. partially contains \#46.          & N & None & 2.6      \\
  22 &    structure  &  Not centrally peaked. Close to \hii region GM1-24$^{7}$.       & N & None & 5.6      \\
  23 &       &  Not centrally peaked. Close to \hii region GM1-24$^{7}$.                & N & None & 1.6      \\
  25 &    & Not centrally peaked. Close to \#32.                                        & N & None & 1.7      \\
  29 &      &  Poor fit. Not centrally peaked. Close to \#6.                            & Y & None & 0.5      \\  
  34 &     &  Poor fit. Not centrally peaked. Missing submm measures.                   & N & None & 14.7     \\
  43 &     &  Poor fit. Not centrally peaked.                                             & N & None & 6.5      \\
\hline              
\end{tabular}
\tablefoot{(1) \cite{Rodriguez82}. (2) \cite{Loughran86}. (3) \cite{Pestalozzi05} and \cite{Caswell10}. (4)\cite{Persi10}. (5) \cite{Tapia09}. (6) \cite{Gezari82}. (7) \cite{Tapia91}. `Poor' means fit with reduced $5 <\chi^2 < 10$. `Strong' and `Weak' means 21-24~$\mu$m flux respectively greater or lower than the minimum flux of IR-bright MDCs defined in Sect.~\ref{section_bright_quiet}. $^{*}$ is for MDCs among the best starless candidates.}
\end{table*}

\section{Discussion}
\label{section_discussions}

We now discuss the nature of the NGC~6334 MDCs (see Sect.~\ref{section_nature}) and use this sample to investigate the probability of the existence of high-mass prestellar cores (see Sect.~\ref{section_pcores}), and give constraints on the high-mass star formation process (see Sects.~\ref{section_HMSF}--\ref{section_scenario}).

\begin{figure*}[htpb]  
\centering
\includegraphics[width=18cm]{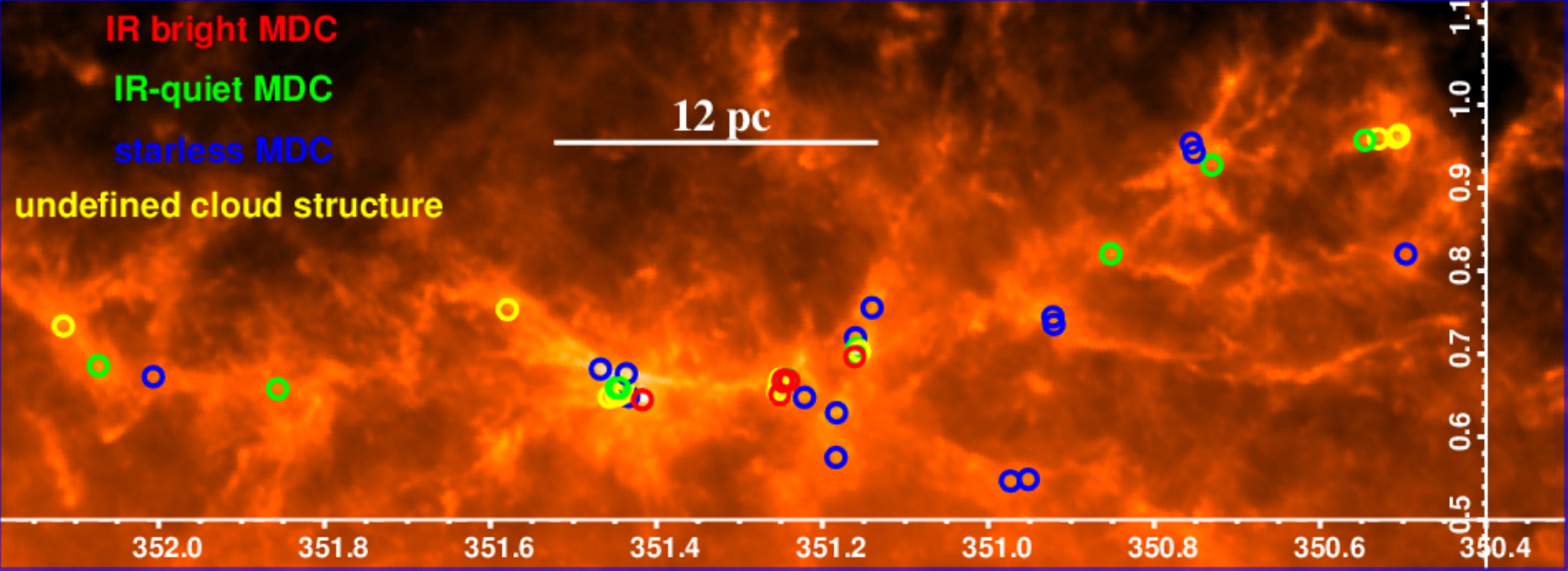}
\caption{Evolutionary subsamples of MDCs in NGC~6334, as defined in Sect.~\ref{section_nature}, plotted on the high-resolution (HPBW$~\sim18.2\arcsec$) $N_{\rm H_2}$ column-density image of the molecular complex. IR-bright MDCs cluster in the central part of the region whereas IR-quiet protostellar MDCs and starless MDC candidates are more widely distributed.}
\label{mix_sag}
\end{figure*} 

\subsection{Nature and evolutionary status of MDCs}
\label{section_nature}

To define the nature of MDCs, we use their properties as derived from our extraction and SED fits as well as their appearance on \emph{Herschel} and other submillimeter images (see Table~\ref{tab_parameters_6334} and Figs.~\ref{MDC_1}--\ref{MDC_43}).
Table~\ref{tab_nature} lists the 46 MDCs along with their SED properties and associated high-mass star formation signposts, which allow for determination of their probable nature.
Figure~\ref{mix_sag} locates MDCs of all types (starless, IR-Quiet, IR-Bright, undefined) on the column density image of the NGC~6334 molecular complex.

\subsubsection{Undefined cloud structures, doubtful MDCs}
\label{section_poc}

In principle, MDCs are local gravitational potential wells and should correspond to centrally concentrated cloud fragments, and thus far-infrared/(sub)millimeter emission peaks. \emph{getsources}, the compact source extraction software we used (see Sect.~\ref{section_extraction_getsources}), is powerful in deblending clusters of sources. It tends, however, to identify secondary sources around dominant ones. While some of these likely represent local gravitational wells, others may just be unbound cloud structures like the gas flows toward hubs/ridges.
We therefore inspected, by eye, the column density, 160\,$\mu$m, and ground-based (sub)millimeter images (see Figs.~\ref{MDC_1}--\ref{MDC_43}) that have better angular resolution than the \emph{Herschel} $250-500~\mu$m bands. From this inspection, we confirm 32 MDC candidates and identify 14 as not centrally peaked, undefined cloud structures. Submillimeter observations at even higher-angular resolutions toward some undefined cloud structures in particular confirm that they are indeed not centrally concentrated (Andr\'e et al., in prep.).

The undefined cloud structures often have SED fits of poor quality with reduced $\chi^2>5$, suggesting they are rather diffuse. Furthermore, MDCs identified as undefined cloud structures are mostly located close to major column density peaks (see Fig.~\ref{mix_sag}), whose intense far-infrared emission could have affected the source extraction. Despite our reservations, however, in the absence of a definite argument to remove these 14 undefined cloud structures from our sample, we keep them in Tables~\ref{tab_nature} and \ref{tab_parameters_6334}. Since their nature and evolutionary status are difficult to define, however, we exclude these undefined cloud structures in our following discussion.

\begin{figure*}[htpb]  
\centering
\includegraphics[width=13.5cm, angle = -90]{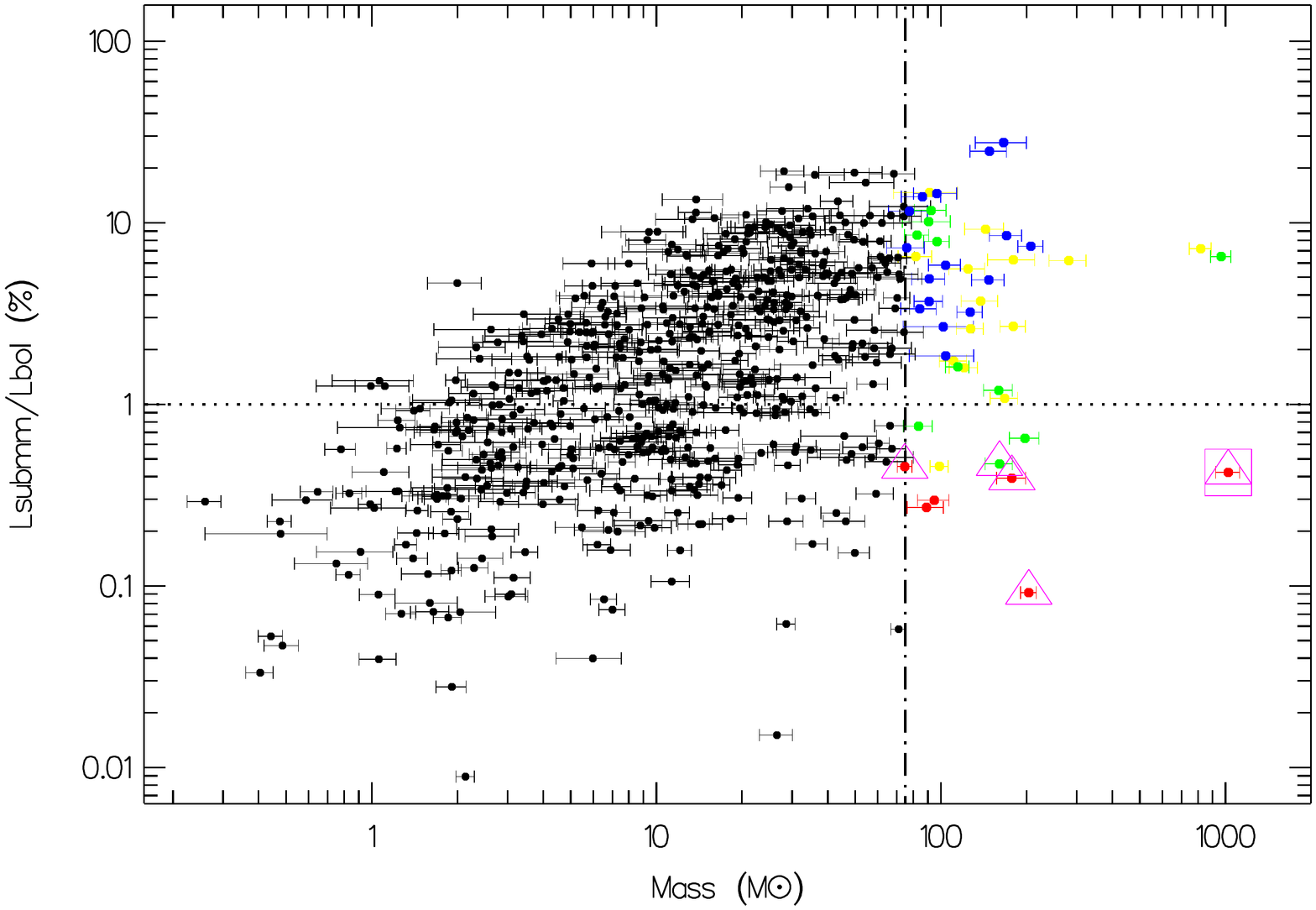}
\vskip -1.4cm

?

?

?

\caption{
$L_{\rm submm}$/ $L_{\rm bol}$ versus mass diagram of NGC~6334 dense cores (black dots). 
Pink squares and triangles denote MDCs associated with radio centimeter compact and CH$_{3}$OH maser 
sources, respectively. Red/green/blue/yellow dots highlight IR-bright MDCs, IR-quiet MDCs, starless MDC candidates, and 
undefined cloud structures, respectively (as defined in Sect.~\ref{section_nature}). 
Mass errors are given by \emph{MPFIT}.}
\label{Lsubmm_Lbol_Mass_6334}
\end{figure*}

\subsubsection{Separation between IR-quiet and IR-bright MDCs}
\label{section_bright_quiet}

To determine the evolutionary statuses of the 32 remaining MDCs, we used the definition of IR-bright versus IR-quiet protostellar MDCs of \cite{Motte07}, followed by \cite{Russeil10} and \cite{Nguyen11}. Based on the predicted mid-infrared emission of a B3-type stellar embryo, the minimum fluxes of IR-bright MDCs located at 1.75~kpc should be $S_{\rm 21\,\mu m}=10\,$Jy \citep{Motte10}, $S_{\rm 22\,\mu m}=12\,$Jy, and 
$S_{\rm 24\,\mu m}=15\,$Jy \citep{Russeil10}. 
With these criteria, IR-bright MDCs should already host at least one stellar embryo of mass larger than $8~\msun$ while IR-quiet MDCs only contain stellar embryos of lower mass, which could further accrete.
IR-quiet MDCs could therefore evolve into IR-bright MDCs of similar masses. For context, the HMPOs of \cite{Beuther02} are IR-bright cloud structures ranging from 1~pc clumps to 0.01~pc cores, among which are several tens of IR-bright MDCs.
In Fig.~\ref{F24_Mass_6334}, we separate IR-bright and IR-quiet MDCs using the mid-infrared flux limits given above. The \emph{Spitzer} 24~$\mu$m fluxes were primarily used but, in saturated areas, the 22~$\mu$m \emph{WISE} or 21~$\mu$m \emph{MSX} fluxes were the alternative choices (see Table~\ref{tab_21_22_24}). Among MDCs of Table~\ref{tab_nature}, six qualify as IR-bright protostellar MDCs: MDCs \#1, \#6, \#10, \#32, \#38, and \#46, as seen in Fig.~\ref{F24_Mass_6334}. Four of these are also associated with CH$_{3}$OH maser sources \citep{Pestalozzi05}.

Given their spatial sizes ($\sim$0.1~pc), MDCs cannot be precursors of single high-mass stars. MDCs with weak or even undetected mid-infrared emission probably host young protostars and/or prestellar cores. MDCs are qualified as "IR-quiet protostellar" if they are associated with compact emission at 70~$\mu$m. Most of these, however, are also detected at mid-infrared (21~$\mu$m to 24~$\mu$m) wavelengths. Similarly, a detection at both 21-24~$\mu$m and 70\,$\mu$m mitigates the misidentification of externally heated diffuse cores without internal stellar embryos as being IR-quiet MDCs. "Starless" MDC candidates are generally sources for which no compact emission at 70~$\mu$m emission can be distinguished above the noise level of \emph{Herschel} images. In a few cases, however, they have weak 70~$\mu$m emission at the level of or below that of the SED fits to longer far-infrared and submillimeter wavelengths (see, e.g., Figs.~\ref{MDC_5}--\ref{MDC_45}, also \citealt{Motte10}).

For each MDC, we compared the ratio of the submillimeter luminosity to the bolometric luminosity, $L_{\rm bol}$, for the IR-quiet versus IR-bright subsamples. The submillimeter luminosity, $L_{\rm submm}=L_{\rm \ge 350\,\mu m}$, is defined as the integrated luminosity below the resulting SED fit between 350~$\mu$m (857\,GHz) and 1200~$\mu$m (250\,GHz). The submillimeter to bolometric luminosity ratio has previously been used in the low-mass regime to distinguish between Class\,0 and Class\,I protostars. \cite{Andre00} found a separation in this ratio amongst their individual protostars (size~$\sim$~0.01--0.1~pc) with a luminosity ratio less than 1\% for Class\,I protostars. 
We cannot, however, directly use the same threshold values taken to separate Class\,0 and Class\,I protostars. Indeed, the physical sizes, stellar luminosities, and cloud environments of MDCs and low-mass protostars are different enough to justify a change of this ratio. In the intermediate- to high-mass regime, \cite{Hennemann10} and \cite{Nguyen11} found a clear separation between IR-bright and IR-quiet MDCs in terms of $L_{\rm \ge 350\,\mu m}$/$L_{\rm bol}$ ratio. Namely, IR-bright MDCs or UCH II regions have small ($<$1\%) ratios and IR-quiet or starless MDCs can have ratios ranging from 1\% to 20\%. 

Figure~\ref{Lsubmm_Lbol_Mass_6334} displays the submillimeter to bolometric luminosity ratio of all NGC~6334 dense cores as a function of their mass.
In this diagram, IR-bright MDCs, as defined above, are well separated from IR-quiet MDCs and lower-mass dense cores. They exhibit high masses ($M\ge 75~\msun$) and low submillimeter to bolometric luminosity ratios ($L_{\rm submm}/ L_{\rm bol}<0.5-1\%$). MDCs with $L_{\rm submm}/L_{\rm bol}$ ratios larger than $1\%$ with $\beta=2$ or 1.5, are all IR-quiet. Among the five MDCs with $L_{\rm submm}/L_{\rm bol}=0.5\%-1\%$, the one associated with compact infrared emission at 8~$\mu$m and 22~$\mu$m, MDC \#46, is an IR-bright protostellar MDC while the others are IR-quiet protostellar MDCs. NGC6334-I(N) or MDC \#14 itself has a $L_{\rm submm}/L_{\rm bol}=0.5\%$ ratio but its mid-infrared emission is small enough for it to be an IR-quiet protostellar MDC, in agreement with the \cite{Sandell00} and \cite{Brogan09} classifications of this object. Therefore, we confirm that there is a good but not perfect agreement between evolutionary classifications made with mid-infrared fluxes and $L_{\rm submm}/L_{\rm bol}$ ratios. 

The present sample of MDCs can help us evaluate the probability that 
high-mass prestellar cores exist in NGC~6334. 
Below, we first investigate if the starless MDCs of NGC~6334 could host 
high-mass prestellar cores (see Sect.~\ref{section_pcores}) 
and then discuss the lifetime of any prestellar phase (see Sect.~\ref{section_HMSF}).

\begin{figure*}[htpb]  
\centering
\includegraphics[width=16.3cm]{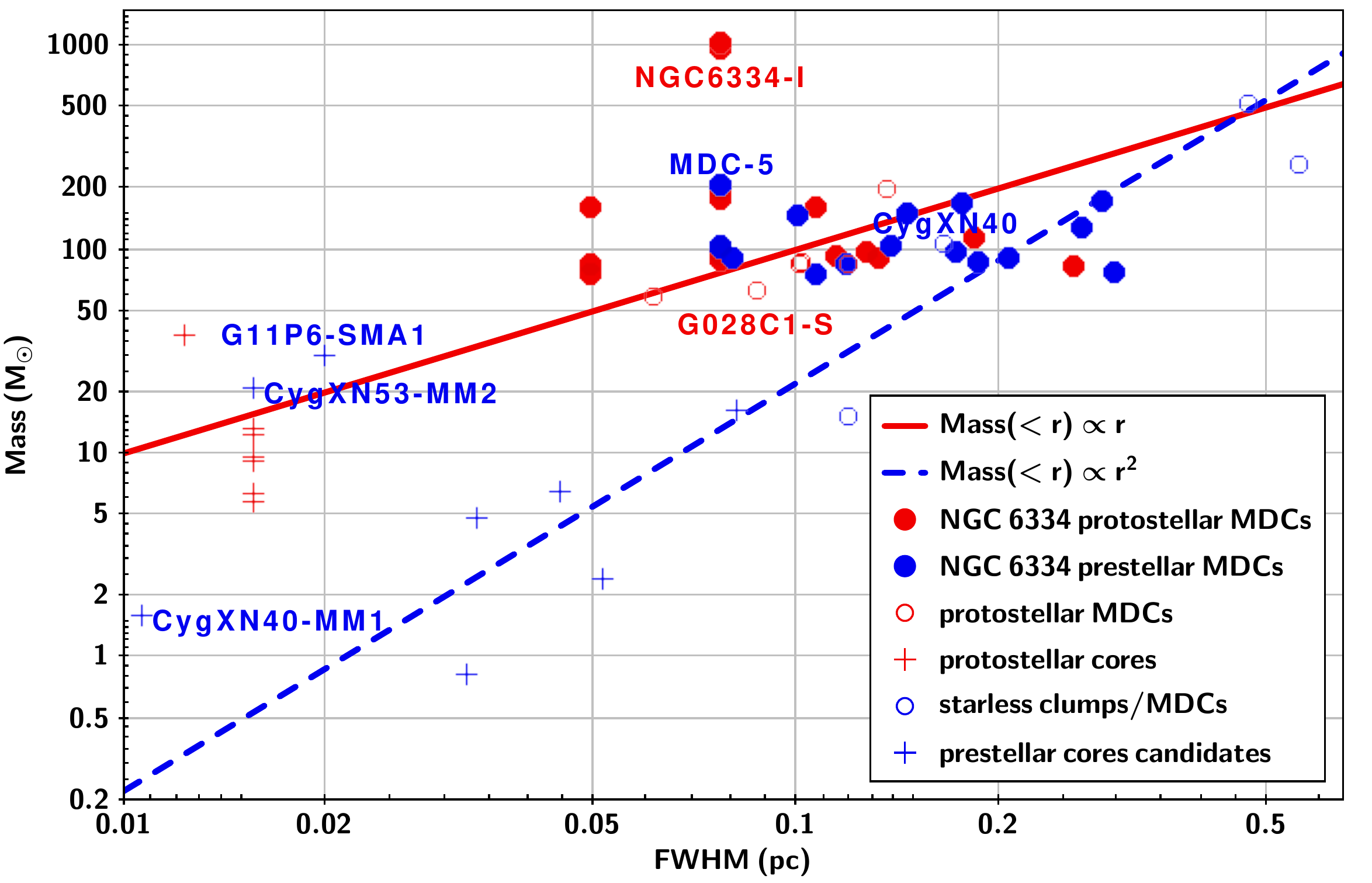}
\caption{Predicting gas concentration within NGC~6334 MDCs through a mass versus size diagram. 
Starless MDC candidates in NGC~6334, displayed in filled blue circles, 
are compared with a handful of massive starless clumps, and prestellar sub-fragments/cores 
\citep[][in blue open circles and crosses]{Bontemps10b,Butler12,Tan13,Duarte14,Wang14}, respectively. 
NGC~6334 protostellar MDCs, displayed in filled red circles, 
are themselves compared to five protostellar MDCs and their embedded high-mass protostars \citep[][in red open circles and crosses]{Motte07,Bontemps10b,Tan16}, respectively. 
The red line and blue dashed line represent mass radial power-laws of 
Mass(<r) $\propto r$ and Mass(<r) $\propto r^{2}$, expected to be followed by NGC~6334 protostellar and starless MDCs respectively.
The sample of 32 MDCs is too small to present any statistically meaningful mass versus size relation.
}
\label{M_r}
\end{figure*}

\subsection{Quest for high-mass prestellar cores in MDCs}
\label{section_pcores}

Our \emph{Herschel} study has identified a new population of MDCs; 16 starless candidates (see Table~\ref{tab_nature}). The present MDC sample can help us evaluate the probability that high-mass prestellar cores do exist in NGC~6334, especially within its starless MDCs. Thanks to our careful SED analysis using multi-wavelength \emph{Herschel} images, the number of starless MDC candidates is much larger than found using the SIMBA 1~mm image only by \cite{Russeil10}, that is, 16 to 1. The dust temperatures of starless MDCs have been measured to be $\sim$15~K (from 9.5~K to 21.1~K, see Fig.~\ref{distrib_subset} middle-right), almost five degrees lower than the 20~K uniform temperature assumed by \cite{Russeil10}. With 20~K instead of 15~K, mass estimates from a flux measurement at 1~mm are underestimated by $\sim$30\%. This difference alone can explain why the number of starless MDC candidates more massive than $75~\msun$ was largely underestimated before \emph{Herschel}. Indeed, if we were to suppose a uniform temperature of 20~K for each MDC, only one starless candidate, MDC \#26, would be considered massive enough. 
Figure~\ref{M_r} locates the starless MDC candidates and protostellar MDCs of NGC~6334 on a mass versus size diagram, where the gas mass concentration of other protostellar and starless samples is displayed \citep{Bontemps10b, Butler12, Tan13, Duarte14, Wang14}. An empirical relation close to Mass(<r) $\propto r^{2}$ is found to link starless clumps/MDCs and their prestellar core candidates.
With a flux-scaling correction consistent with this relation, the number of starless MDCs in NGC~6334 reduces to 13, showing that 16 is definitively an upper limit for the number of starless MDCs in NGC~6334.
 
That said, the natures of the 16 starless MDC candidates of Table~\ref{tab_nature} remain in question. Nine are characterized by poorly defined SEDs (with $5 <\chi^2 < 10$, see Col.~7 of Table~\ref{tab_parameters_6334}), casting doubts on the quality of their flux extraction and mass estimates, and thus their nature. Moreover, all MDCs with poor fits (except MDC \#21) are located against concentrated filamentary structures (see, e.g., Figs.~\ref{MDC_27} and~\ref{MDC_45}), explaining why fluxes extracted from the highest-resolution images deviate from those obtained from their SED fits, largely set by \emph{Herschel} fluxes at moderate-angular resolution. Such complex structures cast doubt on the ability of at least half of the $\sim$0.1~pc starless MDC candidates to concentrate enough mass into a $\sim$0.02~pc high-mass prestellar core. Of note, the location of MDC \#17, close to a dominant column density source, suggests it could instead be a gas flow structure attracted by the dominant source rather than a centrally-concentrated starless MDC hosting a high-mass prestellar core.

Visually speaking, the best starless MDC candidates could be MDCs \#21 (though it has a poor fit), \#44, \#35, \#11, and \#39 since they are the most isolated MDCs, with the least filamentary structures in their background. Their masses range from $80~\msun$ to $170~\msun$ but they have among the largest deconvolved sizes of our sample, 0.2--0.3~pc (see Col.~2 of Table~\ref{tab_parameters_6334}) like in the Rosette molecular cloud (\citealt{Motte10}). Their extracted characteristics lead to densities of $1-5\times 10^{5}$~cm$^{-3}$, more typical of low-mass prestellar cores than MDCs \citep[$10^5$~cm$^{-3}$ versus $10^6$~cm$^{-3}$, see][]{Ward-Thompson99,Motte07}.
In contrast, mass and density-wise ($150-210~\msun$, $n_{\rm H2}>$ $5\times 10^{6}$~cm$^{-3}$), MDCs \#5,  \#13, \#16, and \#17 would be the best starless MDC candidates. As mentioned above however, MDCs \#13 and \#16 have poorly defined SEDs and MDC \#17 may not be centrally concentrated enough. In the end, MDC \#5 is our best candidate starless MDC with its unresolved size, $\le$0.08~pc, and high mass, $\sim$210~$\msun$. Interestingly, these characteristics make it at least five times denser and thus more favorable to host a high-mass prestellar core than the starless MDC candidate of \cite{Tan13}.

To explore even further the probability of NGC~6334 MDCs containing high-mass prestellar cores or high-mass protostars, we used the mass versus size diagram of Fig.~\ref{M_r}. The masses and FWHMs of NGC~6334 MDCs are compared with those of two $\sim$0.1~pc MDC samples for which the $\sim$0.02~pc-scale content is known \citep[see][]{Bontemps10b,Tan13}. 
The gas mass concentration of the six most massive Cygnus~X MDCs follows a classical $M(<r) \propto r$ relation, using masses and sizes estimated by \cite{Motte07} and \cite{Bontemps10b}. This relation suggests these MDCs have a centrally concentrated density distribution of $\rho(r) \propto r^{-2}$ or steeper, similar to those of MDCs or hubs on $\sim$0.1\,--\,1~pc scales \citep[e.g.,][]{Beuther02,Didelon15} or low-mass protostellar envelopes on 0.01\,--\,0.1~pc scales \citep[e.g.,][]{Motte01}. 
By contrast, the starless structures found by submillimeter surveys or within infrared dark clouds are much less concentrated and tend to follow a M($<$r) $\propto$ r$^{2}$ relation \citep{Butler12,Tan13}. Note that the only known starless MDC N40 in Cygnus-X follows a similar relation (see \citealt{Bontemps10b}, Fig.~\ref{M_r}). Given their low density ($1-3\times 10^{5}$~cm$^{-3}$), these MDC fragments at 0.02~pc may better represent intermediate- to low-mass prestellar cores.  
Most starless MDC candidates in NGC~6334 are in fact 1--10 times denser than IRDC fragments (see  Fig.~\ref{distrib_subset} bottom-right). If they follow a $M(<r) \propto r^2$ relation, they could on average form $0.02$~pc cores, each of $1-5~\msun$ mass. Given that eight starless MDCs already display non-centrally peaked morphologies, and that the seven remaining MDCs are not all very dense, the maximum number of high-mass prestellar cores expected in NGC~6334 would be seven. If MDC \#5 is the only good starless MDC of NGC~6334, the minimum number of high-mass prestellar cores could be one. An ALMA project targeting the starless MDCs of NGC~6334 should soon reveal if they do indeed host groups of intermediate- to low-mass prestellar cores or simply a few dominant high-mass prestellar cores. 

\begin{table*}[htbp]
\begin{center}
\caption{Median properties and lifetime estimates for MDCs of different evolutionary states}
\label{lifetime_tab}
\begin{tabular}{| c | c | c | c | c |}
\cline{2-5}
\multicolumn{1}{c|}{}          &     IR-bright MDC     &    IR-quiet MDC     &   starless MDC  candidate   & O-B3 stars \\
\cline{2-4}
\hline
Number                         &      6                &     10              &     16                 &  150         \\
Size (pc)                      &    0.08               &    0.11             &     0.14               &     --         \\
<T$_{\rm dust}$> (K)           &      29               &      16             &     15                 &    --            \\
L$_{\rm bol}$ (\lsun)          &     9500              &     760             &     130                &    --              \\
Mass (\msun)                   &     140               &     106             &     103                &   --              \\
$<n_{\rm H_{2}}>$   (cm$^{-3}$)& 1.4 $\times$ 10$^{7}$   & 3.1 $\times$ 10$^{6}$   &  1.4 $\times$ 10$^{6}$   &     --      \\ 
Free-fall time$^1$            &  $9\times10^3$~yr &  $2\times10^4$~yr  &  $3\times10^4$~yr &    --           \\
Number of high-mass cores$^2$       &  \multicolumn{2}{c|}{32 high-mass protostars}  & 1-7 high-mass prestellar cores    &      --       \\
Statistical lifetime of high-mass cores$^3$  &\multicolumn{2}{c|}{$3\times 10^5$~yr} &  $1-7\times 10^4$~yr & $1.5\times 10^6$~yr  \\
\hline
\end{tabular}
\end{center}
\tablefoot{
$^1$ Free-fall time measured from the median values of the density averaged over the full MDC volume, which is approximately a sphere with a FWHM radius:
$<n_{\rm H_{2}}>_{\rm full}=<n_{\rm H_{2}}>/8$ and
$t_{\rm free-fall} = \sqrt{\frac{3~\pi}{32~G~\mu~m_{\rm H}~<n_{\rm H_{2}}>_{\rm full}}} $.  \\
$^2$ We assumed that protostellar MDCs host $\sim$2 protostars as was found 
for Cygnus-X by \cite{Bontemps10b}, while 1 to 7 starless MDCs could host one high-mass prestellar core. \\
$^3$ The statistical lifetimes of high-mass protostars/prestellar cores is estimated from the relative numbers of high-mass protostars/prestellar cores versus OB stars.}
\end{table*}

\subsection{Constraints on the high-mass star formation process}
\label{section_HMSF}

At the chosen 0.1~pc size, our large sample of 490 dense cores should be complete for $>$$15~\msun$ cloud structures,  according to completeness simulations performed for \emph{getsources} extractions in similar \emph{Herschel} images \citep[e.g.,][]{Konyves15, Marsh16} and the dense core mass distribution shown in Fig.~\ref{distrib_subset} Top. Therefore, the present sample must be complete for the $>$$75~\msun$ MDCs hosting massive young stellar objects and allows us to start building an evolutionary scenario for the formation of high-mass stars. 
While starless MDCs cannot all form high-mass stars (see Sect.~\ref{section_pcores}), IR-quiet MDCs will probably become IR-bright and form high-mass stars. Indeed, IR-quiet MDCs generally have more powerful SiO outflows and stronger mid-IR emission than a cluster of low-mass Class 0s \citep{Motte07, Russeil10}.

Table~\ref{lifetime_tab} lists the median characteristics of each MDC evolutionary phase and provides estimates of their statistical lifetime.
With the relative number of IR-bright and IR-quiet protostellar MDCs as well as starless MDC candidates with respect to OB stars in the NGC~6334 molecular complex, we can infer the statistical lifetimes of each evolutionary phase.
We use the work of \cite{Russeil12}, who performed a statistical census of O-B3 ionizing stars in the NGC 6334-NGC 6357 complex. They estimated a total number of $\sim$150 high-mass stars in NGC~6334, which encompasses $\sim$2/3 of the NGC~6334-6357 gas mass ($\sim$$7\times 10^5~\msun$).
We assume the median age of the 150 O-B3 stars of NGC~6334 to be $1.5\pm0.5\times 10^6$~yr according to the age of the main clusters identified by \cite{Kharchenko13}, \cite{Morales13}, and \cite{Getman14}. 
We also use the fragmentation level of Cygnus~X protostellar MDCs \citep{Bontemps10b} and assumed that IR-quiet and IR-bright protostellar MDCs in NGC~6334 should host, on average, two high-mass protostars. As for starless MDC candidates, given our discussion in Sect.~\ref{section_pcores}, we assume that only one to seven of them could host at most one high-mass prestellar core. Accordingly, we obtain lifetimes of $1.1\times10^{5}$~yr, $1.9\times10^{5}$~yr, and $1-7\times10^{4}$~yr, for the the IR-bright, IR-quiet, and prestellar phases of high-mass stars in NGC~6334, respectively. The complete duration of the protostellar phase, $3\times10^5$~yr, is only twice that of previous estimates\footnote{
        The statistical lifetime of protostellar MDCs has been corrected by \cite{Russeil10} from values given in \cite{Motte07}.} 
\citep[$\sim$ $1.5\times 10^{5}$~yr according to][]{Motte07,Russeil10}. The main difference is the fragmentation level here taken to be 2 while \cite{Motte07} and \cite{Russeil10} assumed that each MDC would form a single high-mass progenitor. The complete duration of the protostellar core phase corresponds to a few free-fall times of MDCs with $\sim$10$^{6}$~cm$^{-3}$ densities averaged over twice their FWHM and a single free-fall time of their $\sim$10$^{5}$~cm$^{-3}$ parental clump (Sect.~\ref{section_ridge}). 
The lifetime of the prestellar core phase itself measured to be $1-7\times 10^4$~yr could either be as short as a single free-fall time or even be nonexistent. 

Other constraints arise with the new population of 16 starless MDC candidates, which we found to account for 50\% of the complete sample of MDCs. This proportion is much larger than in previous MDC studies \citep[e.g.][2\%\footnote{
The SiO emission used by \cite{Motte07} to classify CygX-N40 as a protostellar MDC has since been largely if not totally associated to cloud-cloud collision by \cite{Duarte14}. CygX-N40 thus is the only starless MDC candidate in Cygnus X.
} 
and 6\%]{Motte07,Russeil10}. 
As stated in Sect.~\ref{section_pcores}, however, many starless MDCs listed in Table~\ref{tab_parameters_6334} may not be able to host high-mass prestellar cores. The short statistical lifetime of starless MDCs is discussed by \cite{Motte07}, who suggest that high-mass prestellar cores could be dynamically evolving into protostars. 

Recent studies showed that the formation of ridges and massive stellar clusters are dynamical events, which are tightly linked \citep[e.g.,][]{Nguyen13,Louvet16}. Indeed, both ridges and hubs are observed to form through a global free-fall collapse \citep[e.g.,][]{Schneider10a,Hill11,Hennemann12}, followed by intense formation of high-mass stars quoted as mini-starbursts \citep[e.g.,][]{Motte03,Nguyen11,Louvet14}.
Indeed, the shape and velocity drift of the south-western part of the ridge \citep[i.e., the connecting filament in Fig.~\ref{ridge_hubs}, see also][]{Zernickel13, Andre16} recalls sub-filaments falling onto ridges, like those around DR21 \citep{Schneider10a}.
Interestingly, the lack of high-mass MDCs within the south-western sub-filament suggests it is stretched while falling onto the gravitation potential of the NGC~6334 ridge (see MDC distribution in Fig.~\ref{mix_sag}). 
Moreover, a burst of star formation appears to have occurred recently in the central part of NGC~6334, corresponding to its ridge \citep{Willis13}.
In the context of the kinematical studies of ridges and hubs, the lifetime estimates of the present paper imply that either high-mass prestellar cores form and evolve in a single free-fall time of ridges/hubs or that their masses continue increasing well within the protostellar phase.
The prestellar phase of high-mass star formation may be a starless MDC containing a few low-mass prestellar cores, that subsequently evolves into an IR-quiet protostellar MDC hosting a low-mass protostellar core.

The main physical properties (size, temperature, and mass) of the different evolutionary subsamples exhibit clear differences, as shown in Table~\ref{lifetime_tab}. The median deconvolved size decreases by a factor of two during MDC evolution from starless candidate to IR-quiet and finally IR-bright MDC. While this increasing compactness suggests that MDCs continue contracting, gathering material as they evolve, it also reflects our improved ability to detect cold starless MDCs at longer wavelengths with lower angular resolution. 
We confirmed this decreasing size trend by comparing the 250~$\mu$m deconvolved sizes of all MDCs, and IR-bright protostellar MDCs. The average temperatures themselves rise from $\sim$15~K to $\sim$29~K, from starless or IR-quiet protostellar to IR-bright protostellar MDCs. Protostellar MDCs also seem to have masses increasing by $30\%$ from their IR-quiet to their IR-bright phases. More surprisingly, starless MDCs are on average a factor of four smaller in density than protostellar MDCs: 
$\sim$1.4$\times$ 10$^{5}$~cm$^{-3}$ versus $\sim$5$\times$10$^{6}$~cm$^{-3}$ (see also Fig.~\ref{distrib_subset}), respectively. 
Either protostars grow in stellar mass at the same time as their MDC gas reservoir condenses or starless MDC candidates identified here do not have the ability to form high-mass stars in the near future.

The spatial distribution of the different evolutionary subsamples is shown on the high-resolution column density map in Fig.~\ref{mix_sag}. MDCs are distributed along massive filaments and especially at their junctions, where material may accumulate \citep[e.g.,][]{Schneider12}. IR-bright MDCs are found at the very center of the region, within the NGC~6334 ridge and hubs (see Sect.~\ref{section_ridge} and Fig.~\ref{ridge_hubs}), suggesting, once again, that filaments   and protostars grow in mass simultaneously. 

\begin{figure*}[htpb]  
\centerline{\includegraphics[width=24cm, angle=0]{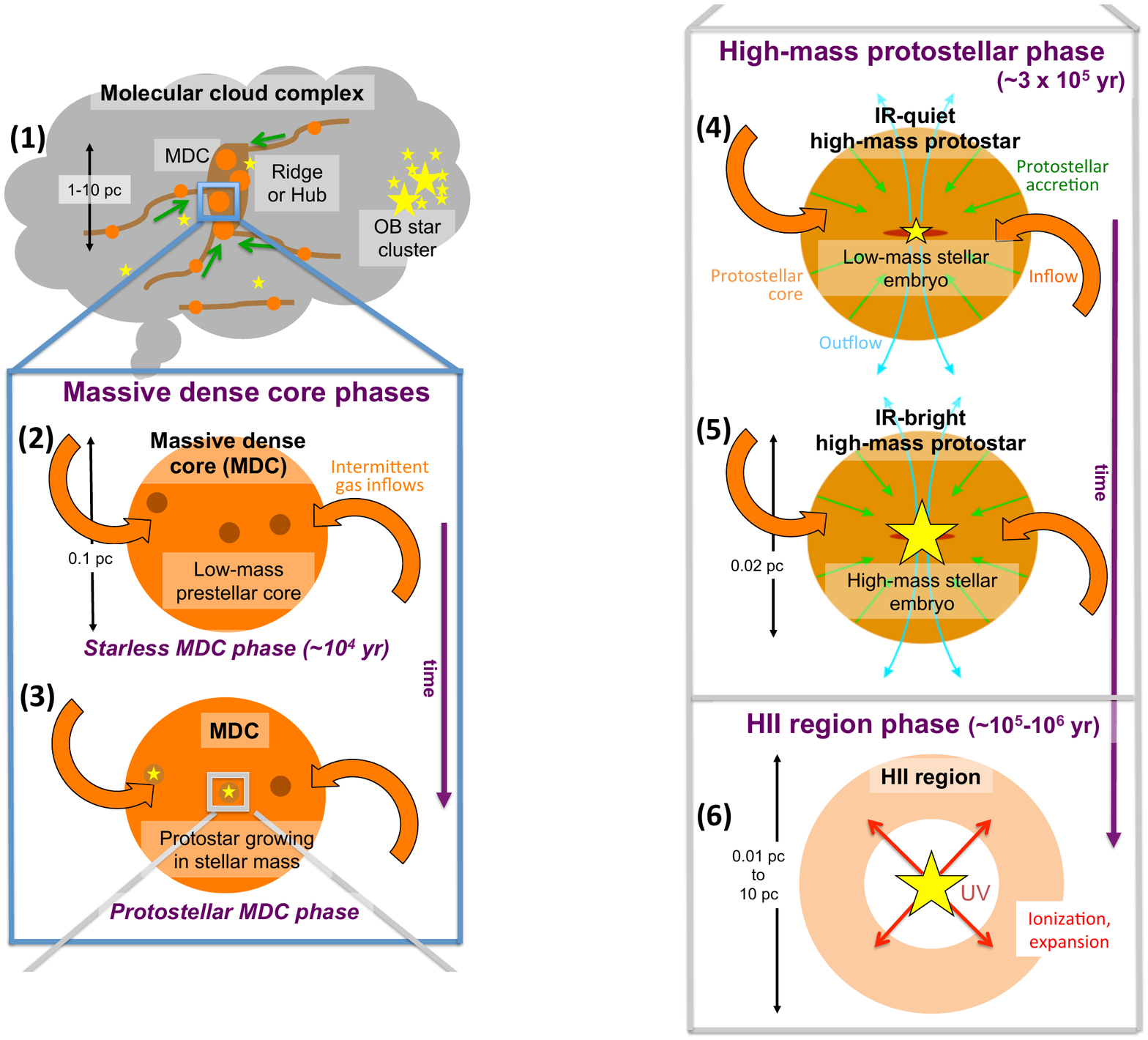}}
\caption{
Schematic evolutionary diagram proposed for the formation of high-mass stars.
\textbf{(1)} Massive filaments and spherical clumps, called ridges and hubs, host massive dense cores (MDCs, 0.1~pc) forming high-mass stars.
\textbf{(2)} During their starless phase, MDCs only harbor low-mass prestellar cores.
\textbf{(3)} IR-quiet MDCs become protostellar when hosting a stellar embryo of low mass.
The local 0.02~pc protostellar collapse is accompanied by the global $0.1-1$~pc collapse of MDCs and ridges/hubs.
\textbf{(4)} Protostellar envelope feeds from these gravitationally-driven inflows, leading to the formation of high-mass protostars. The latter are IR-quiet as long as their stellar embryos remain low-mass.
\textbf{(5)} High-mass protostars become IR-bright for stellar embryos with mass larger than $8~\msun$.
\textbf{(6)} The main accretion phase terminates when the stellar UV field ionizes the protostellar envelope and an \hii region develops.
}
\label{fig_scenario_hmsf}
\end{figure*}

\subsection{Proposed scenario for the formation of high-mass stars}
\label{section_scenario}
Combining the statistical constraints of the present paper on MDCs with results on ridges/hubs leads to a scenario for the formation of high-mass stars, which is sketched in Fig.~\ref{fig_scenario_hmsf} \citep[see also][]{Motte17}.
While the association between the later \hii region phase and gas ionization by the nascent OB star UV field is rather well known \citep[][see Fig.~\ref{fig_scenario_hmsf}, the $6^{\rm th}$ phase]{Churchwell02,Zavagno07}, the earliest phases have only recently been unveiled.  

Ridges and hubs represent the first and largest-scale phase of high-mass star formation (see Fig.~\ref{fig_scenario_hmsf}, $1^{\rm st}$ scheme). 
One of the main results of the \emph{Herschel}/HOBYS survey has been the recognition of ridges and hubs as hyper-massive clumps preferentially forming high-mass stars (e.g., \citealt{Hill11}; \citealt{Nguyen11}, see also present paper). \cite{Schneider10a}, \cite{Peretto13}, and \cite{Didelon15}, among others, showed that gas was continuously inflowing onto ridges or hubs, probably like the northern part of the connecting filament toward the NGC~6334 ridge (see Fig.~\ref{ridge_hubs}).

Ridges and hubs play a major role in driving gas flows and matter concentration from parsec scales down to the 0.1~pc and finally 0.01~pc scales of MDCs and protostellar cores (see Fig.~\ref{fig_scenario_hmsf}, $2^{\rm nd}-3^{\rm rd}$ and $4^{\rm th}-5^{\rm th}$ schemes).
While ridges form rich clusters of high-mass stars \citep[e.g.,][]{Hill11,Hennemann12,Nguyen13}, hubs may just form a single high-mass star \citep[e.g.,][]{Peretto13,Didelon15}. These structures may represent the parsec-scale reservoir, from which gas is accreted onto 0.02~pc-scale cores.

In such a dynamical scenario, ridges/hubs, MDCs, protostellar envelopes, and stellar embryos would most likely form simultaneously. A high-mass prestellar phase may therefore not be necessary for high-mass star formation. The present study demonstrates a possible lack of high-mass prestellar cores, with the largest statistics and the most accurate mass estimates so far obtained. 

At this stage, the most probable process to form high-mass stars thus is associated with converging flows and does not go through any high-mass prestellar core phase (see Fig.~\ref{fig_scenario_hmsf}). Therefore during their starless phase, MDCs could only host a couple of prestellar cores of low to intermediate mass ($2^{\rm nd}$ scheme). These prestellar cores should shortly become protostars, embedded within protostellar MDCs ($3^{\rm rd}$ scheme). Protostars qualify as high-mass and IR-quiet if they are embedded within MDCs ($>$$75~\msun$ within 0.1~pc) but their stellar embryos remain low-mass, $<$$8~\msun$ ($4^{\rm th}$ scheme). When large amounts of inflowing gas reach these low-mass protostars, they could become high-mass IR-bright protostars with a $>$$8~\msun$ stellar embryo ($5^{\rm th}$ scheme), \hii regions ($6^{\rm th}$ scheme), and finally O-B3 type stars.

\section{Conclusions}

To improve our knowledge of the high-mass star formation process, we performed an unbiased study of its earliest phases in the entire NGC~6334 molecular cloud complex. As part of the \emph{Herschel}/HOBYS key program, we especially aimed at finding MDCs hosting the high-mass analogs of low-mass prestellar cores and Class~0 protostars. We used \emph{Herschel} far-infrared and submillimeter images together with mid-infrared and (sub)millimeter ground-based data (see Table~\ref{tab_images_properties}) to obtain a complete census of 32 MDCs, which are the most likely progenitors of high-mass stars at 0.1~pc scales. Our main findings can be summarized as follows:

\begin{enumerate}
\item 
The \emph{Herschel} imaging of NGC~6334 gives a complete view of the cloud structures ranging from 0.05~pc to 30~pc, from dense cores to clouds (see Figs.~\ref{F70_coldens}, and~\ref{appendix_nh2}--\ref{appendix_850}). In the column density image, we outlined the NGC~6334 ridge and hubs, which are $\sim$10$^{5}$~cm$^{-3}$-density, 1~pc clumps, believed to form high-mass stars (see Fig.~\ref{ridge_hubs}).
\item 
The exhaustive dataset gathered for the well-known NGC~6334 complex has been used as a template to determine the necessary analysis steps of HOBYS studies, focused on high-mass young stellar objects. We used \emph{getsources}, an extraction software dedicated to multi-wavelength datasets. \emph{getsources} performed a multi-resolution analysis of \emph{Herschel} and complementary images, extracted 4733 compact sources, and provided their multi-wavelength fluxes.
\item 
We established a detailed procedure, for which including (sub)millimeter fluxes and scaling \emph{Herschel}/SPIRE fluxes to $160~\mu$m or $250~\mu$m FWHM sizes are crucial steps. We visually inspected all far-infrared to millimeter images independently and carefully constrained the complete SED of each compact source (see, e.g., Figs.~\ref{MDC_1}--\ref{MDC_43}). Using criteria developed to identify the precursors of high-mass stars, we ensured extracted sources are genuine dense cloud structures. The resulting sample contains 490 well-characterized dense cores with typical FWHM sizes of $\sim$0.1~pc (see Table~\ref{cores_properties}). 
\item 
We cross-correlated the NGC~6334 dense core sample with catalogs of compact radio centimeter and CH$_{3}$OH maser sources, which are signposts of high-mass star formation. We determined a lower mass limit of $75~\msun$ for high-mass star formation to occur in the NGC~6334 region and listed the multi-wavelength fluxes and physical characteristics of 32 NGC~6334 MDCs and 14 undefined cloud structures (see Tables~\ref{tab_nature}--\ref{tab_parameters_6334} and Tables~\ref{tab_getsources_70}--\ref{tab_getsources_1200}). MDCs have temperatures ranging from 9.5~K to 40~K, masses from $75~\msun$ to $1000~\msun$, and volume-averaged densities from $1~\times 10^5$~cm$^{-3}$ to $7\times 10^7$~cm$^{-3}$ (see Table~\ref{mdcs_properties} and Fig.~\ref{distrib_subset}).
\item 
We used mid-infrared catalogs and images to expand the SEDs of the MDCs beyond far-infrared to millimeter fluxes. Integrating MDC fluxes from $3.6~\mu$m to $1200~\mu$m leads to bolometric luminosities ranging from $10~\lsun$ to $9\times 10^4~\lsun$ (see Table~\ref{mdcs_properties}).
We separated the MDCs into IR-bright (evolved) and IR-quiet (young) protostellar using the flux of their compact counterparts at $21-24~\mu$m (see Fig.~\ref{F24_Mass_6334}). This evolutionary discriminator is consistent with a submillimeter-to-bolometric luminosity ratio limit of $ L_{\rm submm}/L_{\rm bol} \sim 0.5-1\%$ (see Fig.~\ref{Lsubmm_Lbol_Mass_6334}). Starless MDC candidates themselves are cold cloud fragments which are associated with at most a weak and compact $70~\mu$m emission, which falls on the SED fits to their $\ge$160~$\mu$m fluxes. 
\item
Using \emph{Herschel} and complementary (sub)millimeter data, we were able to constrain properly SEDs with cold temperatures. As a consequence, the present study allowed the discovery of 16 starless MDC candidates (see Figs.~\ref{MDC_5}--\ref{MDC_45}). The nature of these MDCs must be further investigated. For various reasons, we doubt their ability to concentrate gas to high density and thus form high-mass stars in the near-future (see Fig.~\ref{M_r}). The high-mass prestellar core phase thus remains elusive and may even not exist.
\item
We used the respective proportion of MDCs found in the different evolutionary subsamples (6 IR-bright, 10 IR-quiet, and 16 starless) to infer their statistical lifetimes. Assumptions on their sub-fragmentation level lead to protostellar and prestellar lifetimes for high-mass star formation of $3\times 10^5$~yr and $1-7\times 10^4$~yr, respectively. We also found a clear increase in temperature and density in MDCs from the starless phase to the IR-quiet and IR-bright phases, suggesting that while heating its parental MDC gas, the protostellar embryo, and  its parental MDC, grow in mass simultaneously (see Table~\ref{lifetime_tab}). During star formation, MDCs tend to become more compact and cluster in denser parts of NGC~6334 (see Fig.~\ref{mix_sag}), underlining the link between the formation of ridges/hubs and high-mass stars.
\item
The present study shows that high-mass prestellar cores are still to be found. This lack of good candidates agrees with previous cloud structure analyses made by, for example, HOBYS. It favors a scenario wherein ridges/hubs, MDCs, and high-mass protostellar embryos form simultaneously, as shown in Fig.~\ref{fig_scenario_hmsf}.
\end{enumerate}

\begin{acknowledgement}
We are grateful to Alexander Men'shchikov for his help and discussions on \emph{getsources}. 
SPIRE has been developed by a consortium of institutes led 
by Cardiff Univ. (UK) and including Univ. Lethbridge (Canada); NAOC (China); 
CEA, LAM (France); IFSI, Univ. Padua (Italy); IAC (Spain); 
Stockholm Observatory (Sweden); Imperial College London, RAL, UCL-MSSL, UKATC, 
Univ. Sussex (UK); and Caltech, JPL, NHSC, Univ. Colorado (USA). 
This development has been supported by national funding agencies: CSA (Canada); 
NAOC (China); CEA, CNES, CNRS (France); ASI (Italy); MCINN (Spain); 
SNSB (Sweden); STFC, UKSA (UK); and NASA (USA). PACS has been developed by 
a consortium of institutes led by MPE (Germany) and including UVIE (Austria); 
KU Leuven, CSL, IMEC (Belgium); CEA, LAM (France); MPIA (Germany); 
INAF-IFSI/OAA/OAP/OAT, LENS, SISSA (Italy); IAC (Spain). 
This development has been supported by the funding agencies BMVIT (Austria), 
ESA-PRODEX (Belgium), CEA/CNES (France), DLR (Germany), ASI/INAF (Italy), and CICYT/MCYT (Spain). 
This research has made use of the SIMBAD database, operated at CDS, Strasbourg, France. 
Part of this work was supported by the ANR (Agence Nationale pour la Recherche) 
project ``PROBeS'', number ANR-08-BLAN-0241.
We acknowledge financial support from ``Programme National de Physique Stellaire'' (PNPS) and program ``Physique et Chime du Milieu Interstellaire'' (PCMI) of CNRS/INSU, France. GJW gratefully acknowledges the receipt of a Leverhulme Emeritus Professorial Fellowship.
\end{acknowledgement}

\bibliographystyle{aa}
\bibliography{jtige.bib}

\clearpage
 
\appendix

\begin{center}
\section{HOBYS catalogs for the 46 MDCs found in NGC~6334}
\end{center}
We present in this appendix the flux catalogs 
for the 46 MDCs discussed in the text.

\noindent
Catalog entries are as follows: 

\begin{enumerate}

\item \emph{Getsources} source number.

\item Name = HOBYS\_J prefix directly followed by a tag created from the J2000 sexagesimal coordinates.

\item Flag for the fluxes reliability and upper limit (see criteria in Sect.~\ref{section_extraction_getsources}). \\
`0' means that the flux is considered unreliable because the source is too weak, $S^{\rm peak}/\sigma<2$ and/or $S^{\rm int}/\sigma<2$. Its error, $2\sigma$, is used as an upper limit on SEDs when $\lambda\ge160$~$\mu$m and unused otherwise (see Sect.~\ref{section_fitting}). \\
`1' means that the flux is considered reliable. It is used for SED fits when $\lambda\ge160$~$\mu$m and when $\lambda\ge70$~$\mu$m for $T_{\rm dust}>32$~K. \\
`2' means that the flux is not considered reliable because the source is too large, \emph{FWHM}~$>0.3$~pc, and/or too elliptical, $e>2$. Its estimated flux is used as an upper limit on SEDs when $\lambda\ge160$~$\mu$m or when $\lambda\ge70$~$\mu$m for $T_{\rm dust}>32$~K and it is unused otherwise.

\item Estimate of the peak flux and its error in Jy/beam.

\item Estimate of the integrated flux and its error in Jy. Flux scaling has not been applied.

\item Major/minor (non-deconvolved) size estimate (FWHM) of the source in arcseconds. 

\item Position angle of the major axis, east of north, in degrees.

\end{enumerate}

\begin{table*}
\centering
\caption{HOBYS catalog of \emph{Herschel} fluxes extracted by \emph{getsources} 
for the 46 MDCs found in NGC~6334}
\label{tab_getsources_coordinates}
\begin{tabular}{| c| c || c | c |}
\hline
 Number          &  Name     & RA$_{\rm 2000}$  & DEC$_{\rm 2000}$                \\                     
\emph{getsources}&  HOBYS\_J  & (h m s)      & (\ensuremath{^\circ}~$^{\prime}$~$^{\prime\prime}$)  \\
     (1)         &    (2)    &              &                              \\
\hline
\hline
  1 & 172053.4-354702 & 17:20:53.47 & -35:47:02.3\\
  2 & 172055.2-354500 & 17:20:55.23 & -35:45:00.0\\
  3 & 172056.4-354528 & 17:20:56.48 & -35:45:28.9\\
  4 & 172053.6-354521 & 17:20:53.66 & -35:45:21.1\\
  5 & 172053.0-354317 & 17:20:53.03 & -35:43:17.0\\
  6 & 171957.5-355752 & 17:19:57.55 & -35:57:52.1\\
  7 & 171955.7-355749 & 17:19:55.74 & -35:57:49.9\\
  8 & 172057.4-354517 & 17:20:57.47 & -35:45:17.8\\
  9 & 172023.0-355443 & 17:20:23.05 & -35:54:43.6\\
  10 & 172018.0-355456 & 17:20:18.04 & -35:54:56.5\\
  11 & 172002.1-361214 & 17:20:02.16 & -36:12:14.5\\
  12 & 172020.8-355432 & 17:20:20.86 & -35:54:32.9\\
  13 & 171907.2-360820 & 17:19:07.22 & -36:08:20.2\\
  14 & 172054.6-354511 & 17:20:54.67 & -35:45:11.0\\
  15 & 171954.7-355739 & 17:19:54.70 & -35:57:39.1\\
  16 & 171905.2-360757 & 17:19:05.24 & -36:07:57.8\\
  17 & 171951.8-355707 & 17:19:51.86 & -35:57:07.7\\
  18 & 172228.8-350933 & 17:22:28.89 & -35:09:33.8\\
  19 & 172022.9-355500 & 17:20:22.97 & -35:55:00.3\\
  20 & 172023.8-355450 & 17:20:23.83 & -35:54:50.4\\
  21 & 172030.9-360054 & 17:20:30.91 & -36:00:54.8\\
  22 & 171700.4-362050 & 17:17:00.41 & -36:20:50.2\\
  23 & 171701.8-362045 & 17:17:01.82 & -36:20:45.7\\
  24 & 171834.8-360846 & 17:18:34.83 & -36:08:46.0\\
  25 & 172019.2-355422 & 17:20:19.28 & -35:54:22.8\\
  26 & 172055.3-354605 & 17:20:55.36 & -35:46:05.7\\
  27 & 171939.6-355652 & 17:19:39.66 & -35:56:52.0\\
  28 & 172049.1-354500 & 17:20:49.10 & -35:45:00.6\\
  29 & 171954.9-355755 & 17:19:54.98 & -35:57:55.5\\
  30 & 171709.0-361903 & 17:17:09.03 & -36:19:03.3\\
  31 & 171745.8-360854 & 17:17:45.81 & -36:08:54.8\\
  32 & 172019.5-355440 & 17:20:19.50 & -35:54:40.4\\
  33 & 172233.4-351316 & 17:22:33.45 & -35:13:16.7\\
  34 & 172054.5-353518 & 17:20:54.57 & -35:35:18.3\\
  35 & 171958.0-361313 & 17:19:58.00 & -36:13:13.5\\
  36 & 172017.7-355906 & 17:20:17.70 & -35:59:06.5\\
  37 & 171748.0-361102 & 17:17:48.01 & -36:11:02.5\\
  38 & 172018.8-355435 & 17:20:18.88 & -35:54:35.3\\
  39 & 171733.9-362613 & 17:17:33.93 & -36:26:13.1\\
  40 & 172019.7-355633 & 17:20:19.72 & -35:56:33.9\\
  41 & 172017.9-355446 & 17:20:17.96 & -35:54:46.8\\
  42 & 172204.1-352454 & 17:22:04.14 & -35:24:54.6\\
  43 & 172059.4-354457 & 17:20:59.48 & -35:44:57.2\\
  44 & 172225.5-351659 & 17:22:25.58 & -35:16:59.9\\
  45 & 171747.9-360928 & 17:17:47.95 & -36:09:28.5\\
  46 & 172024.0-355459 & 17:20:24.00 & -35:54:59.7\\
\hline
\end{tabular}
\end{table*}

\clearpage


\begin{table*}
\centering
\caption{Table~\ref{tab_getsources_coordinates} (continued) - \emph{Herschel}-70~$\mu$m}
\label{tab_getsources_70}
\begin{tabular}{| c || c | c | c | c | c | c |}
\hline
 Number            &   Tag$_{\rm 070}$  & S$^{\rm peak}_{\rm 070}$  & S$^{\rm int}_{\rm 070}$   & FWHM$^{\rm maj}_{\rm 070}$  & FWHM$^{\rm min}_{\rm 070}$  &  PA$_{\rm 070}$ \\
        Id         &                &   (Jy/beam)          &      (Jy)         &  (arcsec)      &   (arcsec)     &  (degrees)  \\
                   &   (3)          &         (4)          &      (5)            &    (6)         &     (6)        &     (7)       \\                            
\hline
\hline 
  1 & 1 & 1560 $\pm$  20 & 11700 $\pm$  60 & 14.7 & 12.1 & 178\\
  2 & 2 & 2.9 $\pm$ 1.0 & 49.8 $\pm$  2.6 & 21.2 & 15.4 & 174\\
  3 & 2 & 0.9 $\pm$  0.2 & 2.9 $\pm$  0.3 & 14.3 & 6.9 & 102\\
  4 & 0 &                     &                    &     &     &  \\
  5 & 0 &                     &                    &     &     &  \\
  6 & 2 & 1160 $\pm$  3 & 14100 $\pm$  20 & 19.8 & 13.4 & 124\\
  7 & 0 &                     &                    &     &     &  \\
  8 & 0 &                     &                    &     &     &  \\
  9 & 0 &                     &                    &     &     &  \\
  10 & 1 & 268 $\pm$  59 & 765 $\pm$  96 & 10.8 & 7.7 & 110\\
  11 & 0 &                     &                    &     &     &  \\
  12 & 0 &                     &                    &     &     &  \\
  13 & 0 &                     &                    &     &     &  \\
  14 & 1 & 98.1 $\pm$  1.8 & 311 $\pm$  4 & 11.9 & 6.1 & 98\\
  15 & 2 & 9.7 $\pm$  3.2 & 213 $\pm$  8 & 24.4 & 16.6 & 61\\
  16 & 0 &                     &                    &     &     &  \\
  17 & 0 &                     &                    &     &     &  \\
  18 & 0 &                     &                    &     &     &  \\
  19 & 0 &                     &                    &     &     &  \\
  20 & 0 &                     &                    &     &     &  \\
  21 & 0 &                     &                    &     &     &  \\
  22 & 0 &                     &                    &     &     &  \\
  23 & 0 &                     &                    &     &     &  \\
  24 & 0 &                     &                    &     &     &  \\
  25 & 0 &                     &                    &     &     &  \\
  26 & 0 &                     &                    &     &     &  \\
  27 & 0 &                     &                    &     &     &  \\
  28 & 0 &                     &                    &     &     &  \\
  29 & 1 & 57.8 $\pm$  18.6 & 113 $\pm$  23 & 8.0 & 7.5 & 125\\
  30 & 0 &                     &                    &     &     &  \\
  31 & 0 &                     &                    &     &     &  \\
  32 & 1 & 99.3 $\pm$  41.3 & 238 $\pm$  46 & 10.3 & 6.5 & 139\\
  33 & 0 &                     &                    &     &     &  \\
  34 & 0 &                     &                    &     &     &  \\
  35 & 0 &                     &                    &     &     &  \\
  36 & 1 & 0.8 $\pm$  0.2 & 0.7 $\pm$  0.3 & 6.3 & 5.9 & 54\\
  37 & 0 &                     &                    &     &     &  \\
  38 & 2 & 336 $\pm$  2 & 6380 $\pm$  14 & 22.9 & 21.3 & 50\\
  39 & 0 &                     &                    &     &     &  \\
  40 & 0 &                     &                    &     &     &  \\
  41 & 1 & 153 $\pm$  45 & 266 $\pm$  49 & 11.1 & 5.9 & 86\\
  42 & 0 &                     &                    &     &     &  \\
  43 & 1 & 0.8 $\pm$ 0.3 & 2.7 $\pm$ 0.3  & 13.2 & 10.4 & 146\\
  44 & 0 &                     &                    &     &     &  \\
  45 & 0 &                     &                    &     &     &  \\
  46 & 1 & 186 $\pm$  12 & 304 $\pm$  15 & 6.3 & 5.9 & 89\\
\hline
\end{tabular}
\end{table*}

\clearpage


\begin{table*}
\centering
\caption{Table~\ref{tab_getsources_coordinates} (continued) - \emph{Herschel}-160~$\mu$m}
\begin{tabular}{| c || c | c | c | c | c | c |}
\hline
 Number            &   Tag$_{\rm 160}$  & S$^{\rm peak}_{\rm 160}$  & S$^{\rm int}_{\rm 160}$  & FWHM$^{\rm maj}_{\rm 160}$  & FWHM$^{\rm min}_{\rm 160}$  &  PA$_{\rm 160}$ \\
  Id         &                &   (Jy/beam)          &      (Jy)         &  (arcsec)      &   (arcsec)     &  (degrees)  \\                     
\hline
\hline
  1 & 1 & 2180 $\pm$ 56 & 9660 $\pm$ 116 & 23.6 & 20.0 & 94\\
  2 & 1 & 167 $\pm$ 29 & 468 $\pm$ 35 & 22.8 & 13.2 & 169\\
  3 & 1 & 111 $\pm$ 44 & 271 $\pm$ 44 & 19.1 & 11.7 & 50\\
  4 & 2 & 26.4 $\pm$ 46.9 & 14.8 $\pm$ 46.9 & 11.7 & 11.7 & 108\\
  5 & 1 & 30.4 $\pm$ 6.7 & 97.2 $\pm$ 10.9 & 21.1 & 17.0 & 19\\
  6 & 1 & 2240 $\pm$ 39 & 7450 $\pm$ 68 & 19.3 & 17.6 & 145\\
  7 & 1 & 720 $\pm$ 41 & 1310 $\pm$ 41 & 16.0 & 13.6 & 35\\
  8 & 2 & 39.5 $\pm$ 25.4 & 48.0 $\pm$ 25.4  & 14.9 & 11.7 & 179\\
  9 & 1 & 148 $\pm$ 18 & 277 $\pm$ 23 & 19.5 & 11.7 & 134\\
  10 & 1 & 937 $\pm$ 19 & 2080 $\pm$ 19 & 18.4 & 16.6 & 112\\
  11 & 2 & 3.8 $\pm$ 2.0 & 32.6 $\pm$ 5.2 & 56.5 & 21.1 & 53\\
  12 & 1 & 497 $\pm$ 20 & 832 $\pm$ 20 & 20.2 & 11.7 & 2\\
  13 & 2 & 0.3 $\pm$ 0.3 & 1.7 $\pm$ 0.6 & 32.8 & 29.0 & 88\\
  14 & 1 & 858 $\pm$ 35 & 1262 $\pm$ 35 & 14.5 & 11.7 & 119\\
  15 & 1 & 225 $\pm$ 43 & 553 $\pm$ 52 & 21.6 & 13.7 & 101\\
  16 & 2 & -1.0 $\pm$ 0.4 & -4.8 $\pm$ 0.6 & 33.1 & 21.4 & 134\\
  17 & 1 & 45.6 $\pm$ 14.1 & 111 $\pm$ 19 & 24.9 & 15.5 & 172\\
  18 & 2 & 0.7 $\pm$ 0.3 & 17.0 $\pm$ 0.9 & 65.5 & 42.8 & 61\\
  19 & 2 & 47.8 $\pm$ 4.9 & 38.4 $\pm$ 4.9 & 11.7 & 11.7 & 137\\
  20 & 1 & 102 $\pm$ 17 & 229 $\pm$ 17 & 20.2 & 15.7 & 129\\
  21 & 1 & 15.3 $\pm$ 1.3 & 163 $\pm$ 4 & 42.1 & 31.6 & 8\\
  22 & 2 & 4.0 $\pm$ 9.0 & -2.7 $\pm$ 9.0 & 11.7 & 11.7 & 64\\
  23 & 1 & 163 $\pm$ 36 & 231 $\pm$ 36 & 15.2 & 11.7 & 68\\
  24 & 1 & 14.2 $\pm$ 0.9 & 108 $\pm$ 2 & 31.7 & 24.8 & 109\\
  25 & 1 & 123 $\pm$ 20 & 204 $\pm$ 20 & 19.3 & 11.7 & 91\\
  26 & 2 & 29.5 $\pm$ 64.0 & 43.7 $\pm$ 64.2  & 17.5 & 11.7 & 168\\
  28 & 2 & 36.2 $\pm$ 24.9 & 64.7 $\pm$ 32.9 & 25.2 & 11.7 & 23\\
  27 & 1 & 16.6 $\pm$ 2.8 & 30.5 $\pm$ 2.8 & 15.6 & 12.7 & 173\\
  29 & 1 & 458 $\pm$ 38 & 918 $\pm$ 38 & 17.0 & 14.7 & 110\\
  30 & 1 & 8.9 $\pm$ 2.4 & 34.4 $\pm$ 4.3 & 27.1 & 16.8 & 86\\
  31 & 1 & 2.4 $\pm$ 0.8 & 8.2 $\pm$ 1.0 & 28.7 & 15.3 & 133\\
  32 & 1 & 818 $\pm$ 22 & 1290 $\pm$ 22 & 19.2 & 11.7 & 82\\
  33 & 2 & 2.5 $\pm$ 0.7 & 6.9 $\pm$ 0.9 & 28.5 & 11.7 & 158\\
  34 & 2 & 2.9 $\pm$ 1.9 & 5.4 $\pm$ 2.4  & 24.4 & 13.7 & 174\\
  35 & 2 & 7.9 $\pm$ 1.6 & 26.8 $\pm$ 3.4 & 47.4 & 14.9 & 175\\
  36 & 1 & 37.5 $\pm$ 11.7 & 61.0 $\pm$ 11.7 & 19.4 & 11.7 & 76\\
  37 & 2 & 2.1 $\pm$ 0.8 & 6.2 $\pm$ 1.2 & 40.8 & 11.9 & 86\\
  38 & 1 & 655 $\pm$ 20 & 2870 $\pm$ 56 & 25.1 & 18.0 & 77\\
  39 & 1 & 1.9 $\pm$ 0.8 & 7.0 $\pm$ 1.5  & 19.5 & 18.1 & 172\\
  40 & 1 & 37.9 $\pm$ 5.7 & 98.7 $\pm$ 8.2 & 25.1 & 13.1 & 69\\
  41 & 1 & 530 $\pm$ 20 & 588 $\pm$ 20 & 11.7 & 11.7 & 165\\
  42 & 1 & 2.4 $\pm$ 0.4 & 15.5 $\pm$ 1.0 & 34.0 & 19.6 & 78\\
  43 & 1 & 13.8 $\pm$ 5.2 & 24.0 $\pm$ 5.2  & 16.6 & 13.9 & 161\\
  44 & 2 & 0.6 $\pm$ 0.2 & 11.2 $\pm$ 0.9 & 56.3 & 53.4 & 110\\
  45 & 2 & 5.0 $\pm$ 1.2 & 18.9 $\pm$ 1.7 & 31.8 & 14.7 & 100\\
  46 & 1 & 535 $\pm$ 17 & 710 $\pm$ 17 & 12.5 & 11.7 & 75\\
\hline
\end{tabular}
\end{table*}


\clearpage

\begin{table*}
\centering
\caption{Table~\ref{tab_getsources_coordinates} (continued) - \emph{Herschel}-250~$\mu$m}
\begin{tabular}{| c || c | c | c | c | c | c |}
\hline
 Number            &   Tag$_{\rm 250}$  & S$^{\rm peak}_{\rm 250}$  & S$^{\rm int}_{\rm 250}$   & FWHM$^{\rm maj}_{\rm 250}$  & FWHM$^{\rm min}_{\rm 250}$  &  PA$_{\rm 250}$ \\
  Id         &                &   (Jy/beam)          &      (Jy)         &  (arcsec)      &   (arcsec)     &  (degrees)  \\                           
\hline
\hline
  1 & 1 & 2660 $\pm$ 38 & 3320 $\pm$ 39 & 20.5 & 18.2 & 50\\
  2 & 1 & 666 $\pm$ 37 & 781 $\pm$ 32 & 19.9 & 18.2 & 159\\
  3 & 1 & 445 $\pm$ 42 & 452 $\pm$ 37 & 19.3 & 18.2 & 62\\
  4 & 1 & 139 $\pm$ 41 & 135 $\pm$ 36 & 18.2 & 18.2 & 12\\
  5 & 1 & 118 $\pm$ 13 & 128 $\pm$ 13 & 18.2 & 18.2 & 32\\
  6 & 1 & 1530 $\pm$ 4 & 1910 $\pm$ 3 & 22.0 & 18.2 & 51\\
  7 & 1 & 605 $\pm$ 4 & 689 $\pm$ 3 & 19.1 & 18.2 & 22\\
  8 & 1 & 114 $\pm$ 43 & 112 $\pm$ 38  & 18.7 & 18.2 & 137\\
  9 & 1 & 263 $\pm$ 6 & 269 $\pm$ 6 & 18.2 & 18.2 & 156\\
  10 & 1 & 441 $\pm$ 5 & 515 $\pm$ 4 & 19.0 & 18.2 & 59\\
  11 & 1 & 13.6 $\pm$ 2.2 & 61.3 $\pm$ 4.0 & 49.8 & 29.3 & 53\\
  12 & 1 & 411 $\pm$ 6 & 456 $\pm$ 5 & 19.3 & 18.2 & 162\\
  13 & 1 & 4.1 $\pm$ 1.9 & 8.6 $\pm$ 2.0 & 31.6 & 24.2 & 63\\
  14 & 1 & 675 $\pm$ 40 & 667 $\pm$ 36 & 18.2 & 18.2 & 117\\
  15 & 1 & 339 $\pm$ 4 & 477 $\pm$ 3 & 33.1 & 18.2 & 100\\
  16 & 1 & 5.2 $\pm$ 1.7 & 8.4 $\pm$ 1.7 & 33.2 & 18.9 & 90\\
  17 & 1 & 93.0 $\pm$ 4.1 & 123 $\pm$ 4 & 24.8 & 19.0 & 175\\
  18 & 1 & 5.6 $\pm$ 1.4 & 21.0 $\pm$ 2.3 & 48.1 & 26.8 & 75\\
  19 & 1 & 133 $\pm$ 6 & 139 $\pm$ 5 & 18.2 & 18.2 & 20\\
  20 & 1 & 224 $\pm$ 6 & 245 $\pm$ 5 & 20.2 & 18.2 & 100\\
  21 & 1 & 31.7 $\pm$ 2.3 & 117.8 $\pm$ 4 & 43.2 & 30.5 & 4\\
  22 & 1 & 88.7 $\pm$ 11.3 & 89.5 $\pm$ 10.0 & 19.5 & 18.2 & 31\\
  23 & 1 & 264 $\pm$ 13 & 274 $\pm$ 11 & 18.2 & 18.2 & 75\\
  24 & 1 & 28.9 $\pm$ 1.4 & 80.5 $\pm$ 2.0 & 32.2 & 25.0 & 93\\
  25 & 1 & 337 $\pm$ 6 & 351 $\pm$ 5 & 19.8 & 18.2 & 111\\
  26 & 1 & 187 $\pm$ 38 & 194 $\pm$ 34  & 20.1 & 18.2 & 105\\
  27 & 1 & 31.9 $\pm$ 2.7 & 47.4 $\pm$ 2.4 & 31.6 & 18.9 & 139\\
  28 & 1 & 133 $\pm$ 37 & 149 $\pm$ 35 & 22.5 & 18.2 & 30\\
  29 & 1 & 120 $\pm$ 4 & 252 $\pm$ 3 & 28.8 & 24.6 & 160\\
  30 & 1 & 23.6 $\pm$ 3.0 & 42.1 $\pm$ 3.4 & 27.3 & 20.4 & 76\\
  31 & 2 & 5.2 $\pm$ 2.6 & 10.1 $\pm$ 2.5 & 34.2 & 21.8 & 139\\
  32 & 1 & 402 $\pm$ 6 & 400 $\pm$ 5 & 18.2 & 18.2 & 68\\
  33 & 1 & 12.8 $\pm$ 1.3 & 19.0 $\pm$ 1.6 & 25.8 & 19.9 & 162\\
  34 & 1 & 12.8 $\pm$ 5.5 & 15.8 $\pm$ 4.9  & 24.5 & 20.3 & 92\\
  35 & 1 & 21.3 $\pm$ 1.4 & 59.1 $\pm$ 2.1 & 42.8 & 21.7 & 164\\
  36 & 1 & 80.0 $\pm$ 5.9 & 105 $\pm$ 5 & 25.4 & 18.2 & 90\\
  37 & 1 & 9.8 $\pm$ 2.7 & 16.2 $\pm$ 2.9 & 31.7 & 18.2 & 78\\
  38 & 1 & 348 $\pm$ 5 & 365 $\pm$ 9 & 18.2 & 18.2 & 104\\
  39 & 1 & 6.7 $\pm$ 1.6 & 12.3 $\pm$ 1.6  & 34.1 & 23.9 & 167\\
  40 & 1 & 61.1 $\pm$ 5.0 & 84.1 $\pm$ 5.0 & 29.0 & 18.2 & 79\\
  41 & 1 & 416 $\pm$ 4 & 403 $\pm$ 4 & 18.2 & 18.2 & 68\\
  42 & 1 & 9.0 $\pm$ 0.9 & 37.3 $\pm$ 1.7 & 37.7 & 33.4 & 52\\
  43 & 1 & 71.9 $\pm$ 21 & 78 $\pm$ 19  & 19.4 & 18.2 & 54\\
  44 & 1 & 4.1 $\pm$ 1.0 & 15.9 $\pm$ 1.8 & 49.1 & 31.8 & 137\\
  45 & 1 & 22.2 $\pm$ 2.6 & 35.6 $\pm$ 2.7 & 27.0 & 18.2 & 176\\
  46 & 1 & 331 $\pm$ 6 & 376 $\pm$ 5 & 20.1 & 18.2 & 103\\
\hline
\end{tabular}
\end{table*}

\clearpage


\begin{table*}
\centering
\caption{Table~\ref{tab_getsources_coordinates} (continued) - \emph{Herschel}-350~$\mu$m}
\begin{tabular}{| c || c | c | c | c | c | c |}
\hline
 Number            &   Tag$_{\rm 350}$  & S$^{\rm peak}_{\rm 350}$   & S$^{\rm int}_{\rm 350}$   & AFWHM$_{\rm 350}$  & FWHM$^{\rm min}_{\rm 350}$  &  PA$_{\rm 350}$ \\
  Id         &                &   (Jy/beam)          &      (Jy)         &  (arcsec)      &   (arcsec)     &  (degrees)  \\                           
\hline
\hline
  1 & 1 & 1600 $\pm$ 30 & 1610 $\pm$ 28 & 24.9 & 24.9 & 33\\
  2 & 1 & 544 $\pm$ 22 & 603 $\pm$ 20 & 24.9 & 24.9 & 113\\
  3 & 1 & 389 $\pm$ 28 & 393 $\pm$ 26 & 24.9 & 24.9 & 34\\
  4 & 1 & 163 $\pm$ 29 & 166 $\pm$ 27 & 24.9 & 24.9 & 28\\
  5 & 1 & 111 $\pm$ 11 & 107 $\pm$ 10 & 24.9 & 24.9 & 177\\
  6 & 1 & 743 $\pm$ 7 & 735 $\pm$ 6 & 24.9 & 24.9 & 15\\
  7 & 1 & 359 $\pm$ 7 & 358 $\pm$ 6 & 24.9 & 24.9 & 69\\
  8 & 1 & 94.0 $\pm$ 21.8 & 124 $\pm$ 20  & 30.4 & 24.9 & 156\\
  9 & 1 & 176 $\pm$ 5 & 165 $\pm$ 4 & 24.9 & 24.9 & 160\\
  10 & 1 & 266 $\pm$ 4 & 292 $\pm$ 4 & 26.5 & 24.9 & 27\\
  11 & 1 & 14.4 $\pm$ 2.2 & 42.4 $\pm$ 3.2 & 50.8 & 32.6 & 52\\
  12 & 1 & 218 $\pm$ 5 & 217 $\pm$ 4 & 24.9 & 24.9 & 71\\
  13 & 1 & 5.9 $\pm$ 2.4 & 7.6 $\pm$ 2.2 & 34.1 & 24.9 & 71\\
  14 & 1 & 435 $\pm$ 29 & 447 $\pm$ 26 & 24.9 & 24.9 & 164\\
  15 & 1 & 276 $\pm$ 7 & 337 $\pm$ 6.0 & 32.0 & 24.9 & 102\\
  16 & 1 & 9.2 $\pm$ 2.3 & 11.4 $\pm$ 2.1 & 27.7 & 25.6 & 96\\
  17 & 1 & 71.9 $\pm$ 7.1 & 85.9 $\pm$ 6.5 & 36.0 & 24.9 & 107\\
  18 & 1 & 7.4 $\pm$ 1.0 & 14.5 $\pm$ 1.2 & 51.3 & 26.1 & 58\\
  19 & 1 & 87.9 $\pm$ 4.7 & 82.6 $\pm$ 4.2 & 24.9 & 24.9 & 150\\
  20 & 1 & 117 $\pm$ 4 & 114 $\pm$ 4 & 24.9 & 24.9 & 132\\
  21 & 1 & 30.9 $\pm$ 1.1 & 73.6 $\pm$ 1.4 & 39.2 & 33.3 & 166\\
  22 & 1 & 69.0 $\pm$ 2.7 & 66.5 $\pm$ 2.4 & 24.9 & 24.9 & 173\\
  23 & 1 & 151 $\pm$ 3 & 146 $\pm$ 3 & 24.9 & 24.9 & 63\\
  24 & 1 & 24.9 $\pm$ 1.2 & 70.6 $\pm$ 2.2 & 44.7 & 30.3 & 70\\
  25 & 1 & 194 $\pm$ 4 & 158 $\pm$ 4 & 24.9 & 24.9 & 97\\
  26 & 1 & 88.6 $\pm$ 29.9 & 137 $\pm$ 27  & 37.2 & 24.9 & 101\\
  27 & 1 & 28.8 $\pm$ 2.3 & 42.7 $\pm$ 2.1 & 43.1 & 24.9 & 150\\
  28 & 1 & 102 $\pm$ 28 & 109 $\pm$ 26 & 24.9 & 24.9 & 67\\
  29 & 1 & 92.5 $\pm$ 6.5 & 91.2 $\pm$ 5.9 & 24.9 & 24.9 & 143\\
  30 & 1 & 23.8 $\pm$ 1.7 & 31.9 $\pm$ 1.6 & 41.2 & 24.9 & 99\\
  31 & 1 & 7.0 $\pm$ 2.3 & 8.9 $\pm$ 2.1 & 30.6 & 25.1 & 114\\
  32 & 1 & 123 $\pm$ 5 & 117 $\pm$ 4 & 24.9 & 24.9 & 47\\
  33 & 1 & 16.5 $\pm$ 1.0 & 20.2 $\pm$ 1.0 & 28.9 & 24.9 & 173\\
  34 & 1 & 14.9 $\pm$ 2.8 & 14.9 $\pm$ 2.5 & 27.7 & 24.9 & 50\\
  35 & 1 & 17.5 $\pm$ 1.3 & 44.1 $\pm$ 1.9 & 40.8 & 33.8 & 157\\
  36 & 1 & 76.1 $\pm$ 3.5 & 98.9 $\pm$ 3.2 & 29.4 & 24.9 & 131\\
  37 & 1 & 8.8 $\pm$ 1.9 & 10.5 $\pm$ 1.8 & 31.6 & 24.9 & 97\\
  38 & 1 & 203 $\pm$ 5 & 197 $\pm$ 6 & 24.9 & 24.9 & 135\\
  39 & 1 & 6.8 $\pm$ 0.9 & 9.6 $\pm$ 0.8  & 30.7 & 27.5 & 165\\
  40 & 1 & 39.4 $\pm$ 4.6 & 43.6 $\pm$ 4.2 & 29.7 & 24.9 & 75\\
  41 & 1 & 157 $\pm$ 4 & 148 $\pm$ 4 & 24.9 & 24.9 & 63\\
  42 & 1 & 7.6 $\pm$ 0.8 & 15.7 $\pm$ 0.9 & 47.9 & 27.9 & 53\\
  43 & 1 & 77.0 $\pm$ 15.5 & 83.7 $\pm$ 14.2  & 26.1 & 24.9 & 27.5\\
  44 & 1 & 5.8 $\pm$ 0.7 & 17.0 $\pm$ 1.1 & 46.0 & 41.5 & 144\\
  45 & 1 & 20.0 $\pm$ 2.2 & 23.1 $\pm$ 2.0 & 27.3 & 24.9 & 167\\
  46 & 1 & 178 $\pm$ 4 & 179 $\pm$ 4 & 24.9 & 24.9 & 146\\
\hline
\end{tabular}
\end{table*}

\clearpage


\begin{table*}
\centering
\caption{Table~\ref{tab_getsources_coordinates} (continued) - \emph{Herschel}-500~$\mu$m}
\begin{tabular}{| c || c | c | c | c | c | c |}
\hline
 Number            &   Tag$_{\rm 500}$  & S$^{\rm peak}_{\rm 500}$  & S$^{\rm int}_{\rm 500}$  & FWHM$^{\rm maj}_{\rm 500}$  & FWHM$^{\rm min}_{\rm 500}$  &  PA$_{\rm 500}$ \\
 Id         &                &   (Jy/beam)          &      (Jy)         &  (arcsec)      &   (arcsec)     &  (degrees)  \\                           
\hline
\hline
  1 & 1 & 548 $\pm$ 4 & 526 $\pm$ 4 & 36.3 & 36.3 & 44\\
  2 & 1 & 310 $\pm$ 5 & 299 $\pm$ 4 & 36.3 & 36.3 & 48\\
  3 & 1 & 205 $\pm$ 5 & 193 $\pm$ 5 & 36.3 & 36.3 & 164\\
  4 & 1 & 67.6 $\pm$ 5.0 & 64.6 $\pm$ 4.6 & 36.3 & 36.3 & 161\\
  5 & 1 & 69.6 $\pm$ 5.2 & 65.6 $\pm$ 4.7 & 36.3 & 36.3 & 28\\
  6 & 1 & 226 $\pm$ 2 & 209 $\pm$ 1 & 36.3 & 36.3 & 169\\
  7 & 1 & 142 $\pm$ 2 & 134 $\pm$ 1 & 36.3 & 36.3 & 116\\
  8 & 1 & 19.2 $\pm$ 5.0 & 18.1 $\pm$ 4.6  & 36.3 & 36.3 & 9\\
  9 & 1 & 53.7 $\pm$ 1.3 & 52.3 $\pm$ 1.2 & 36.3 & 36.3 & 172\\
  10 & 1 & 93.0 $\pm$ 1.2 & 86.0 $\pm$ 1.1 & 36.3 & 36.3 & 39\\
  11 & 1 & 10.4 $\pm$ 1.6 & 17.2 $\pm$ 1.9 & 48.2 & 39.0 & 58\\
  12 & 1 & 75.5 $\pm$ 1.3 & 74.9 $\pm$ 1.2 & 36.3 & 36.3 & 60\\
  13 & 1 & 7.1 $\pm$ 1.9 & 8.3 $\pm$ 1.7 & 44.2 & 36.3 & 79\\
  14 & 1 & 111 $\pm$ 5 & 105 $\pm$ 4 & 36.3 & 36.3 & 147\\
  15 & 1 & 30.5 $\pm$ 1.6 & 29.0 $\pm$ 1.4 & 36.3 & 36.3 & 107\\
  16 & 1 & 9.4 $\pm$ 1.8 & 9.5 $\pm$ 1.6 & 36.4 & 36.3 & 87\\
  17 & 1 & 41.2 $\pm$ 1.5 & 45.9 $\pm$ 1.3 & 38.3 & 36.3 & 64\\
  18 & 1 & 6.4 $\pm$ 1.1 & 9.2 $\pm$ 1.1 & 53.1 & 36.3 & 70\\
  19 & 2 & 8.0 $\pm$ 17.8 & 7.8 $\pm$ 16.3 & 36.3 & 36.3 & 80\\
  20 & 1 & 72.0 $\pm$ 1.3 & 69.6 $\pm$ 1.2 & 36.3 & 36.3 & 152\\
  21 & 1 & 19.6 $\pm$ 0.6 & 24.8 $\pm$ 0.6 & 41.9 & 36.3 & 95\\
  22 & 1 & 22.9 $\pm$ 1.7 & 23.5 $\pm$ 1.8 & 36.3 & 36.3 & 33\\
  23 & 1 & 34.8 $\pm$ 1.7 & 33.1 $\pm$ 1.5 & 36.3 & 36.3 & 50\\
  24 & 1 & 15.8 $\pm$ 0.9 & 24.7 $\pm$ 1.1 & 45.2 & 38.3 & 65\\
  25 & 1 & 37.5 $\pm$ 1.3 & 35.2 $\pm$ 1.2 & 36.3 & 36.3 & 93\\
  26 & 1 & 43.2 $\pm$ 5.0 & 54.0  $\pm$ 4.6  & 43.2 & 36.3 & 170\\
  27 & 1 & 22.9 $\pm$ 1.4 & 32.0 $\pm$ 1.3 & 51.1 & 36.3 & 154\\
  28 & 1 & 40.2 $\pm$ 5.0 & 38.5 $\pm$ 4.6 & 36.3 & 36.3 & 150\\
  29 & 1 & 26.4 $\pm$ 1.5 & 24.7 $\pm$ 1.4 & 36.3 & 36.3 & 110\\
  30 & 1 & 17.2 $\pm$ 0.7 & 21.3 $\pm$ 0.7 & 44.0 & 36.3 & 89\\
  31 & 1 & 6.6 $\pm$ 1.4 & 7.1 $\pm$ 1.3 & 36.3 & 36.3 & 91\\
  32 & 1 & 74.5 $\pm$ 1.3 & 70.2 $\pm$ 1.1 & 36.3 & 36.3 & 97\\
  33 & 1 & 12.8 $\pm$ 0.9 & 13.0 $\pm$ 0.8 & 36.3 & 36.3 & 72\\
  34 & 1 & 9.8 $\pm$ 1.1  & 9.1 $\pm$ 1.1  & 37.5 & 36.3 & 59\\
  35 & 1 & 10.1 $\pm$ 1.0 & 15.9 $\pm$ 1.1 & 47.2 & 40.7 & 159\\
  36 & 1 & 38.8 $\pm$ 1.0 & 44.2 $\pm$ 0.9 & 36.6 & 36.3 & 159\\
  37 & 1 & 5.9 $\pm$ 1.2 & 7.0 $\pm$ 1.1 & 48.9 & 36.3 & 106\\
  38 & 1 & 55.0 $\pm$ 1.3 & 51.6 $\pm$ 1.2 & 36.3 & 36.3 & 105\\
  39 & 1 & 4.5 $\pm$ 0.7 & 6.2 $\pm$ 0.6 & 42.0 & 38.9 & 161\\
  40 & 1 & 22.5 $\pm$ 1.1 & 23.0 $\pm$ 1.0 & 36.3 & 36.3 & 57\\
  41 & 1 & 73.2 $\pm$ 1.2 & 67.5 $\pm$ 1.1 & 36.3 & 36.3 & 27\\
  42 & 1 & 7.1 $\pm$ 0.7 & 8.9 $\pm$ 0.7 & 40.5 & 36.3 & 74\\
  43 & 1 & 28.3 $\pm$ 4.9 & 31.5 $\pm$ 4.5 & 41.3 & 36.3 & 13\\
  44 & 1 & 4.8 $\pm$ 0.9 & 7.3 $\pm$ 1.0 & 44.2 & 43.0 & 4\\
  45 & 1 & 10.7 $\pm$ 1.4 & 11.0 $\pm$ 1.3 & 36.3 & 36.3 & 175\\
  46 & 1 & 71.7 $\pm$ 1.3 & 69.4 $\pm$ 1.1 & 36.3 & 36.3 & 130\\
\hline
\end{tabular}
\end{table*}

\clearpage


\begin{table*}
\centering
\caption{HOBYS catalog of additional GLIMPSE fluxes at 3.6, 4.5, 5.8 and 8.0~$\mu$m 
for the 46 MDCs found in NGC~6334. Stars ($^{*}$) on fluxes indicate when we did aperture photometry.}
\begin{tabular}{| c || c | c | c | c | c|}
\cline{2-6}
\multicolumn{1}{c|}{}  &  \multicolumn{5}{c|}{GLIMPSE}        \\
\hline
 Number       &  GLIMPSE ID  & S$^{\rm int}_{3.6}$   &  S$^{\rm int}_{4.5}$   &  S$^{\rm int}_{5.8}$   &   S$^{\rm int}_{8.0}$     \\ 
   Id         &              &     (Jy)              &            (Jy)          &            (Jy)          &            (Jy)    \\                      
\hline
\hline
  1 && 0.89$^{*}$ & 1.85$^{*}$ & 7.7$^{*}$ & 39.0$^{*}$\\
  2 &&    &    &    &   \\
  3 &&    &    &    &   \\
  4 &&    &    &    &   \\
  5 &&    &    &    &   \\
  6 && 0.43$^{*}$ & 0.85$^{*}$ & 0.79$^{*}$ & 0.7$^{*}$\\
  7 &&    &    &    &   \\
  8 &&    &    &    &   \\
  9 &&    &    &    &   \\
  10 && 0.07$^{*}$ & 0.12$^{*}$ & 1.1$^{*}$ & 2.7$^{*}$\\
  11 &&    &    &    &   \\
  12 &&    &    &    &   \\
  13 &&    &    &    &   \\
  14 &&    &    &    &   \\
  15 &&    &    &    &   \\
  16 &&    &    &    &   \\
  17 &&    &    &    &   \\
  18 & G352.1150+00.7366 & 0.002 & 0.006 & 0.01 & 0.01\\
  19 &&    &    &    &   \\
  20 &&    &    &    &   \\
  21 &&    &    &    &   \\
  22 &&    &    &    &   \\
  23 &&    &    &    &   \\
  24 & G350.8518+00.8209 + G350.8504+00.8198 & 0.003 & 0.007 & 0.015 & 0.02\\
  25 &&    &    &    &   \\
  26 &&    &    &    &   \\
  27 &&    &    &    &   \\
  28 &&    &    &    &   \\
  29 & G351.1556+00.7030 &    &    &    & 0.02\\
  30 &&    &    &    &   \\
  31 &&    &    &    &   \\
  32 && 0.04$^{*}$ & 0.22$^{*}$ & 0.87$^{*}$ & 2.7$^{*}$\\
  33 &&    &    &    &   \\
  34 &&    &    &    &   \\
  35 &&    &    &    &   \\
  36 &&    &    &    &   \\
  37 & G350.7304+00.9294 & 3.0E-4 & 0.0015 & 0.005 & 0.007\\
  38 && 0.015$^{*}$ & 0.05$^{*}$ & 0.3$^{*}$ & 1.1$^{*}$\\
  39 &&    &    &    &   \\
  40 &&    &    &    &   \\
  41 &&    &    &    &   \\
  42 &&    &    &    &   \\
  43 &&    &    &    &   \\
  44 &&    &    &    &   \\
  45 &&    &    &    &   \\
  46 && 0.22$^{*}$ & 1.0$^{*}$ & 3.9$^{*}$ & 2.7$^{*}$\\
\hline 
\end{tabular}
\end{table*}

\clearpage


\begin{table*}
\centering
\caption{HOBYS catalog of additional fluxes at \emph{Spitzer}-24~$\mu$m, \emph{WISE}-22~$\mu$m or \emph{MSX}-21~$\mu$m for the 46 MDCs found in NGC~6334. 
Stars ($^{*}$) on fluxes indicate when we did aperture photometry.}
\label{tab_21_22_24}
\begin{tabular}{| c || c | c | c | c | c | c|}
\cline{2-7}
\multicolumn{1}{c|}{}  &  \multicolumn{2}{c|}{MIPSGAL} &   \multicolumn{2}{|c|}{ \emph{WISE}} &    \multicolumn{2}{|c|}{ \emph{MSX}}             \\
\hline
 Number            &  MIPSGAL ID  & S$^{\rm int}_{24}$   &  \emph{WISE} ID      &  S$^{\rm int}_{22}$  &   \emph{MSX} ID      &  S$^{\rm int}_{21}$           \\ 
        Id         &              &     (Jy)            &              &       (Jy)           &              &       (Jy)     \\                      
\hline
\hline
  1 & saturated &  &&     &  MSX6C G351.4170+00.6440 & 821.0\\
  2 &&   &&     &&  \\
  3 &&   &&     &&  \\
  4 &&   &&     &&  \\
  5 &&   &&     &&  \\
  6 & saturated &   &&     &  MSX6C G351.1612+00.6973 & 348.0\\
  7 & saturated &   &&     && \\
  8 &&   &&     && \\
  9 &&   &&     &&  \\
  10 & saturated &   &&     &  MSX6C G351.2434+00.6664 & 90.0\\
  11 &&   &&     &&  \\
  12 &&   &&     && \\
  13 &&   &&     &&  \\
  14 & MG351.4449+00.6597 & 0.72  &&     && \\
  15 &&   &&     &&  \\
  16 &&   &&     &&  \\
  17 &&   &&     &&  \\
  18 & MG352.1151+00.7366 & 0.03  &&     && \\
  19 &&   &&     &&  \\
  20 &&   &&     &&  \\
  21 &&   &&     &&  \\
  22 & saturated &   &&    && \\
  23 & saturated &   &&    && \\
  24 && 0.027$^{*}$  &&     && \\
  25 & saturated &   &&     && \\
  26 &&   &&    && \\
  27 &&   &&     &&  \\
  28 &&   &&    && \\
  29 &&   &&     && \\
  30 &&   &&     && \\
  31 &&   &&     &&  \\
  32 & saturated &   &&     & MSX6C G351.2434+00.6664 & 90.0\\
  33 &&   &&     &&  \\
  34 &&   &&     &&  \\
  35 &&   &&     &&  \\
  36 &&   &&     && \\
  37 & MG350.7303+00.9294 & 0.07  &&     && \\
  38 & saturated &   &&     & MSX6C G351.2514+00.6731 & 30.0\\
  39 &&   &&     &&  \\
  40 &&   &&    && \\
  41 & saturated &   &&     && \\
  42 &&   &&     &&  \\
  43 &&   &&     &&  \\
  44 &&   &&     &&  \\
  45 &&   &&     &&  \\
  46 &&   &  J172023.54-355500.2 &  22.0   &&    \\
\hline 
\end{tabular}
\end{table*}

\clearpage


\begin{table*}
\centering
\caption{Table~\ref{tab_getsources_coordinates} (continued) - SCUBA2-450~$\mu$m}
\begin{tabular}{| c || c | c | c | c | c | c |}
\hline
 Number            &   Tag$_{\rm 450}$  & S$^{\rm peak}_{\rm 450}$   & S$^{\rm int}_{\rm 450}$   & FWHM$^{\rm maj}_{\rm 450}$  & FWHM$^{\rm min}_{\rm 450}$  &  PA$_{\rm 450}$ \\
 Id         &                &   (Jy/beam)          &      (Jy)         &  (arcsec)      &   (arcsec)     &  (degrees)  \\                            
\hline
\hline
  1 & 1 & 69.9 $\pm$ 3.2 & 411 $\pm$ 7 & 19.4 & 17.6 & 101\\
  2 & 1 & 21.8 $\pm$ 9.0 & 93.9 $\pm$ 15.2 & 19.4 & 12.8 & 179\\
  3 & 2 & 5.5 $\pm$ 9.5 & 12.8 $\pm$ 10.0 & 14.1 & 8.5 & 46\\
  4 & 2 & 3.2 $\pm$ 7.4 & 5.0 $\pm$ 7.4 & 10.5 & 8.8 & 54\\
  5 & 1 & 7.6 $\pm$ 0.3 & 25.3 $\pm$ 0.6 & 17.3 & 10.3 & 14\\
  6 & 0 &                     &                    &     &     &  \\
  7 & 0 &                     &                    &     &     &  \\
  8 & 2  & 1.5 $\pm$ 3.8 & 2.9 $\pm$ 3.8  & 15.4 & 8.5 & 104\\
  9 & 0 &                     &                    &     &     &  \\
  10 & 0 &                     &                    &     &     &  \\
  11 & 0 &                     &                    &     &     &  \\
  12 & 0 &                     &                    &     &     &  \\
  13 & 0 &                     &                    &     &     &  \\
  14 & 1 & 27.2 $\pm$ 8.7 & 54.4 $\pm$ 9.5 & 14.6 & 9.8 & 165\\
  15 & 0 &                     &                    &     &     &  \\
  16 & 0 &                     &                    &     &     &  \\
  17 & 0 &                     &                    &     &     &  \\
  18 & 0 &                     &                    &     &     &  \\
  19 & 0 &                     &                    &     &     &  \\
  20 & 0 &                     &                    &     &     &  \\
  21 & 0 &                     &                    &     &     &  \\
  22 & 0 &                     &                    &     &     &  \\
  23 & 0 &                     &                    &     &     &  \\
  24 & 0 &                     &                    &     &     &  \\
  25 & 0 &                     &                    &     &     &  \\
  26 & 2  & 0.31 $\pm$ 3.0 & -7.8 $\pm$ 3.0  & 8.5 & 8.5 & 0\\
  27 & 0 &                     &                    &     &     &  \\
  28 & 1 & 7.9 $\pm$ 2.1 & 19.6 $\pm$ 3.8 & 15.2 & 8.5 & 48\\
  29 & 0 &                     &                    &     &     &  \\
  30 & 0 &                     &                    &     &     &  \\
  31 & 0 &                     &                    &     &     &  \\
  32 & 0 &                     &                    &     &     &  \\
  33 & 0 &                     &                    &     &     &  \\
  34 & 0 &                     &                    &     &     &  \\
  35 & 0 &                     &                    &     &     &  \\
  36 & 0 &                     &                    &     &     &  \\
  37 & 0 &                     &                    &     &     &  \\
  38 & 0 &                     &                    &     &     &  \\
  39 & 0 &                     &                    &     &     &  \\
  40 & 0 &                     &                    &     &     &  \\
  41 & 0 &                     &                    &     &     &  \\
  42 & 0 &                     &                    &     &     &  \\
  43 & 1 & 1.6 $\pm$ 0.6 & 4.2 $\pm$ 0.6  & 11.9 & 9.6 & 175\\
  44 & 0 &                     &                    &     &     &  \\
  45 & 0 &                     &                    &     &     &  \\
  46 & 0 &                     &                    &     &     &  \\
\hline
\end{tabular}
\end{table*}

\clearpage


\begin{table*}
\centering
\caption{Table~\ref{tab_getsources_coordinates} (continued) - SCUBA2-850~$\mu$m}
\begin{tabular}{| c || c | c | c | c | c | c |}
\hline
 Number            &   Tag$_{\rm 850}$  & S$^{\rm peak}_{\rm 850}$ & S$^{\rm int}_{\rm 850}$ & FWHM$^{\rm maj}_{\rm 850}$  & FWHM$^{\rm min}_{\rm 850}$  &  PA$_{\rm 850}$ \\
 Id         &                &   (Jy/beam)          &      (Jy)         &  (arcsec)      &   (arcsec)     &  (degrees)  \\                          
\hline
\hline
  1 & 1 & 23.7 $\pm$ 0.5 & 36.8 $\pm$ 0.8 & 16.6 & 15.0 & 49\\
  2 & 1 & 9.2 $\pm$ 0.7 & 15.4 $\pm$ 0.7 & 22.7 & 15.0 & 171\\
  3 & 1 & 8.1 $\pm$ 0.6 & 12.5 $\pm$ 0.6 & 18.8 & 15.0 & 71\\
  4 & 1 & 1.7 $\pm$ 0.5 & 2.2 $\pm$ 0.5 & 19.1 & 15.0 & 17\\
  5 & 1 & 1.8 $\pm$ 0.5 & 2.5 $\pm$ 0.7 & 16.9 & 15.0 & 57\\
  6 & 0 &                     &                    &     &     &  \\
  7 & 0 &                     &                    &     &     &  \\
  8 & 1  & 1.3 $\pm$ 0.6 & 2.3 $\pm$ 0.6  & 21.2 & 15 & 159\\
  9 & 0 &                     &                    &     &     &  \\
  10 & 0 &                     &                    &     &     &  \\
  11 & 0 &                     &                    &     &     &  \\
  12 & 0 &                     &                    &     &     &  \\
  13 & 0 &                     &                    &     &     &  \\
  14 & 1 & 7.6 $\pm$ 0.6 & 8.7 $\pm$ 0.6 & 15.1 & 15.0 & 148\\
  15 & 0 &                     &                    &     &     &  \\
  16 & 0 &                     &                    &     &     &  \\
  17 & 0 &                     &                    &     &     &  \\
  18 & 0 &                     &                    &     &     &  \\
  19 & 0 &                     &                    &     &     &  \\
  20 & 0 &                     &                    &     &     &  \\
  21 & 0 &                     &                    &     &     &  \\
  22 & 0 &                     &                    &     &     &  \\
  23 & 0 &                     &                    &     &     &  \\
  24 & 0 &                     &                    &     &     &  \\
  25 & 0 &                     &                    &     &     &  \\
  26 & 2 & 0.49 $\pm$ 0.63 & 0.50 $\pm$ 0.63  & 15 & 15 & 167\\
  27 & 0 &                     &                    &     &     &  \\
  28 & 2 & 1.7 $\pm$ 0.8 & 1.7 $\pm$ 1.0 & 18.7 & 15.0 & 52\\
  29 & 0 &                     &                    &     &     &  \\
  30 & 0 &                     &                    &     &     &  \\
  31 & 0 &                     &                    &     &     &  \\
  32 & 0 &                     &                    &     &     &  \\
  33 & 0 &                     &                    &     &     &  \\
  34 & 0 &                     &                    &     &     &  \\
  35 & 0 &                     &                    &     &     &  \\
  36 & 0 &                     &                    &     &     &  \\
  37 & 0 &                     &                    &     &     &  \\
  38 & 0 &                     &                    &     &     &  \\
  39 & 0 &                     &                    &     &     &  \\
  40 & 0 &                     &                    &     &     &  \\
  41 & 0 &                     &                    &     &     &  \\
  42 & 0 &                     &                    &     &     &  \\
  43 & 1 & 0.79 $\pm$ 0.17 & 0.79 $\pm$ 0.17  & 15.0 & 15.0 & 13\\
  44 & 0 &                     &                    &     &     &  \\
  45 & 0 &                     &                    &     &     &  \\
  46 & 0 &                     &                    &     &     &  \\
 \hline
\end{tabular}
\end{table*}

\clearpage


\begin{table*}
\centering
\caption{Table~\ref{tab_getsources_coordinates} (continued) - APEX-870~$\mu$m}
\begin{tabular}{| c || c | c | c | c | c | c |}
\hline
 Number            &   Tag$_{\rm 870}$  & S$^{peak}_{\rm 870}$   & S$^{\rm int}_{\rm 870}$  & FWHM$^{\rm maj}_{\rm 870}$  & FWHM$^{\rm min}_{\rm 870}$  &  PA$_{\rm 870}$ \\
 Id         &                &   (Jy/beam)          &      (Jy)         &  (arcsec)      &   (arcsec)     &  (degrees)  \\                          
\hline
\hline
  1 & 1 & 44.0 $\pm$ 1.1 & 55.5 $\pm$ 1.3 & 21.0 & 18.2 & 28\\
  2 & 1 & 17.4 $\pm$ 1.0 & 21.2 $\pm$ 1.0 & 20.2 & 18.2 & 170\\
  3 & 1 & 11.1 $\pm$ 1.1 & 12.4 $\pm$ 1.1 & 22.5 & 18.2 & 61\\
  4 & 1 & 9.2 $\pm$ 1.1 & 15.2 $\pm$ 1.1 & 29.2 & 18.2 & 178\\
  5 & 1 & 2.7 $\pm$ 0.5 & 3.9 $\pm$ 0.5 & 24.1 & 18.2 & 17\\
  6 & 1 & 18.0 $\pm$ 1.0 & 22.8 $\pm$ 1.0 & 18.7 & 18.2 & 118\\
  7 & 1 & 7.4 $\pm$ 0.9 & 10.6 $\pm$ 0.9 & 21.5 & 18.2 & 77\\
  8 & 0 &  & &  &  & \\
  9 & 1 & 4.4 $\pm$ 0.4 & 5.7 $\pm$ 0.4 & 20.0 & 18.2 & 159\\
  10 & 1 & 9.2 $\pm$ 0.4 & 15.6 $\pm$ 0.4 & 22.9 & 21.4 & 124\\
  11 & 1 & 0.4 $\pm$ 0.12 & 2.0 $\pm$ 0.2 & 58.2 & 32.5 & 128\\
  12 & 1 & 3.3 $\pm$ 0.4 & 5.3 $\pm$ 0.4 & 26.0 & 18.2 & 2\\
  13 & 2 & 0.18 $\pm$ 0.11 & 0.5 $\pm$ 0.13 & 38.2 & 21.6 & 72\\
  14 & 1 & 23.5 $\pm$ 1.1 & 39.8 $\pm$ 1.6 & 23.9 & 18.2 & 175\\
  15 & 1 & 5.2 $\pm$ 0.9 & 9.8 $\pm$ 0.9 & 29.7 & 21.2 & 125\\
  16 & 1 & 0.5 $\pm$ 0.13 & 0.9 $\pm$ 0.15 & 26.0 & 19.3 & 10\\
  17 & 1 & 1.4 $\pm$ 0.5 & 2.1 $\pm$ 0.5 & 25.2 & 20.2 & 15\\
  18 & 1 & 0.4 $\pm$ 0.08 & 1.9 $\pm$ 0.16 & 52.4 & 33.2 & 51\\
  19 & 1 & 4.0 $\pm$ 0.4 & 6.8 $\pm$ 0.4 & 27.3 & 18.2 & 3\\
  20 & 1 & 2.5 $\pm$ 0.4 & 2.7 $\pm$ 0.4 & 18.2 & 18.2 & 126\\
  21 & 1 & 0.6 $\pm$ 0.13 & 3.6 $\pm$ 0.2 & 48.6 & 37.6 & 19\\
  22 & 1 & 1.7 $\pm$ 0.3 & 2.0 $\pm$ 0.3 & 20.6 & 18.2 & 39\\
  23 & 1 & 4.2 $\pm$ 0.3 & 5.3 $\pm$ 0.3 & 18.2 & 18.2 & 18\\
  24 & 1 & 0.9 $\pm$ 0.07 & 2.9 $\pm$ 0.12 & 31.1 & 26.7 & 93\\
  25 & 1 & 3.9 $\pm$ 0.4 & 4.1 $\pm$ 0.4 & 18.2 & 18.2 & 116\\
  26 & 1 & 2.7 $\pm$ 1.1 & 3.4 $\pm$ 1.1  & 19.8 & 18.2 & 134\\
  27 & 1 & 0.7 $\pm$ 0.3 & 1.1 $\pm$ 0.3 & 28.5 & 19.5 & 124\\
  28 & 0 &  & &  &  & \\
  29 & 2 & 1.0 $\pm$ 0.8 & 3.3 $\pm$ 0.8 & 31.6 & 27.5 & 112\\
  30 & 1 & 0.8 $\pm$ 0.13 & 1.5 $\pm$ 0.16 & 27.9 & 19.1 & 81\\
  31 & 2 & 0.3 $\pm$ 0.18 & 0.6 $\pm$ 0.19 & 27.9 & 20.3 & 163\\
  32 & 1 & 6.0 $\pm$ 0.4 & 6.5 $\pm$ 0.4 & 18.2 & 18.2 & 51\\
  33 & 1 & 0.8 $\pm$ 0.12 & 1.3 $\pm$ 0.12 & 23.8 & 18.6 & 173\\
  34 & 2 & 0.3 $\pm$ 0.11 & 0.4 $\pm$ 0.11  & 26.0 & 20.4 & 111\\
  35 & 1 & 0.6 $\pm$ 0.06 & 4.3 $\pm$ 0.15 & 50.4 & 44.0 & 101\\
  36 & 1 & 1.2 $\pm$ 0.3 & 1.9 $\pm$ 0.3 & 26.5 & 19.6 & 74\\
  37 & 1 & 0.6 $\pm$ 0.17 & 1.1 $\pm$ 0.3 & 22.6 & 21.4 & 156\\
  38 & 1 & 10.0 $\pm$ 0.4 & 11.6 $\pm$ 0.8 & 18.2 & 18.2 & 4\\
  39 & 2 & 0.17 $\pm$ 0.07 & 0.40 $\pm$ 0.09  & 25.8 & 22.6 & 88\\
  40 & 1 & 0.6 $\pm$ 0.2 & 0.7 $\pm$ 0.2 & 21.4 & 18.2 & 84\\
  41 & 1 & 6.8 $\pm$ 0.4 & 7.4 $\pm$ 0.4 & 18.2 & 18.2 & 163\\
  42 & 1 & 0.3 $\pm$ 0.09 & 1.8 $\pm$ 0.16 & 52.2 & 35.9 & 71\\
  43 & 2 & 1.7 $\pm$ 0.2 & 1.2 $\pm$ 0.2 & 18.2 & 18.2 & 2\\
  44 & 2 & 0.4 $\pm$ 0.08 & 0.8 $\pm$ 0.13 & 47.0 & 22.4 & 6\\
  45 & 1 & 0.9 $\pm$ 0.15 & 1.6 $\pm$ 0.19 & 23.1 & 18.9 & 2\\
  46 & 1 & 4.5 $\pm$ 0.4 & 5.7 $\pm$ 0.4 & 19.7 & 18.2 & 135\\
\hline
\end{tabular}
\end{table*}

\clearpage


\begin{table*}
\centering
\caption{Table~\ref{tab_getsources_coordinates} (continued) - SIMBA-1200~$\mu$m}
\label{tab_getsources_1200}
\begin{tabular}{| c || c | c | c | c | c | c |}
\hline
 Number            &   Tag$_{\rm 1200}$  & S$^{\rm peak}_{\rm 1200}$   & S$^{\rm int}_{1200}$   & FWHM$^{\rm maj}_{\rm 1200}$  & FWHM$^{\rm min}_{\rm 1200}$  &  PA$_{\rm 1200}$ \\
  Id         &                &   (Jy/beam)          &      (Jy)         &  (arcsec)      &   (arcsec)     &  (degrees)  \\                          
\hline
\hline

  1 & 1 & 15.1 $\pm$ 0.6 & 20.2 $\pm$ 0.6 & 26.8 & 24.0 & 1\\
  2 & 1 & 7.3 $\pm$ 0.6 & 9.3 $\pm$ 0.6 & 24.4 & 24.0 & 26\\
  3 & 1 & 7.1 $\pm$ 0.6 & 8.0 $\pm$ 0.6 & 24.2 & 24.0 & 46\\
  4 & 1 & 1.8 $\pm$ 0.6 & 3.2 $\pm$ 0.6 & 32.6 & 27.6 & 0\\
  5 & 1 & 0.8 $\pm$ 0.3 & 0.7 $\pm$ 0.3 & 24.0 & 24.0 & 34\\
  6 & 1 & 3.3 $\pm$ 0.13 & 4.0 $\pm$ 0.13 & 24.0 & 24.0 & 169\\
  7 & 1 & 2.8 $\pm$ 0.13 & 3.1 $\pm$ 0.13 & 24.0 & 24.0 & 142\\
  8 & 1 & 2.4  $\pm$ 0.7 & 3.3 $\pm$ 0.7  & 27.8 & 24.0 & 151\\
  9 & 1 & 1.6 $\pm$ 0.1 & 2.4 $\pm$ 0.1 & 31.7 & 24.0 & 80\\
  10 & 1 & 3.5 $\pm$ 0.08 & 3.8 $\pm$ 0.08 & 24.0 & 24.0 & 107\\
  11 & 1 & 0.3 $\pm$ 0.04 & 1.5 $\pm$ 0.09 & 51.4 & 45.3 & 96\\
  12 & 1 & 3.6 $\pm$ 0.09 & 5.1 $\pm$ 0.09 & 30.5 & 24.0 & 89\\
  13 & 1 & 0.13 $\pm$ 0.07 & 0.2 $\pm$ 0.07 & 32.2 & 24.0 & 68\\
  14 & 1 & 2.8 $\pm$ 0.6 & 3.1 $\pm$ 0.6 & 24.0 & 24.0 & 136\\
  15 & 1 & 1.3 $\pm$ 0.13 & 1.7 $\pm$ 0.13 & 25.0 & 24.0 & 101\\
  16 & 1 & 0.17 $\pm$ 0.07 & 0.3 $\pm$ 0.07 & 32.5 & 28.1 & 13\\
  17 & 1 & 0.4 $\pm$ 0.2 & 0.5 $\pm$ 0.16 & 29.0 & 24.0 & 8\\
  18 & 2 & 0.1 $\pm$ 0.05 & 0.3 $\pm$ 0.06 & 60.8 & 27.7 & 68\\
  19 & 1 & 1.0 $\pm$ 0.1 & 1.1 $\pm$ 0.1 & 24.0 & 24.0 & 90\\
  20 & 1 & 0.8 $\pm$ 0.1 & 1.0 $\pm$ 0.1 & 24.0 & 24.0 & 98\\
  21 & 1 & 0.17 $\pm$ 0.04 & 0.5 $\pm$ 0.06 & 40.6 & 34.2 & 42\\
  22 & 1 & 1.4 $\pm$ 0.12 & 1.6 $\pm$ 0.12 & 24.0 & 24.0 & 9\\
  23 & 1 & 1.7 $\pm$ 0.12 & 1.9 $\pm$ 0.12 & 24.0 & 24.0 & 48\\
  24 & 1 & 0.3 $\pm$ 0.03 & 0.8 $\pm$ 0.04 & 42.4 & 36.5 & 128\\
  25 & 1 & 1.1 $\pm$ 0.09 & 1.6 $\pm$ 0.09 & 35.1 & 24.0 & 93\\
  26 & 0 &  &  &  &  & \\
  27 & 2 & 0.15 $\pm$ 0.08 & 0.2 $\pm$ 0.08 & 33.3 & 24.0 & 159\\
  28 & 2 & 0.7 $\pm$ 0.5 & 1.0 $\pm$ 0.5 & 33.0 & 24.0 & 9\\
  29 & 1 & 1.7 $\pm$ 0.13 & 1.8 $\pm$ 0.13 & 24.0 & 24.0 & 153\\
  30 & 1 & 0.3 $\pm$ 0.07 & 0.4 $\pm$ 0.08 & 41.4 & 24.0 & 91\\
  31 & 2 & 0.09 $\pm$ 0.07 & 0.13 $\pm$ 0.07 & 29.6 & 25.7 & 29\\
  32 & 1 & 2.6 $\pm$ 0.09 & 2.8 $\pm$ 0.09 & 24.0 & 24.0 & 23\\
  33 & 1 & 0.2 $\pm$ 0.05 & 0.4 $\pm$ 0.05 & 33.8 & 28.3 & 134\\
  34 & 2 & 0.11 $\pm$ 0.09  & 0.13 $\pm$ 0.09  & 26.5 & 24.0 & 62\\
  35 & 1 & 0.18 $\pm$ 0.02 & 0.5 $\pm$ 0.03 & 45.7 & 27.3 & 157\\
  36 & 1 & 0.4 $\pm$ 0.08 & 0.6 $\pm$ 0.08 & 30.6 & 24.9 & 165\\
  37 & 2 & 0.1 $\pm$ 0.07 & 0.18 $\pm$ 0.08 & 37.2 & 27.9 & 163\\
  38 & 1 & 1.2 $\pm$ 0.08 & 1.9 $\pm$ 0.12 & 28.9 & 24.0 & 172\\
  39 & 2 & 0.018 $\pm$ 0.033 & 0.035 $\pm$ 0.034  & 31.4 & 24.0 & 103\\
  40 & 2 & 0.13 $\pm$ 0.09 & 0.19 $\pm$ 0.09 & 33.2 & 24.0 & 67\\
  41 & 1 & 2.6 $\pm$ 0.08 & 2.8 $\pm$ 0.08 & 24.0 & 24.0 & 170\\
  42 & 1 & 0.09 $\pm$ 0.03 & 0.3 $\pm$ 0.04 & 47.8 & 31.1 & 91\\
  43 & 2 & 0.90 $\pm$ 0.04 & 0.42 $\pm$ 0.04  & 24.0 & 24.0 & 159\\
  44 & 1 & 0.07 $\pm$ 0.01 & 0.18 $\pm$ 0.02 & 42.2 & 28.4 & 161\\
  45 & 1 & 0.2 $\pm$ 0.06 & 0.3 $\pm$ 0.06 & 30.9 & 24.0 & 172\\
  46 & 1 & 1.1 $\pm$ 0.1 & 1.2 $\pm$ 0.1 & 24.0 & 24.0 & 94\\
\hline
\end{tabular}
\end{table*}

\clearpage

\begin{center}
\section{Multi-wavelength images and spectral energy distribution}
\end{center}

We present in this appendix the multi-wavelength images and spectral energy distributions (SEDs) 
for the 46 MDCs of NGC~6334, which are discussed in the main body of the text.


\clearpage 

\onecolumn
\vspace*{\fill}
{\bf \centerline{\fontsize{120}{120}\selectfont IR-bright MDCs}}
\vfill
\clearpage
\twocolumn

\begin{figure*}[]  
\begin{center}
\includegraphics[width=13cm, angle = 0]{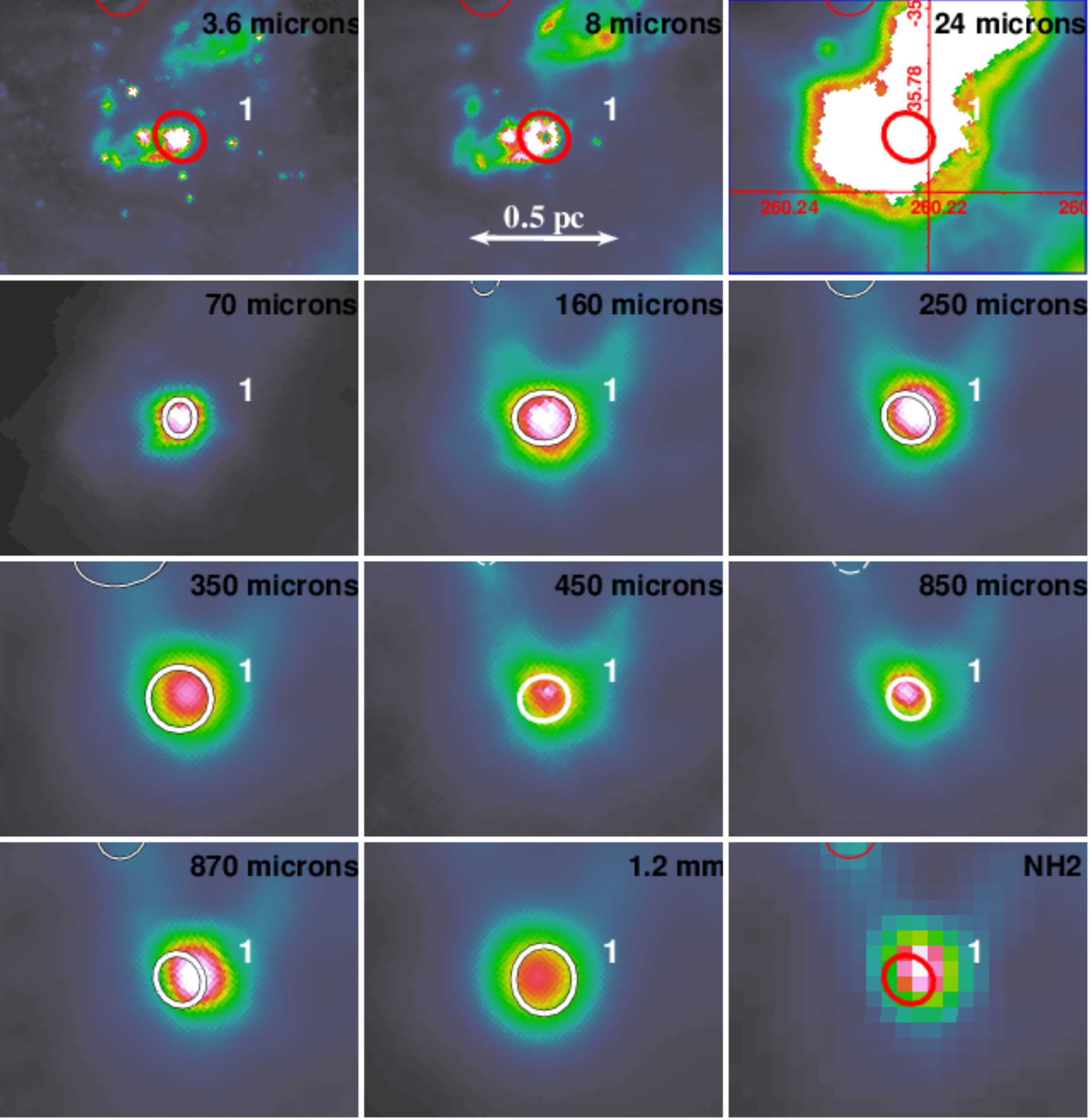}
\includegraphics[width=12cm,angle=0]{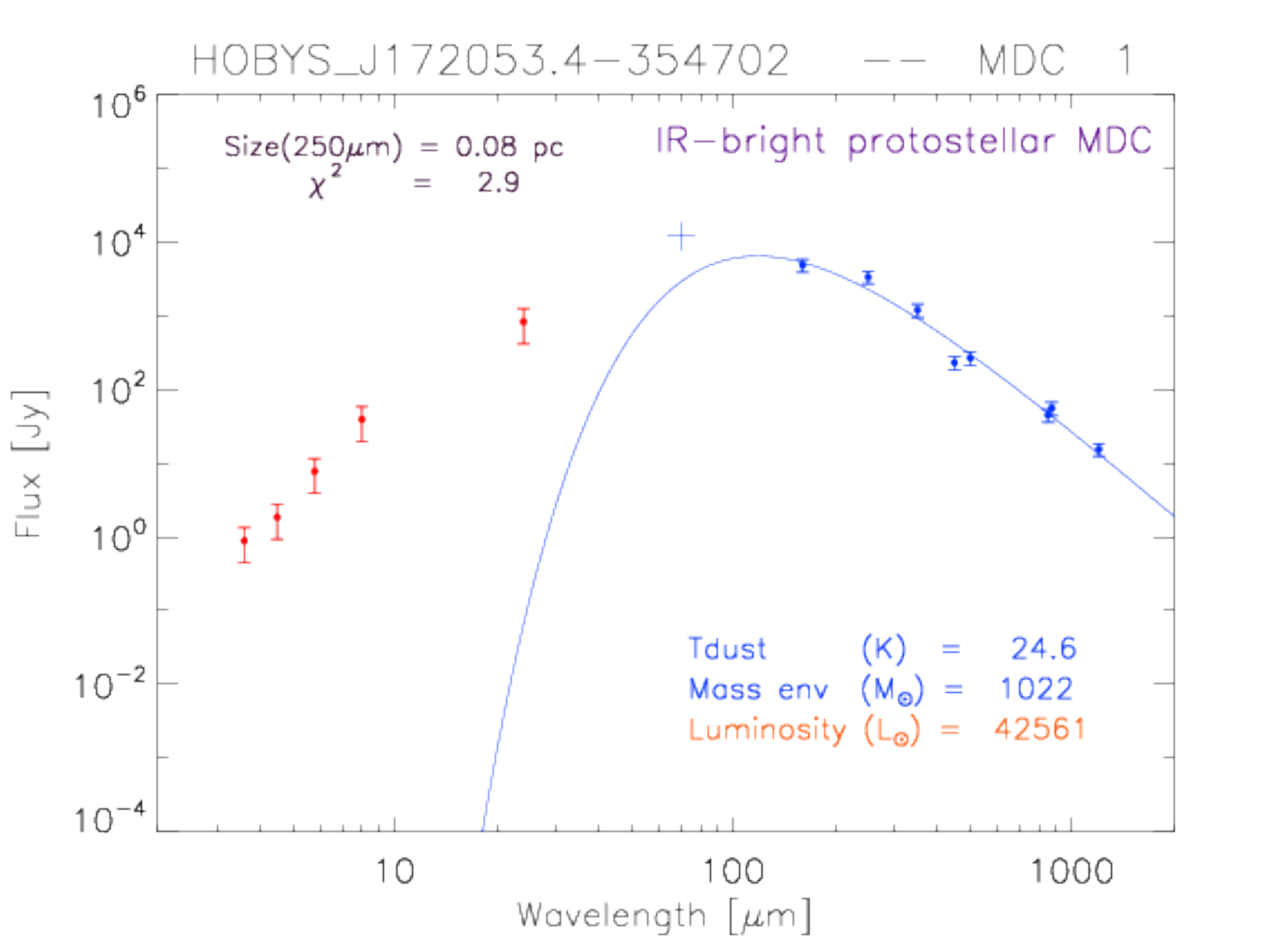}
\vskip -0.3cm
\caption{Maps are: 3.6~$\mu$m, 8~$\mu$m, 24~$\mu$m, 
70~$\mu$m, 160~$\mu$m, 250~$\mu$m, 350~$\mu$m, 
450~$\mu$m, 500~$\mu$m, 850~$\mu$m, 870~$\mu$m, 
1.2~mm and high-resolution N$_{H_{2}}$ column density.
White ellipses represent the \emph{getsources} FWHM integration size measured for $\ge$70~$\mu$m wavelengths
and white numbers are MDC IDs.
Red ellipses show the reference FWHM size (at either 160~$\mu$m or 250~$\mu$m), plotted in $<$70~$\mu$m images or images where source flux is not reliable (see Sect~\ref{section_extraction_getsources}). Dashed ellipses outline extracted sources, whose fluxes are used as upper limits in SEDs. Linear color scales are adjusted in each image to highlight sources of interest and their close surroundings. Large white areas observed at 24~$\mu$m toward MDCs \#1,  \#6, \#7, \#10, \#22,  \#23,  \#25, \#32, \#38, and \#41 correspond to regions of flux saturation with \emph{Spitzer} (see also Table~\ref{tab_21_22_24}).
SED: Flux density versus $\lambda$: Blue fit and fluxes are for the MDC envelope 
and red fluxes are associated IR sources. Crosses at 70~$\mu$m indicate extracted 
70~$\mu$m fluxes not used in the fitting procedure.}
\label{MDC_1}
\end{center}
\end{figure*}

\end{document}